\documentclass[%
 reprint,
 amsmath,amssymb,
 aps,prd,nofootinbib,showpacs,twocolumn,
 mathrsfs,amsfonts,dsfont
]{revtex4-1}
\usepackage{float}
\usepackage{graphicx}
\usepackage{dcolumn}
\usepackage{bm}
\usepackage{wasysym}
\usepackage{verbatim}
\usepackage{color}
\usepackage[linktocpage]{hyperref}
\usepackage{hyperref}
\usepackage{fullpage}
\usepackage{ dsfont }
\usepackage{braket}
\usepackage{scrextend}
\usepackage{wrapfig}

\newcommand{\lsim}{\mathrel{\hbox{\rlap{\lower.55ex\hbox{$\sim$}} \kern-.3em \raise.4ex \hbox{$<$}}}}
\newcommand{\gsim}{\mathrel{\hbox{\rlap{\lower.55ex\hbox{$\sim$}} \kern-.3em \raise.4ex \hbox{$>$}}}}

\newcommand{\mpl}{m_{\mbox{\tiny{Pl}}}}
\newcommand{\Beq}{\begin{equation}\begin{aligned}}
\newcommand{\Eeq}{\end{aligned}\end{equation}}

\newcommand{\bk}{{\textbf{\textit{k}}}}

\makeatletter
\def\l@subsubsection#1#2{}
\makeatother

\begin{document}

\title{Self-resonance after inflation: oscillons, transients and radiation domination}
\author{Kaloian D. Lozanov${}^{1,2}$ and Mustafa A. Amin${}^3$}

\affiliation{${}^1$Max Planck Institute for Astrophysics, Karl-Schwarzschild-Str. 1, 85748 Garching, Germany}

\affiliation{${}^2$Institute of Astronomy, University of Cambridge, CB3 0HA Cambridge, U.K.}

\affiliation{${}^3$Physics \& Astronomy Department, Rice University, Houston, Texas 77005-1827, U.S.A.}

\date{\today}
\begin{abstract}

Homogeneous oscillations of the inflaton after inflation can be unstable to small spatial perturbations even without coupling to other fields. We show that for inflaton potentials $\propto |\phi|^{2n}$ near $|\phi|=0$ and flatter beyond some $|\phi|=M$, the inflaton condensate oscillations can lead to self-resonance, followed by its complete fragmentation. We find that for non-quadratic minima ($n>1$), shortly after backreaction, the equation of state parameter, $w\rightarrow1/3$. If $M\ll \mpl$, radiation domination is established within less than an {\it e}-fold of expansion after the end of inflation. In this case self-resonance is efficient and the condensate fragments into transient, localised spherical objects which are unstable and decay, leaving behind them a virialized field with mean kinetic and gradient energies much greater than the potential energy. This end-state yields $w=1/3$. When $M\sim\mpl$ we observe slow and steady, self-resonace that can last many {\it e}-folds  before backreaction eventually shuts it off, followed by fragmentation and $w\rightarrow 1/3$. We provide analytical estimates for the duration to $w\rightarrow 1/3$ after inflation, which can be used as an upper bound (under certain assumptions) on the duration of the transition between the inflationary and the radiation dominated states of expansion. This upper bound can reduce  uncertainties in CMB observables such as the spectral tilt $n_{\rm{s}}$, and the tensor-to-scalar ratio $r$. For quadratic minima ($n=1$), $w\rightarrow0$ regardless of the value of $M$. This is because when $M\ll\mpl$, long-lived oscillons form within an $e$-fold after inflation, and collectively behave as pressureless dust thereafter. For $M\sim\mpl$, the self-resonance is inefficient and the condensate remains intact (ignoring long-term gravitational clustering) and keeps oscillating about the quadratic minimum, again implying $w=0$. 

\end{abstract}

\maketitle

\tableofcontents

\section{Introduction}

\begin{figure*}[t] 
   \centering
   \includegraphics[width=\textwidth]{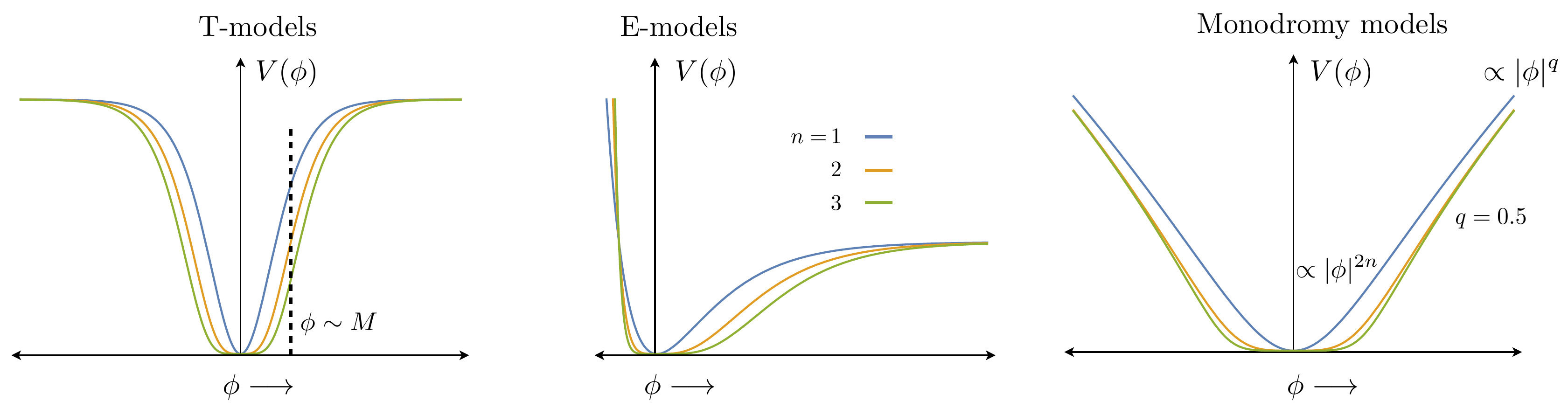}
   \caption{The qualitatively different models used in our analysis. In all cases, the potential behaves as $|\phi|^{2n}$ close to the origin, and changes behavior (flattens at least on one side) for $\phi\gtrsim M$. The T-model and Monodromy models are symmetric about the origin, whereas the E-model is not. In the  T and E-models, the potential asymptotes to a constant for large field values (at least on one side). For the Monodromy models, the potential asymtotes to a general (shallower than quadratic: $q<2$) power law. }
   \label{fig:Models}
\end{figure*}

Observations of the Cosmic Microwave Background (CMB) \cite{Ade:2015lrj,WMAP9} have imposed strong constraints on inflationary cosmology \cite{Starobinsky:1980te,PhysRevD.23.347,LINDE1982389,Albrecht:1982wi}. Simple, single-field driven models of slow-roll inflation with single power-law potentials are disfavored by the data, whereas plateau-like shapes are still allowed \cite{Ade:2015lrj}. Such plateaus (shallower than quadratic power laws) favored during inflation, can be expected to have (quadratic or steeper) power-law minima. See Fig. \ref{fig:Models}.

Inflation must eventually end, leading to the era of reheating where the energy of the inflaton field is (eventually) transferred to Standard Model fields (for reviews, see \cite{Bassett:2005xm,Allahverdi:2010xz,Amin2014}). The end of reheating sets the stage for the production of light elements during Big Bang Nucleosynthesis (BBN) \cite{Wagoner:1966pv,Cyburt:2015mya}. While many models exist for inflation and the generic requirements and implications for reheating have been discussed in the literature, a unique physical model is yet to emerge. 

A number of basic questions, especially regarding the end of inflation, remain model-dependent,\footnote{See \cite{Amin:2015ftc, Amin:2017wvc,Ozsoy:2015rna} for attempts at some model-independent approaches.} including: How efficient was the energy transfer from the inflaton to daughter fields? What were the scale and amplitude of the inhomogeneities generated during reheating? What was the equation of state of the universe following inflation? What was the duration to radiation domination after inflation? What was the duration to thermalization? The answers to these questions are interesting in their own right, and also in the context of how they can potentialy impact other physical processes during this period including dark matter generation \cite{Gelmini:2006pw}, baryogenesis \cite{Allahverdi:2010im} as well as our interpretation of inflationary observables \cite{Martin:2014nya,Munoz:2014eqa,Lozanov:2016hid}. 

The dynamics after inflation can be quite complex and dynamically rich. For example, parametric resonance can play an important role during the early stages of reheating \cite{Kofman:1994rk,Shtanov:1994ce,Kofman1997}, giving rise to copious particle production in fields coupled to the inflaton. In particular, {\it self-resonance}, where the homogeneous inflaton condensate pumps energy into its own fluctuations, can lead to interesting nonlinear effects even in absence of couplings to other fields (e.g.,\cite{Amin:2011hj,Amin:2010xe,Amin:2010dc}). Such explosive particle production, formation of non-topological and topological solitons (e.g.,\cite{Amin:2011hj,Antusch:2017flz}), as well as relics such as black holes (e.g., \cite{Konoplich:1999qq}) and primordial gravitational waves (e.g.,\cite{Khlebnikov:1996mc,Easther:2006gt,Dufaux:2007pt}), amongst others, make reheating an exciting dynamical playground. We find a number of aspects of these rich dynamics even in the simple, single-field models that we consider in this paper.

In this work, we study the post-inflationary evolution of the inflaton field in a class of observationally favored single-field models of inflation. We show that self-resonance can occur as the field oscillates in a potential $V\propto|\phi|^{2n}$, $n\geq1$, near the origin (in the ``bowl" of the potential) and flatter away from it (for $|\phi|>M$, in the ``wings" of the potential). See Fig. \ref{fig:Models}. The dynamics can be dramatically different depending on whether $M\sim \mpl$ or $M\ll \mpl$. Along with $M$, whether $n=1$ or $n>1$ also makes crucial differences to the qualitative and quantitative dynamics of inflaton field (the impatient reader can refer directly to Fig. \ref{fig:ST} in the Conclusions section).

We carefully examine the nature of self-resonance and duration to backreaction using a linear analysis of growth of inflaton peturbations in an expanding background. We confirm the results of this analysis using detailed 3+1 dimensional simulations. Using the same simulations we investigate the nonlinear field dynamics including the formation of long and short-lived localized field configurations, their fragmentation and the virialization of the field.

A better understanding of reheating can influence the predictions for CMB observables from inflation. Specifically, the uncertainties in the effective equation of state after inflation ($w_{\rm{int}}$), that determines the expansion of the universe, and the duration to radiation domination ($\Delta N_{\rm rad}$), are transferred to the predictions of inflationary models for CMB observables such as the spectral tilt, $n_{\rm s}$, and the tensor-to-scalar ratio, $r$.\footnote{An alternative and equally interesting approach is to constrain $w_{\rm{int}}$, $\Delta N_{\rm{rad}}$ and the reheating temperature, $T_{\rm{th}}$, by using observational bounds on $n_{\rm{s}}$ and $r$ \cite{Martin:2010kz,Mielczarek:2010ag,Creminelli:2014oaa,Dai:2014jja,Munoz:2014eqa,Creminelli:2014fca,Martin:2014nya,Cook:2015vqa,Eshaghi:2016kne,Cai:2015soa,Ueno:2016dim}. We shall not follow this second approach here in detail but do provide an appendix related to this approach.}

Given the importance of $w_{\rm int}$ and $\Delta N_{\rm rad}$, we calculate these quantitates from our full nonlinear simulations and provide $\Delta N_{\rm rad}$ as a function of the parameters of the model. Note that while the equation of state for homogeneous oscillating fields has been known for a long time \cite{Turner:1983he}, limited work exists in the literature on characterizing the nonlinear equation of state in a systematic fashion for fully fragmented fields \cite{Podolsky:2005bw, Lozanov:2016hid}. Moreover the expression for $\Delta N_{\rm rad}$ we provide can serve as an upper bound on the duration to radiation domination (at least under the assumption of perturbative decay of the inflaton to other relativistic fields). It is this upper bound that can significantly reduce uncertainties in inflationary observables.

This paper is partly a detailed follow-up to our recently published shorter paper \cite{Lozanov:2016hid} in {\it Physical Review Letters}, but includes new results regarding the nonlinear field dynamics. For example, the larger class of models considered, the discussion on the formation of transients, a significantly more detailed analysis of both the linear fluctuations and the nonlinear power spectra go significantly beyond the PRL. Furthermore, this paper will be followed by another paper \cite{Lozanov:2017b}, where the {\it passive} gravitational effects from self-resonance will be addressed. In \cite{Lozanov:2017b} we will discuss the likelihood of primordial blackhole formation (from oscillons and transients), as well as the expected frequency and amplitude of primordial gravitational waves from inflaton fragmentation.

The paper is organized as follows. In Sec. \ref{sec:Models} we briefly review the models we study in this paper. Then, in Sec. \ref{sec:LinAnal}, we move on to self-resonance after the end of inflation focussing on the growth of perturbations and the beginning of backreaction. We present our numerical studies of the nonlinear dynamics in Sec. \ref{sec:NonLinDyn}, including the evolution of the equation of state. The implications of our investigations for CMB observables and the reheating temperature are given in Sec. \ref{sec:ObsImpl}. In Sec. \ref{sec:Discussion}, we discuss some relevant quality checks on our numerics, as well as some underlying theoretical assumptions, approximations and caveats. We conclude in Sec. \ref{sec:Concl}, with a summary of our results. An additional appendix regarding the reheating temperature is provided at the end of the manuscript.

We use natural units, where $\hbar=c=1$, the reduced Planck mass $\mpl=1/\sqrt{8\pi G}$ and $+---$ signature for the metric. The background FRW metric is assumed to have the form $ds^2=dt^2-a^2(t)d{\bf x}\cdot d{\bf x}$ where $a(t)$ is the scale factor.  


\section{Models}
\label{sec:Models}
The main focus of this paper is on single-field models of inflation, minimally coupled to gravity. The action, and equations of motion are
\Beq
\label{eq:ActionEoM}
&S=\int d^4x \sqrt{-g}\left[-\frac{\mpl^2}{2}R+\frac{1}{2}\partial_\mu\phi\partial^\mu\phi+V(\phi)\right]\,,\\
&\nabla^\mu\nabla_\mu \phi+V'(\phi)=0\,,\\
&G^\mu_\nu=\frac{1}{\mpl^2}\left[\partial^\mu\phi\partial_\nu\phi-\delta^\mu_\nu\left(\frac{1}{2}\partial_\gamma\phi\partial^\gamma\phi-V(\phi)\right)\right]\,.
\Eeq
where $R$ is the Ricci scalar, $\nabla_\mu$ is the covariant derivative, $g$ is the determinant of the metric and $G^\mu_\nu$ is the Einstein tensor. Couplings to other fields will be introduced as needed at a later time. 

Our focus will be on inflaton potentials $V(\phi)$ that have an observationally favored ``plateau" or shallower than quadratic region at large field values (see Fig. \ref{fig:Models}). We consider three different parametrizations: the $\alpha$-attractor T-models \cite{Kallosh:2013hoa}
\Beq
\label{eq:PotentialT}
V(\phi)&=\Lambda^4\tanh^{2n}\left({\dfrac{|\phi|}{M}}\right)\,,\\
&=\begin{cases}
                    \Lambda^4 \left|\dfrac{\phi}{M}\right|^{2n}\,&|\phi|\ll M\,,\\
                     \Lambda^4\,  &  |\phi|\gg M\,,
                   \end{cases}
\Eeq 
the $\alpha$-attractor E-models \cite{Kallosh:2013hoa}
\Beq
\label{eq:PotentialE}
V(\phi)&=\Lambda^4\left|1-\exp\left(-\frac{2\phi}{M}\right)\right|^{2n}\,,\\
&=\begin{cases}
                    \Lambda^4 \left|\dfrac{2\phi}{M}\right|^{2n}\,&|\phi|\ll M\,,\\
                     \Lambda^4\,  &  \phi\gg M\,,
                   \end{cases}
\Eeq 
and Monodromy type potentials \cite{Silverstein:2008sg,McAllister:2014mpa}
\Beq
\label{eq:PotentialMon}
V(\phi)&=\Lambda^4\left[\left(1+\left|\frac{\phi}{M}\right|^{2n}\right)^{\frac{q}{2n}}-1\right]\,\\
&=\begin{cases}
                    \Lambda^4 \dfrac{q}{2n}\left|\dfrac{\phi}{M}\right|^{2n}\,&|\phi|\ll M\,,\\
                     \Lambda^4\left|\dfrac{\phi}{M}\right|^{q}\,  &  |\phi|\gg M\,.
                   \end{cases}
\Eeq 
We shall consider values of the parameter $n\geq1$, for which $\partial_\phi V$ and $\partial_\phi^2 V$ are well-defined when $\phi=0$. That is, the elementary quanta of the inflaton always have a well-defined mass.\footnote{See \cite{Johnson:2008se} for the stability analysis of some models with $n<1$.} The mass scale in eqs. (\ref{eq:PotentialT},\ref{eq:PotentialE}) is related to the conventional $\alpha$-parameter \cite{Kallosh:2013xya,Kallosh:2013hoa,Kallosh:2013yoa,Kallosh:2013wya,Kallosh:2013maa,Galante:2014ifa,Linde:2015uga,Carrasco:2015rva,Carrasco:2015pla,Roest:2015qya,Scalisi:2015qga,Kallosh:2016gqp} through 
\Beq
M=\sqrt{6\alpha}\mpl\,,
\Eeq whereas the additional parameter in eq. \eqref{eq:PotentialMon}, $q$, should not exceed $1$ to be consistent with CMB measurements \cite{Ade:2015lrj}. 

Our models are characterized by $n,M,\Lambda$ (and $q$ for Monodromy models). We can take advantage of the observations of the CMB \cite{Ade:2015lrj} and eliminate some of these parameters. For example, using the amplitude of the scalar perturbations, and spectral tilt, we can eliminate $\Lambda$ (assuming slow-roll inflation). This is purely for convenience to reduce the number of parameters to be varied in our analysis. We will return back to a careful consideration of the interplay between observational constraints and model parameters of the potential in Sec. \ref{sec:ObsImpl}. The most recent constraints on the tensor-to-scalar ratio, $r(k_\star=0.002\,{\rm Mpc}^{-1})\lesssim 0.1$ \cite{Ade:2015lrj}, bound $M$ from above, $M\lesssim 10\mpl$. We shall use this constraint on $M$ when we vary our parameters.\footnote{Embedding in Supergravity of the T and E models has stability issues for $M<\sqrt{2}\mpl$ \cite{Kallosh:2013yoa}, which in principle narrows the parameter space significantly. However, we shall ignore this lower bound and consider inflation with $V$ as given in eqs. \eqref{eq:PotentialT}, \eqref{eq:PotentialE} and  \eqref{eq:PotentialMon} at the phenomenological level, and take $M<10\mpl$ as the only constraint.}\footnote{We have chosen to include the additional factor of $2$ in the parametrization in eq. \eqref{eq:PotentialE} so that for a given $M$, we get the same slow-roll parameters (see Sec. \ref{sec:ObsImpl}) for the T and E models.\footnote{It could be seen most easily from ${\rm{lim}}_{x\gg1}\tanh^{2n}{(x)}=(1-e^{-2x})^{2n}$.} However, the behavior after the end of slow-roll inflation is affected by this parametrization --  it depends on the transition scale in the potential which is $M$ for the T and $M/2$ for the E models.}

\section{Particle Production: Linear Analysis}
\label{sec:LinAnal}
\begin{figure*}[t!] 
   \centering
   \hspace{-0.15in}   
   \raisebox{-0.5\height}{\includegraphics[height=1.33in]{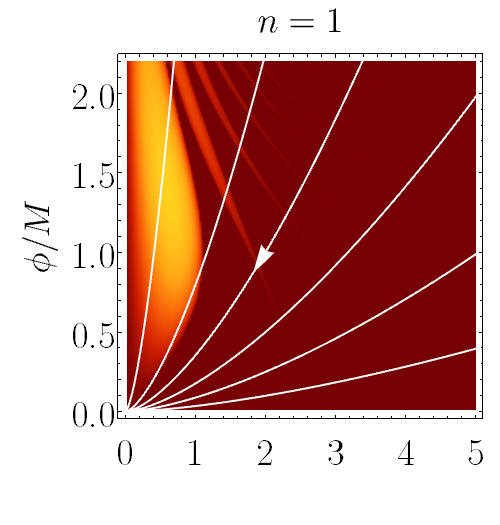}} 
   \hspace{-0.1in}
   \raisebox{-0.41\height}{\includegraphics[width=0.4in]{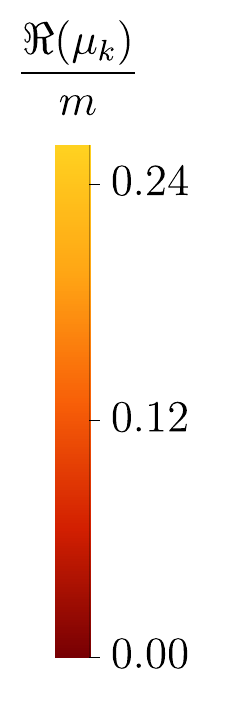}}
   \hspace{-0.11in}
   \raisebox{-0.5\height}{\includegraphics[height=1.33in]{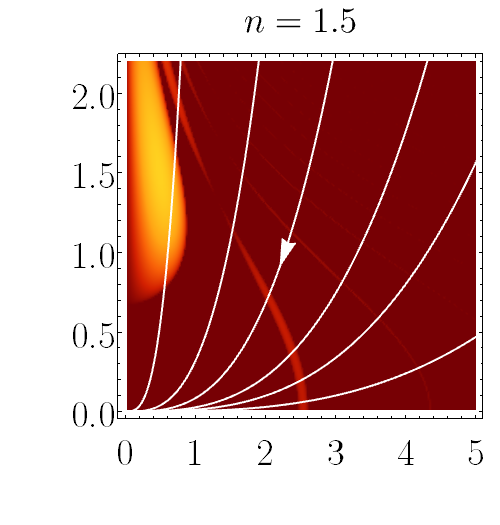}}
   \hspace{-0.1in}
   \raisebox{-0.41\height}{\includegraphics[width=0.4in]{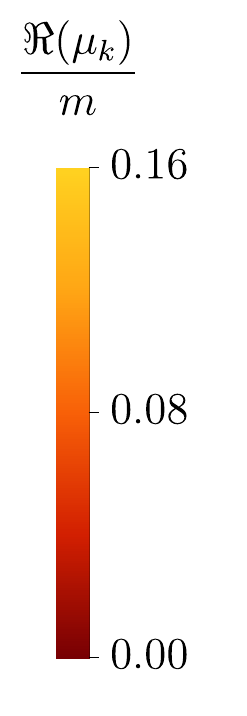}}
   \hspace{-0.11in}
   \raisebox{-0.5\height}{\includegraphics[height=1.33in]{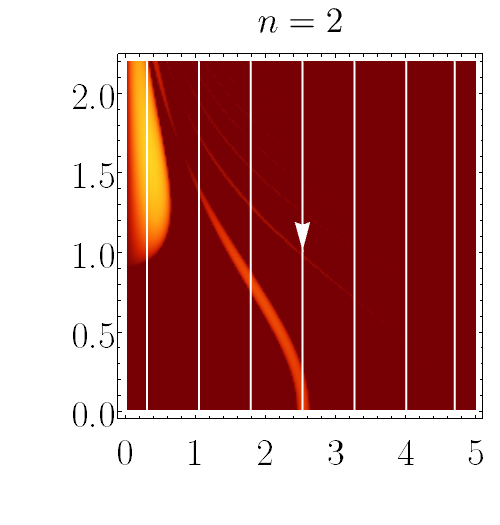}}
   \hspace{-0.1in}
   \raisebox{-0.41\height}{\includegraphics[width=0.4in]{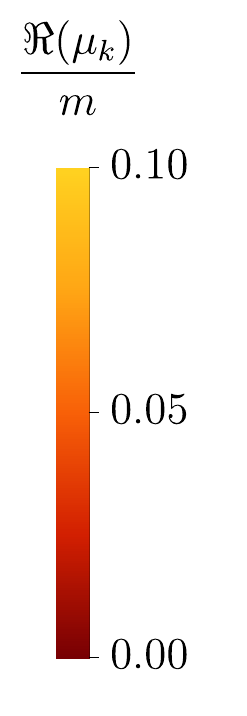}} 
   \hspace{-0.11in}
   \raisebox{-0.5\height}{\includegraphics[height=1.33in]{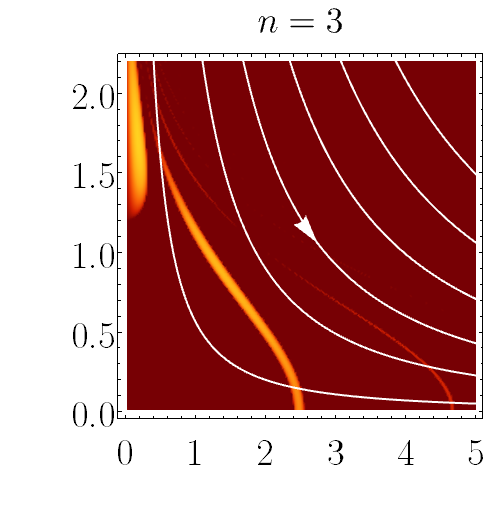}} 
   \hspace{-0.1in}
   \raisebox{-0.41\height}{\includegraphics[width=0.4in]{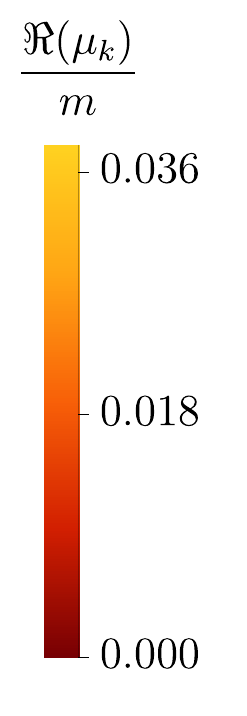}}\\
   \hspace{-0.15in}   
   \raisebox{-0.5\height}{\includegraphics[height=1.33in]{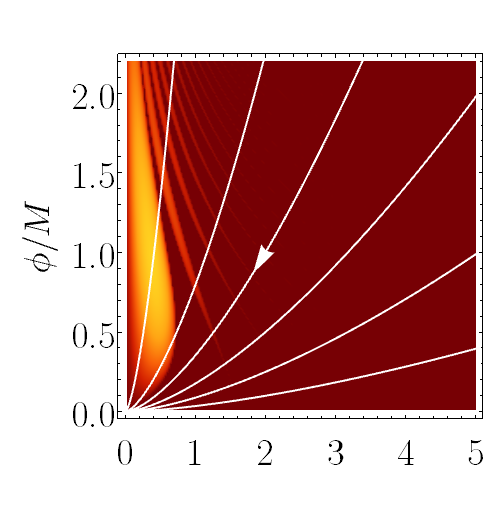}} 
   \hspace{-0.1in}
   \raisebox{-0.41\height}{\includegraphics[width=0.4in]{LegMon1_PRL_AR.pdf}}
   \hspace{-0.1in}
   \raisebox{-0.5\height}{\includegraphics[ height=1.33in]{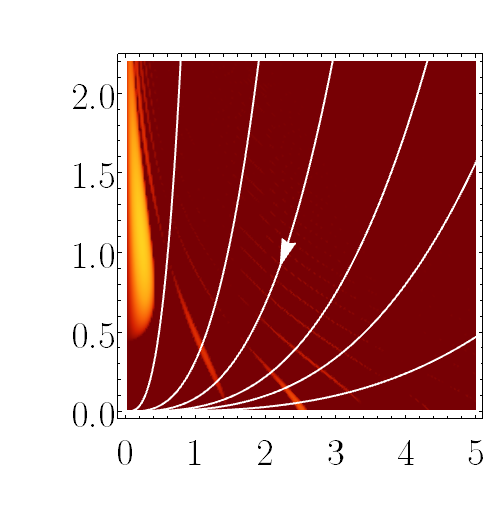}}
   \hspace{-0.1in}
   \raisebox{-0.41\height}{\includegraphics[width=0.4in]{LegMon15_PRL_AR.pdf}}
   \hspace{-0.11in}
   \raisebox{-0.5\height}{\includegraphics[height=1.33in]{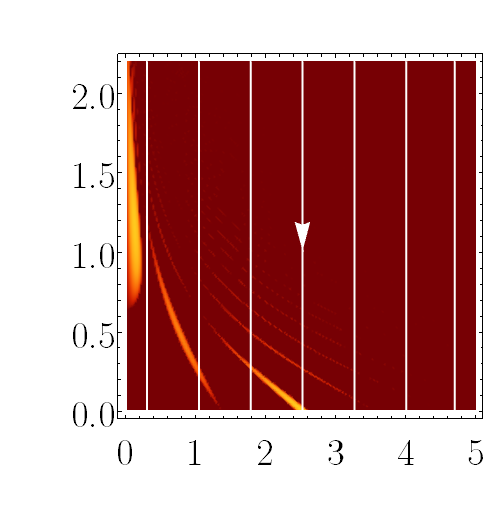}}
   \hspace{-0.1in}
   \raisebox{-0.41\height}{\includegraphics[width=0.4in]{LegMon2_PRL_AR.pdf}} 
   \hspace{-0.11in}
   \raisebox{-0.5\height}{\includegraphics[height=1.33in]{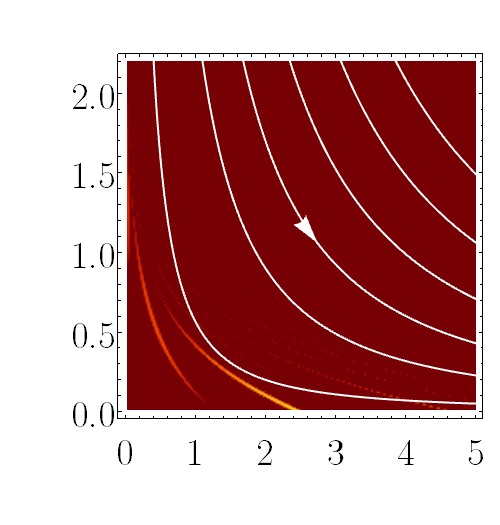}} 
   \hspace{-0.1in}
   \raisebox{-0.41\height}{\includegraphics[width=0.4in]{LegMon3_PRL_AR.pdf}}\\
   \hspace{-0.15in}   
   \raisebox{-0.5\height}{\includegraphics[height=1.33in]{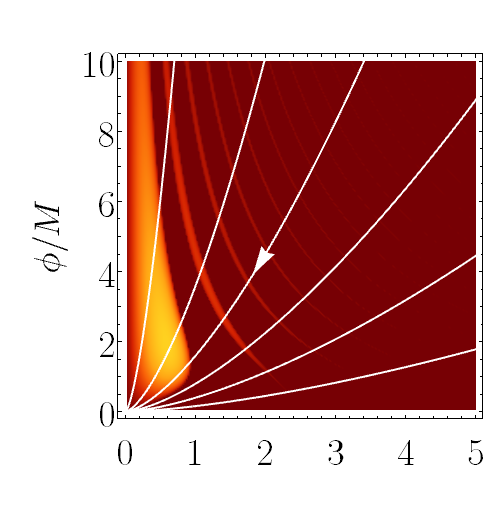}} 
   \hspace{-0.1in}
   \raisebox{-0.41\height}{\includegraphics[width=0.4in]{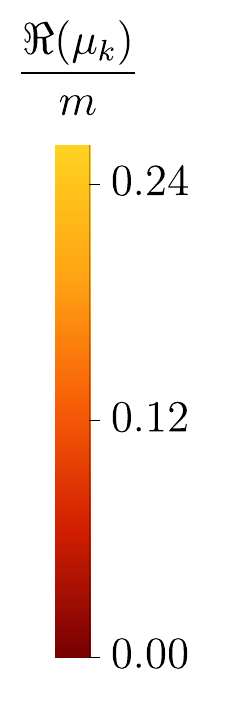}}
   \hspace{-0.11in}
   \raisebox{-0.5\height}{\includegraphics[height=1.33in]{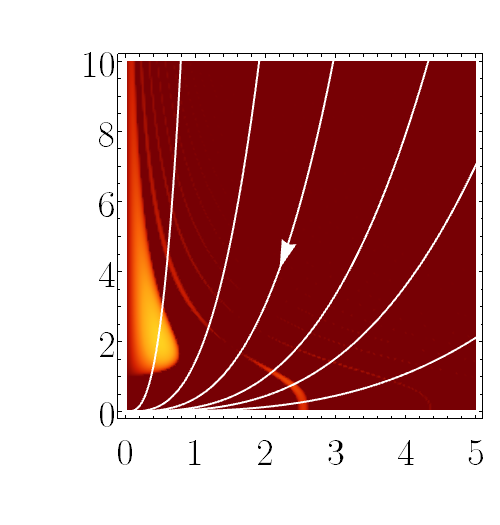}}
   \hspace{-0.1in}
   \raisebox{-0.41\height}{\includegraphics[width=0.4in]{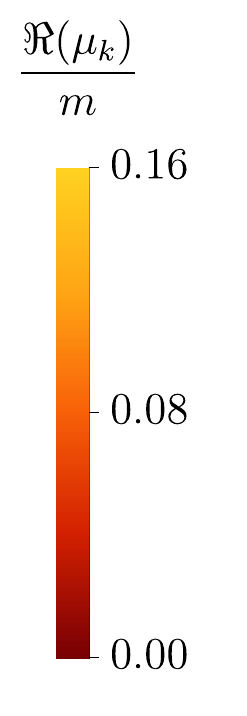}}
   \hspace{-0.11in}
   \raisebox{-0.5\height}{\includegraphics[height=1.33in]{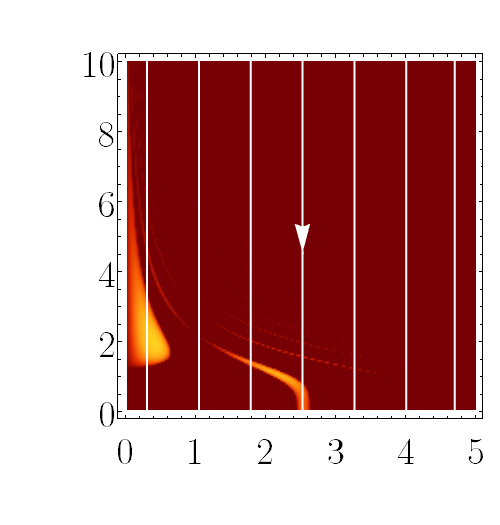}}
   \hspace{-0.1in}
   \raisebox{-0.41\height}{\includegraphics[width=0.4in]{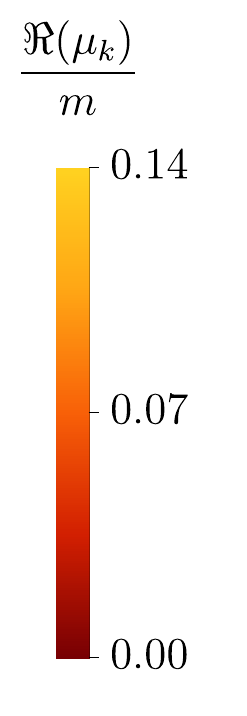}} 
   \hspace{-0.11in}
   \raisebox{-0.5\height}{\includegraphics[height=1.33in]{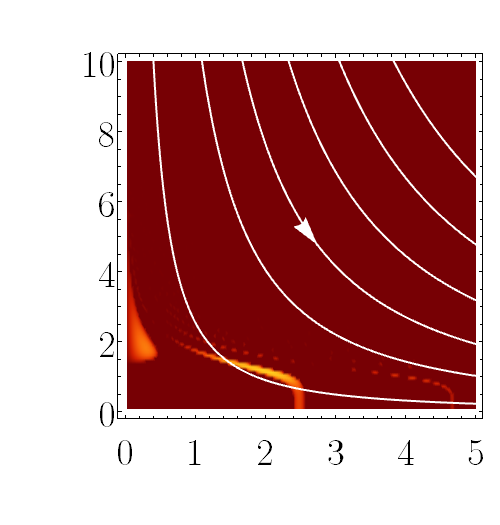}} 
   \hspace{-0.1in}
   \raisebox{-0.41\height}{\includegraphics[width=0.4in]{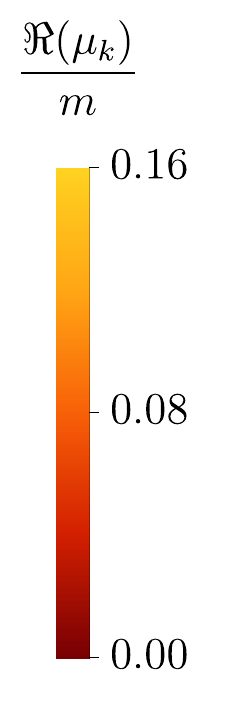}}\\
   \hspace{-0.15in}   
   \raisebox{-0.5\height}{\includegraphics[height=1.33in]{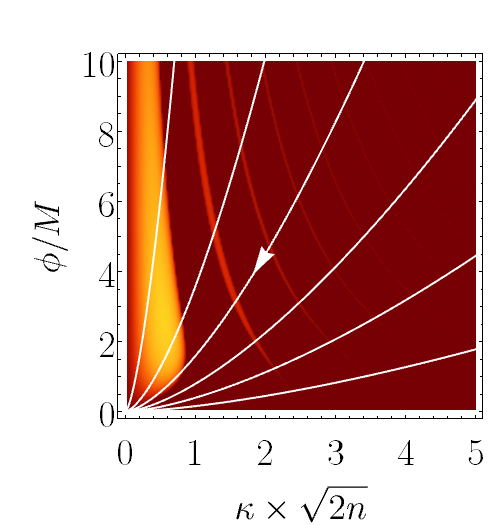}} 
   \hspace{-0.1in}
   \raisebox{-0.41\height}{\includegraphics[width=0.4in]{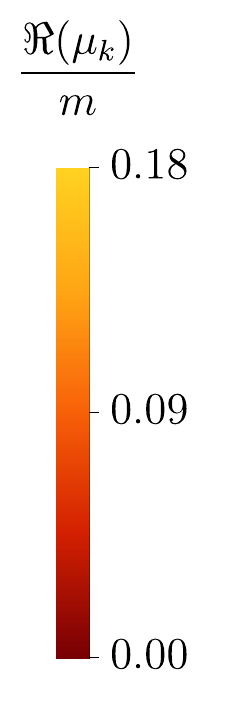}}
   \hspace{-0.11in}
   \raisebox{-0.5\height}{\includegraphics[height=1.33in]{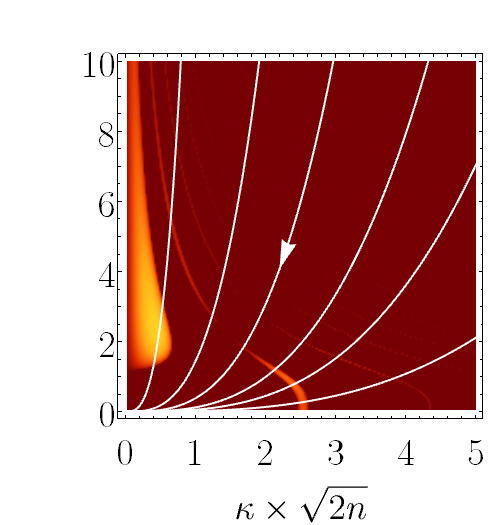}}
   \hspace{-0.1in}
   \raisebox{-0.41\height}{\includegraphics[width=0.4in]{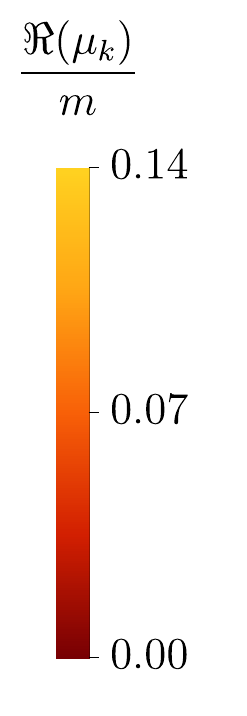}}
   \hspace{-0.11in}
   \raisebox{-0.5\height}{\includegraphics[height=1.33in]{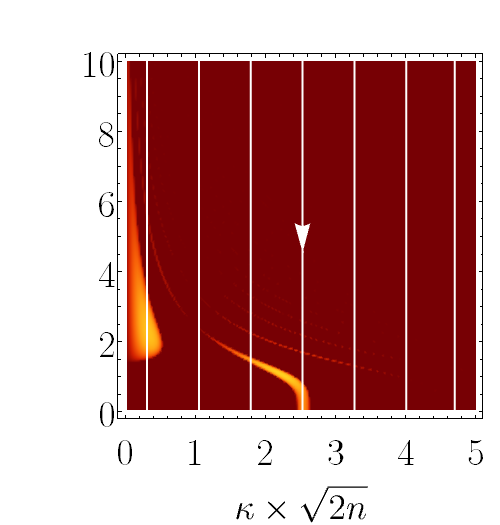}}
   \hspace{-0.1in}
   \raisebox{-0.41\height}{\includegraphics[width=0.4in]{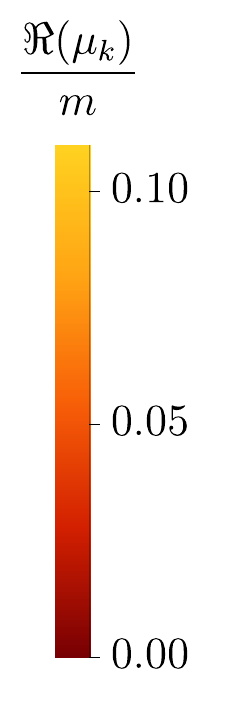}} 
   \hspace{-0.11in}
   \raisebox{-0.5\height}{\includegraphics[height=1.33in]{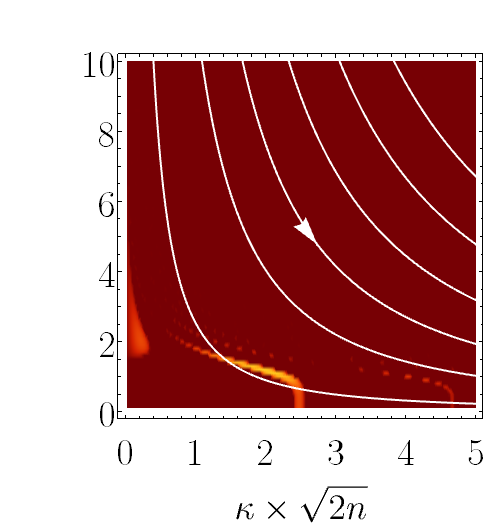}} 
   \hspace{-0.1in}
   \raisebox{-0.41\height}{\includegraphics[width=0.4in]{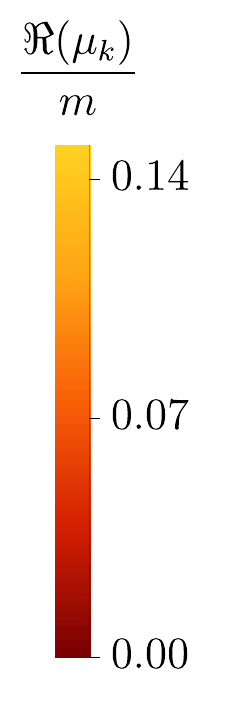}}\\
   \caption{The instability regions and Floquet exponents for the T (first row), E (second row), Monodromy $q=0.5$ (third row) and $q=1$ (fourth row) models. On the horizontal axis is the dimensionless physical wavenumber $\kappa=k/(am)$ and on the vertical axis the amplitude of inflaton oscillations. The effective mass, $m$, is defined in eq. \eqref{eq:EffMass}, and determines the characteristic frequency of oscillations. As the universe expands, a given co-moving mode $k$ flows across the chart as the white lines indicate. The factor of $\sqrt{2n}$ on the horizontal axis is chosen to make the narrow instability bands appear at roughly the same place for different $n$, see also Fig. \ref{fig:FloqPowerLaws}. Although, the broad low-momentum instability bands seem to vanish for large $n$, they never go away. It can be shown by a different rescaling of the horizontal axis. In the T and E models slow-roll inflation ends at $\phi_{\rm{end}}\sim M$ and the amplitude of inflaton oscillations lies in the range $\phi<M$. In the Monodromy models $\phi_{\rm{end}}\lesssim \mpl$ and the initial amplitude of inflaton oscillations can exceed $M$.}
   \label{fig:Floq}
\end{figure*}

\begin{figure*}[t!] 
   \centering
   \hspace{-0.15in}   
   \raisebox{-0.5\height}{\includegraphics[height=1.33in]{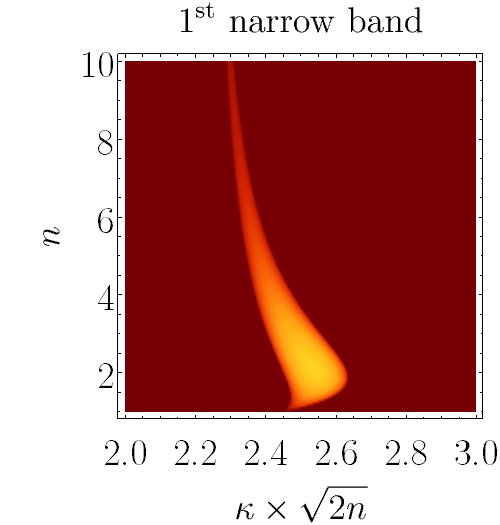}} 
   \hspace{-0.1in}
   \raisebox{-0.45\height}{\includegraphics[width=0.4in]{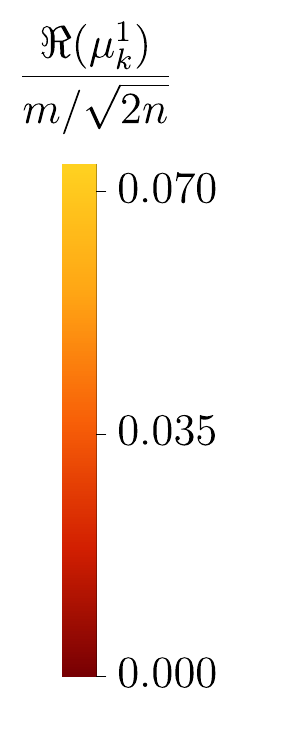}}
   \hspace{-0.11in}
   \raisebox{-0.5\height}{\includegraphics[height=1.33in]{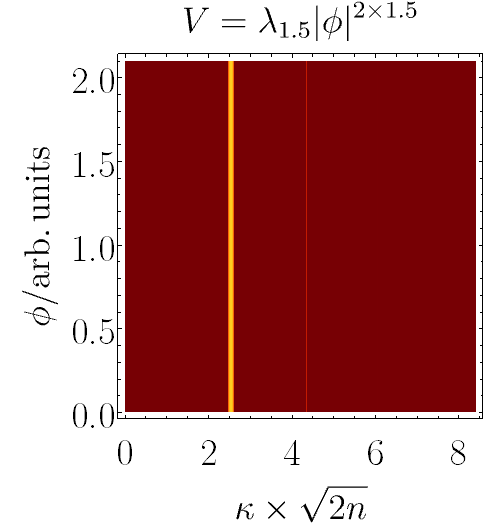}}
   \hspace{-0.1in}
   \raisebox{-0.45\height}{\includegraphics[width=0.4in]{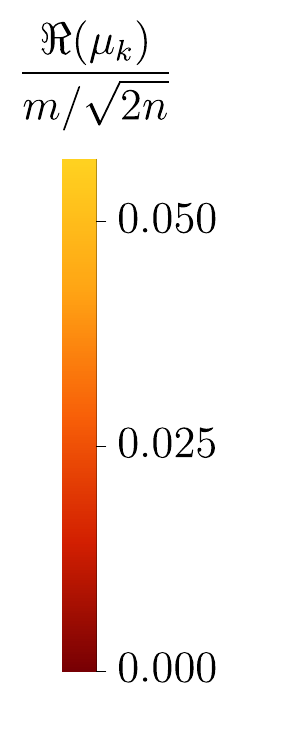}}
   \hspace{-0.11in}
   \raisebox{-0.5\height}{\includegraphics[height=1.33in]{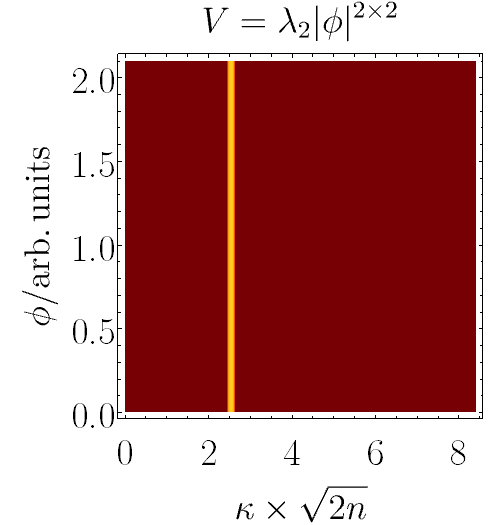}}
   \hspace{-0.1in}
   \raisebox{-0.45\height}{\includegraphics[width=0.4in]{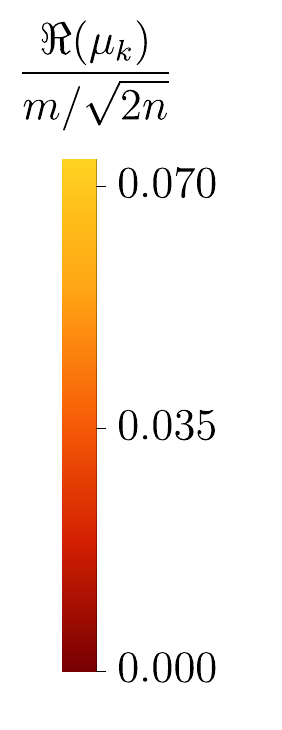}} 
   \hspace{-0.11in}
   \raisebox{-0.5\height}{\includegraphics[height=1.33in]{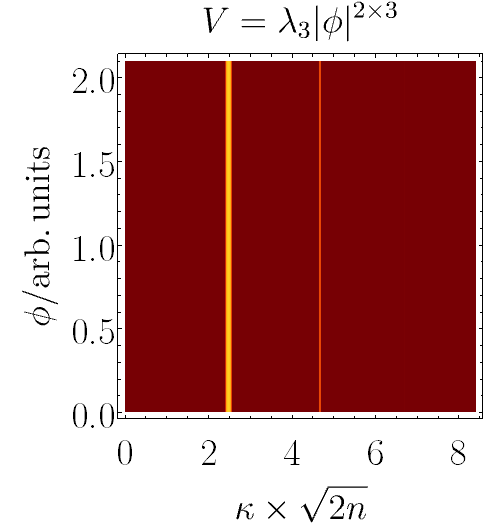}} 
   \hspace{-0.1in}
   \raisebox{-0.45\height}{\includegraphics[width=0.4in]{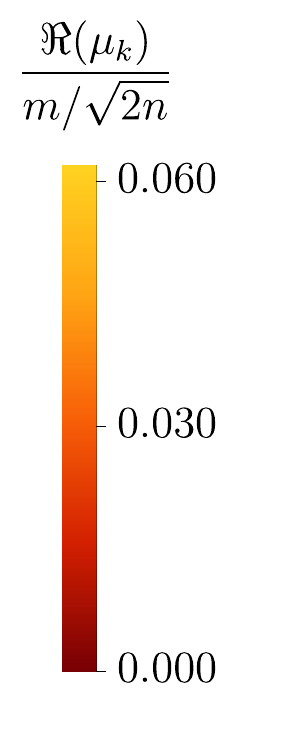}}\\
   \caption{The first panel depicts the first narrow instability band for a monomial potential of the form $V=\lambda_{n}|\phi|^{2n}$(consistent with T, E and Monodromy models when $\phi\ll M$) as a function of $n$ and $\kappa$. The next three panels give the Floquet charts for $n=1.5$, $2$, $3$. Thanks to our definition of  $\kappa\equiv k/am$, with $m$ being the effective mass, the instability bands are exactly vertical and are independent of the amplitude of inflaton oscillations (the values on the vertical axis). The additional rescaling by $\sqrt{2n}$ makes the first narrow instability band appear at roughly the same place and have roughly the same height, both of which are maximal for $n=2$, see also Fig. \ref{fig:1stband}. For $n=1.5$ and $3$ we have a series of decreasing in height and width instability bands for larger $\kappa$. Interestingly, the second, third, etc. narrow instability bands vanish for $n=2$.}
   \label{fig:FloqPowerLaws}
\end{figure*}
For small perturbations around a homogeneous condensate $\phi(t,{\bf x})=\bar{\phi}(t)+\delta\phi(t,{\bf x})$. After inflation $\bar{\phi}(t)$ begins to oscillate around the minimum of $V$. This may lead to non-adiabatic production of inflaton particles of definite co-moving momentum, a.k.a. preheating \cite{Kofman1997}. It can be most easily understood in terms of the linearized field equations for the fluctuations in Fourier space (see eq. \eqref{eq:ActionEoM}):
\Beq
\label{eq:linearEoM}
\partial_t^2\delta\phi_\bk+\left[k^2+\partial_{\bar{\phi}}^2V(\bar{\phi})\right]\delta\phi_\bk=0\,,
\Eeq
where we have ignored expansion as well as metric perturbations for simplicity. We shall reintroduce expansion shortly, but continue to ignore metric perturbations in this section. Since $\partial_t^2\bar{\phi}+\partial_{\bar{\phi}}V=0$, $\partial_{\bar{\phi}}^2V$ is a periodic function of time. Floquet theory \cite{magnus2004hill} tells us that the general solution to eq. \eqref{eq:linearEoM} is of the form
\Beq
\delta\phi_k=\mathcal{P}_{k+}(t)\exp(\mu_{k}t)+\mathcal{P}_{k-}(t)\exp(-\mu_{k}t)\,.
\Eeq
$\mathcal{P}_{k\pm}(t)$ are also periodic functions and are determined by the initial conditions. $\mu_k$ are the Floquet exponents. If $\Re(\mu_k)\neq0$, then there is an `unstable' solution, exponentially growing with time which is a manifestation of non-adiabatic (or resonant) particle production. On a plane with the amplitude of the oscillations on one axis, and the wavenumber on another, sharp boundaries exist between the stable and unstable regions. The unstable regions are often referred to as instability bands.

To better understand the dynamics in the stable and unstable regions, let us define an effective mass:
\Beq
\label{eq:EffMass}
m^2&\equiv\begin{cases}
                    2n\Lambda^2\left(\frac{\Lambda}{M}\right)^{\!\!2}\left(\frac{\bar{\phi}}{M}\right)^{\!\!2(n-1)}\,&{\text{T}}\,,\\
                    2^{2n+1}n\Lambda^2\left(\frac{\Lambda}{M}\right)^{\!\!2}\left(\frac{\bar{\phi}}{M}\right)^{\!\!2(n-1)}\,  & \text{E} \,, \\
                    q\Lambda^2\left(\frac{\Lambda}{M}\right)^{\!\!2}\left(\frac{\bar{\phi}}{M}\right)^{\!\!2(n-1)}\,&{\text{Monodromy}}\,,\\
                   \end{cases}
\Eeq
which is what $\partial_{\bar{\phi}}V/\bar{\phi}$ tends to when $\bar{\phi}\ll M$ and is what sets the period of $\bar{\phi}$. We have also defined a dimensionless physical wavenumber $\kappa\equiv k/m$. In Fig. \ref{fig:Floq}, we show the instability regions for the inflaton potentials from eqs. \eqref{eq:PotentialT}, \eqref{eq:PotentialE} and \eqref{eq:PotentialMon}, for $n=1$, $1.5$, $2$, $3$ for the T, E-models and Monodromy models (for $q=0.5$, $1$).

The $n=1$ case features a broad low-$\kappa$ instability region going all the way down to $\bar{\phi}=0$ and a series of high-$\kappa$ narrow bands, vanishing towards the bottom of the plot. For $n=1$, this is common for all potentials that  flatten below quadratic  away from the minimum ($q<2$, \cite{Amin:2011hj}). For $n>1$ the broad low-$\kappa$ band is absent for $\bar{\phi}\lesssim M$. The narrow bands near the bottom of the charts are reminiscent of those for $V\propto|\phi|^{2n}$, see Fig. \ref{fig:FloqPowerLaws}.  

Let us include the effects of expansion qualitatively. First, note that when we include expansion, $\kappa\equiv k/(am)$ where $a$ is the scale factor. Also recall the general result for the amplitude decay of the inflaton field oscillating in $V\propto|\phi|^{2n}$, $\bar{\phi}\propto a^{-3/(n+1)}$. Hence a given Fourier mode, $k$, flows through a number of Floquet bands as shown in Fig. \ref{fig:Floq} (see the white ``flow lines"). The mode will grow if the expansion rate, $H\equiv\dot{a}/a$, is much less than $|\Re(\mu_{k})|$. Empirically, strong resonance occurs for $|\Re(\mu_{k})|/H\gsim 10$.
\subsection{Instability Bands}
\label{ssec:IB} 
We divide the instability bands into a broad-$\kappa$ band near the $\kappa=0$ axis and the narrow resonance bands away from the $\kappa=0$ axis. We investigate these two classes of bands below.
\subsubsection{Broad-$\kappa$ band}
 Focus on the band hugging the $\kappa=0$ axis. For this broad, low-$\kappa$ band, we find $[|\Re(\mu_{k})|/H]_{\text{max}}=f(n)\mpl/M$, where $f(n)$ is of order unity. Hence, the expansion of the universe allows for broad self-resonance only for $M\ll\mpl$ for all $n$.
\subsubsection{Narrow-$\kappa$ bands}
\label{sssec:NKB}
We now focus on bands which do not hug the $\kappa=0$ axis. For $M\gsim\mpl$, the amplitude of inflaton oscillations is rapidly redshifted by the expansion of the universe to $\bar{\phi}\ll M$ region, i.e., to the bottom of the Floquet charts.\footnote{Recall that the decay rate of the amplitude of $\bar{\phi}$ is set by $H$, whereas the typical frequency of oscillations is $m$. Since $H/m\sim\bar{\phi}/\mpl\sim M/\mpl$ at the end of inflation, then $M\sim\mpl$ leads to a rapid decay of the amplitude, within few oscillations.} As seen from the top-left panel in Fig. \ref{fig:Floq}, there are no narrow instability bands for $n=1$ (i.e., they get infinitesimally narrow at small background amplitudes). Hence, we do not expect significant particle production. This is anticipated since the condensate is oscillating about a quadratic minimum, and behaves as a free scalar. However, for $n>1$, we have a series of narrow resonance bands as a consequence of the intrinsic nonlinearity. They decrease in height and width for higher $\kappa$, see Figs. \ref{fig:Floq} and \ref{fig:FloqPowerLaws} (which focusses on the first narrow instability band). Hence, the first narrow band plays a dominant role in particle production. 

The particle production can be understood in terms of the white flow lines in Fig. \ref{fig:Floq}. The flow lines cross the first narrow band from right to left ($n<2$), left to right ($n>2$) or never leave it ($n=2$). While it is obvious that the narrow resonance will persist until nonlinear effects become important in the $n=2$ case, after a closer look one can argue that the same holds for $n<2$ and $n>2$. In these two cases 
\Beq
|\dot{\kappa}|\approx \frac{|4-2n|}{n+1} H\kappa\,,
\Eeq
 and since $H$ is decreasing, at some point a given $k$-mode will spend enough time within the first narrow band for self-resonance to become efficient. 
 \subsection{Backreaction Time}
The linear growth of perturbations eventually leads to backreaction onto the condensate. For the case when $M\ll\mpl$, since $[\Re(\mu_k)/H]_{\rm max}\gg 1$, backreaction time is short (within 1-2 $e$-folds after inflation). The case with $M\sim \mpl$ and $n=1$ has no-backreaction. Hence, we are left with $M\sim \mpl$ and $n>1$ case to investigate further. It is in this case that the first narrow instability band discussed above is relevant.
\begin{figure}[t] 
   \centering
   \includegraphics[width=3.3in]{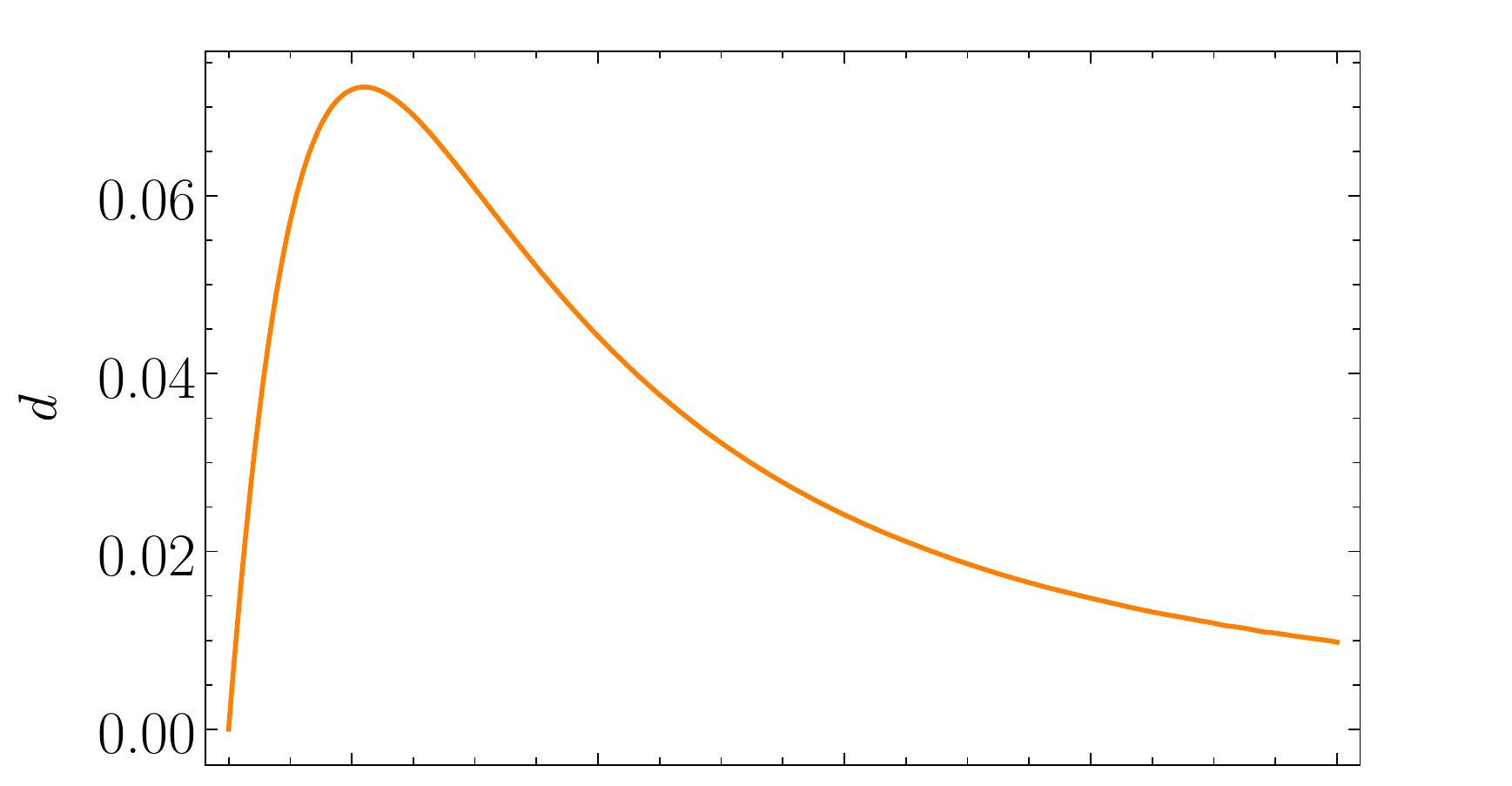}\\
   \includegraphics[width=3.3in]{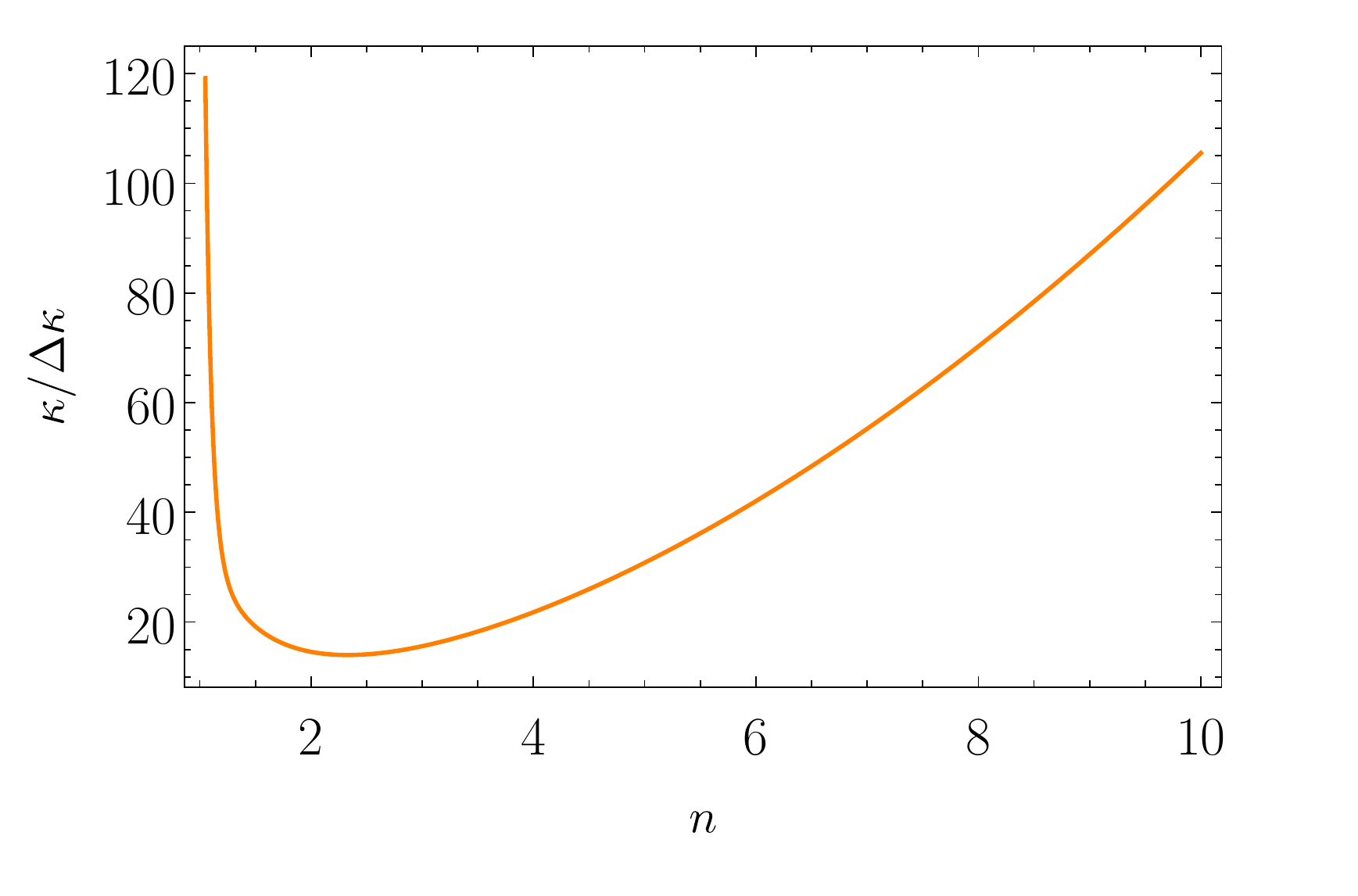}
   \vspace{-0.2in} 
   \caption{The dimensionless ``strength" $d\equiv {|\Re(\mu_k^1)|_{\rm{max}}}/(m/\sqrt{2n})$ and inverse fractional width: $\kappa/\Delta\kappa$ of the first narrow instability band for $V=\lambda_{n}|\phi|^{2n}$  (consistent with T, E and Monodromy models when $\phi\ll M$). Curiously, $d\times\kappa/\Delta\kappa\rightarrow1$.}
   \label{fig:1stband}
\end{figure}
 
 The growth of perturbations from the first narrow instability band might take a long time, but eventually leads to backreaction on the condensate. We can estimate the moment when backreaction begins as follows. Roughly speaking, two conditions have to be met:
 \begin{enumerate}
 \item $|\Re(\mu_k^{1})|_{\rm{max}}\Delta t_{\text{res}}\gg1$ for sufficient particle production.
 \item $|\dot{\kappa}|\Delta t_{\text{res}}\ll\Delta\kappa$ for the given $k$-mode to have spent enough time, $\Delta t_{\text{res}}$, within the first narrow resonance band of width $\Delta\kappa$ and height $|\Re(\mu_k^{1})|_{\rm{max}}$. 
  \end{enumerate}
Using these two conditions, we get $|\Re(\mu_k^1)|_{\rm{max}}\gg (|4-2n|/n+1)H_{\rm br}\kappa/\Delta\kappa$ at the time of backreaction. Let us parametrize this inequality via a small dimensionless number $\delta\ll 1$:
\Beq
\label{eq:br1}
|\Re(\mu_k^1)|_{\rm{max}}=\frac{1}{\delta}\frac{|4-2n|}{n+1}\frac{\kappa}{\Delta\kappa}H_{\text{br}}\,,
\Eeq
where $\delta$ cannot be predicted from the linear analysis. However, as we will see from our nonlinear analysis in the next section, we find $\delta \approx 0.126$ independent of $n$, which makes this parametrization useful. Note that care should be taken for the $n=2$ case, which we turn to shortly.

We can rewrite eq. \eqref{eq:br1} as an equation for the number of $e$-folds of expansion after the end of inflaiton before backreaction takes place. To do so, recall that $H\sim m\bar{\phi}/\sqrt{2n}\mpl$ and that 
$\bar{\phi}\approx\left({a_{\rm{end}}}/{a}\right)^{3/(n+1)}M\,,$  where the initial amplitude of oscillations is $\sim M$ at the end of inflation (for $M\sim\mpl$). Together, they allow us to represent $a_{\rm br}$ in terms of $H_{\rm br}$. Using this $a_{\rm br}$, along with eq. \eqref{eq:br1}, we arrive at the predicted number of {\it e}-folds of expansion from the end of inflation to the beginning of backreaction\footnote{We use $M/2$ in the place of $M$ for the E-models here and in eq. \eqref{eq:DeltaNbrQuart}.}
\Beq 
\label{eq:DeltaNbr}
\Delta N_{\rm{br}}^{\rm{pred}}&\equiv\ln\left(\frac{a_{\text{br}}}{a_{\text{end}}}\right)\\
                                                   &\approx\frac{n+1}{3}\ln\left[\frac{1}{d\delta}\frac{\kappa}{\Delta\kappa}\frac{M}{\mpl}\frac{|4-2n|}{n+1}\right]\,.
\Eeq 
 The dimensionless ratios $\Delta\kappa/\kappa$ (fractional width of the resonance band) and its dimensionless ``strength": 
\Beq
\label{eq:1stbandHeight}
d\equiv \frac{|\Re(\mu_k^1)|_{\rm{max}}}{m/\sqrt{2n}}\,,
\Eeq
for the first narrow instability band are given in Fig. \ref{fig:1stband} (also see Fig. \ref{fig:FloqPowerLaws}). Curiously, the product of this strength and inverse of the fractional width is unity: $d\times \kappa/\Delta\kappa\approx 1$.

For power-laws in the region of $n=2$, our condition 2. is changed to $H\Delta t_{\text{res}}\ll1$, leading to 
\Beq
|\Re(\mu_k^1)|_{\rm{max}}\approx\frac{H_{\text{br}}}{\delta}\,,
\Eeq
whence
\Beq
\label{eq:DeltaNbrQuart} 
\Delta N_{\rm{br}}^{\rm{pred}}\equiv\ln\left(\frac{a_{\text{br}}}{a_{\text{end}}}\right)\approx\ln\left[\frac{1}{d\delta}\frac{M}{\mpl}\right]\,.
\Eeq 
This semi-analytic linear analysis suggests that self-resonance, i.e., inflaton particle production out of the coherent oscillations of the inflaton condensate, can occur after inflation for all values of $M$. For $M\ll \mpl$, the broad, low-$\kappa$ band plays the dominant role in the fragmentation. For $M\sim \mpl$ (and $n>1$), it is the narrow instability band that eventually leads to backreaction and fragmentation.

In the following section we investigate numerically the linear and nonlinear stages of the post-inflationary evolution. We find that when strong resonance takes place ($M\ll\mpl$), the condensate fragments completely into long-lived (oscillons, $n=1$) or short-lived (transients, $n>1$) objects within an $e$-fold after the end of inflation. On the other hand, when $M\sim\mpl$ (and $n>1$), we still eventually have fragmentation, but no transients are formed. In this $M\sim \mpl$ case, the $n=1$ case does not even fragment (ignoring gravitational clustering).

\begin{figure*}[t] 
   \centering
   \includegraphics[height=2.in]{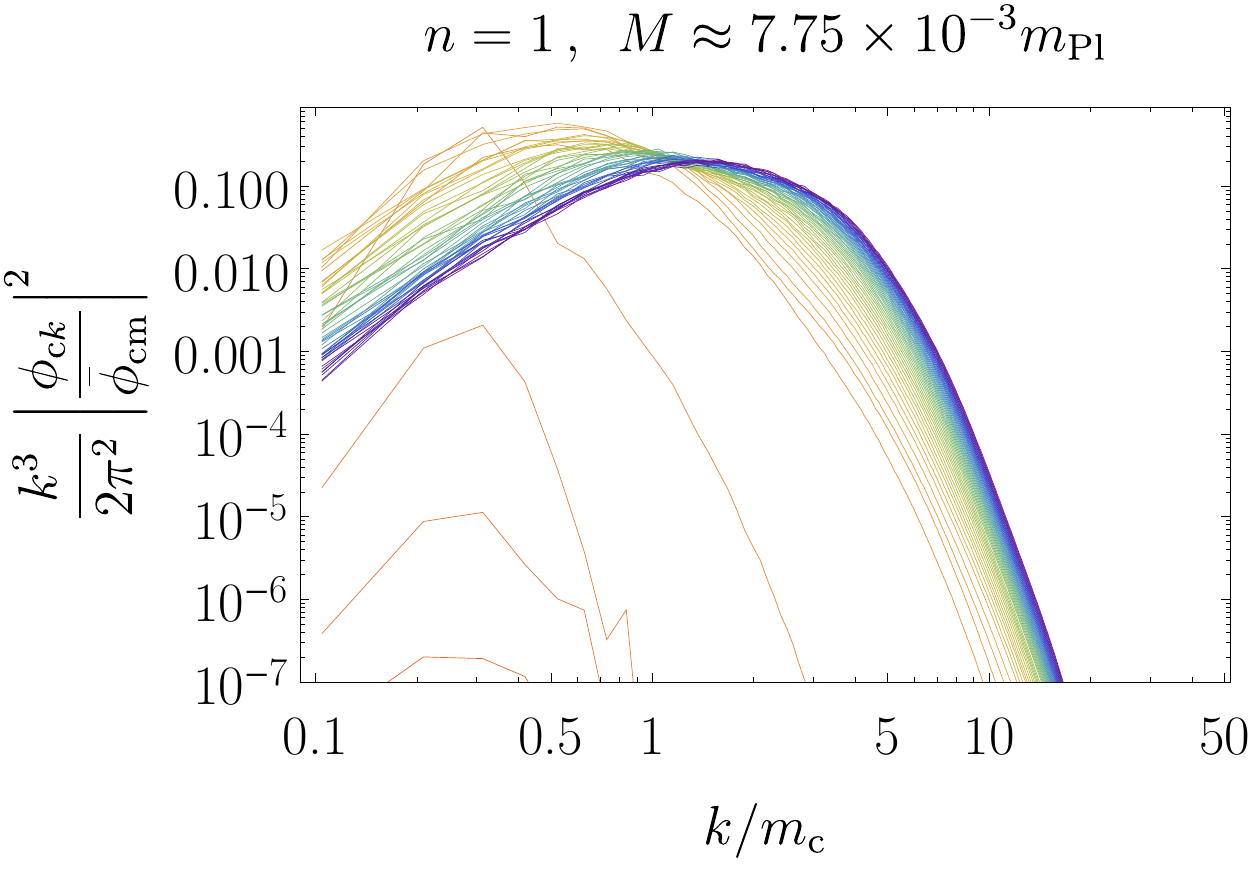}
   \hspace{.5in} 
   \includegraphics[clip, trim=1.6cm 0cm 0cm 0cm, height=2.0in]{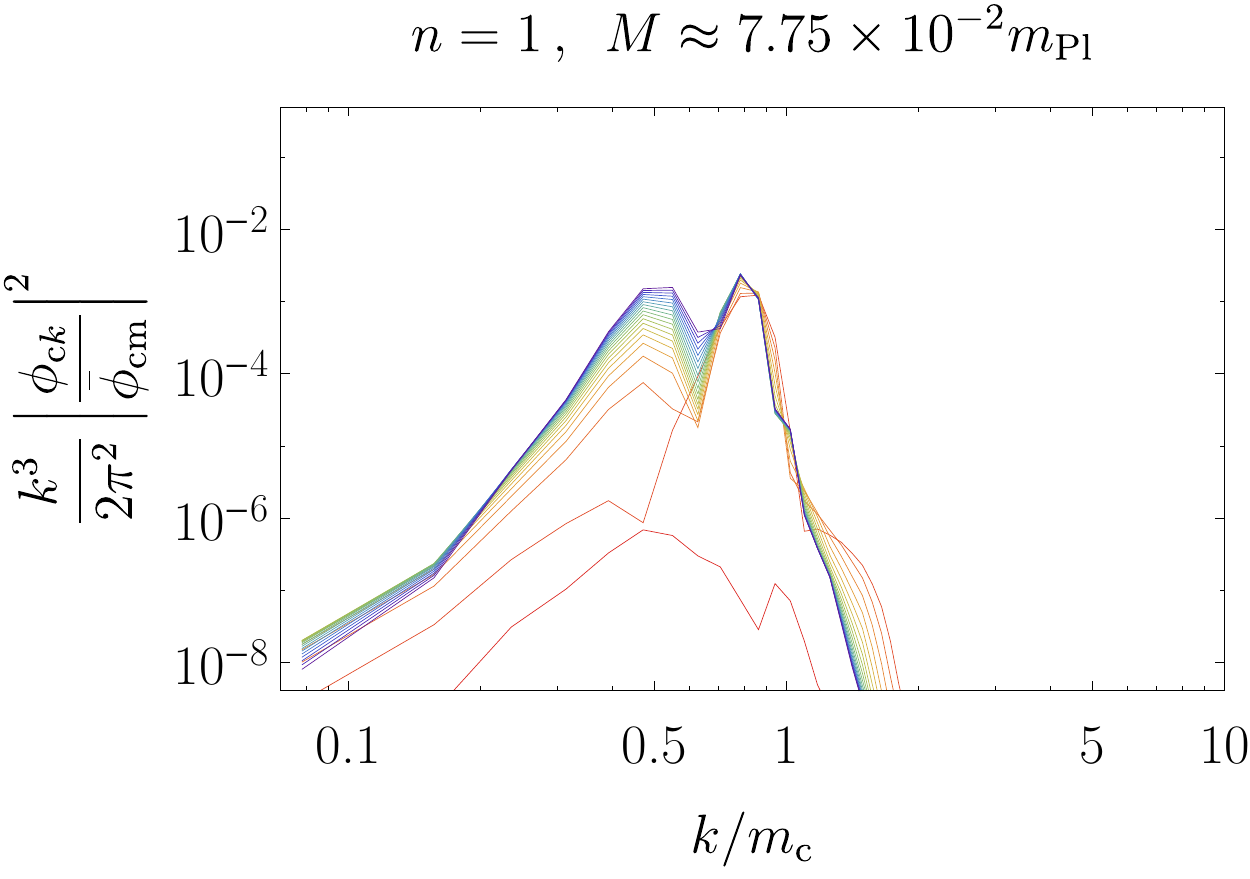} 
   \caption{The time evolution of the power spectra of the inflaton field perturbations, with time running from red to purple. In both panels, we see initial particle production due to the broad low-momentum instability band. In the left panel, where $M$ is sufficiently small, the growth is eventually shut off by backreaction and fragmentation.  The broad peak in the power spectrum is slowly shifted towards higher co-moving wavenumbers as the universe expands at late times, indicating the formation of stable objects of fixed physical size -- oscillons. In the right panel, where $M$ is not small enough, the particle production is quenched by the rapid expansion of the universe and does not lead to backreaction or fragmentation. The subscript `c' stands for conformal -- the Fourier modes, $\phi_{{\rm c}k}$, are rescaled by $a^{3/(n+1)}$ whereas $\bar{\phi}_{{\rm{cm}}}\approx \mathcal{O}[1]\bar{\phi}_{\rm in}$, and $m_{\rm{c}}\equiv m(\bar{\phi}_{{\rm{cm}}})=m$ for $n=1$. With these scalings, when the peak of the rescaled (by an inflaton oscillation amplitude) power spectrum reaches unity, the variance becomes comparable to the mean  (as in the left panel) and indicates the start of backreaction. The data above is for the T-model. }
   \label{fig:pspn1}
\end{figure*}

\begin{figure*}[t] 
   \centering
   \includegraphics[width=5.5in]{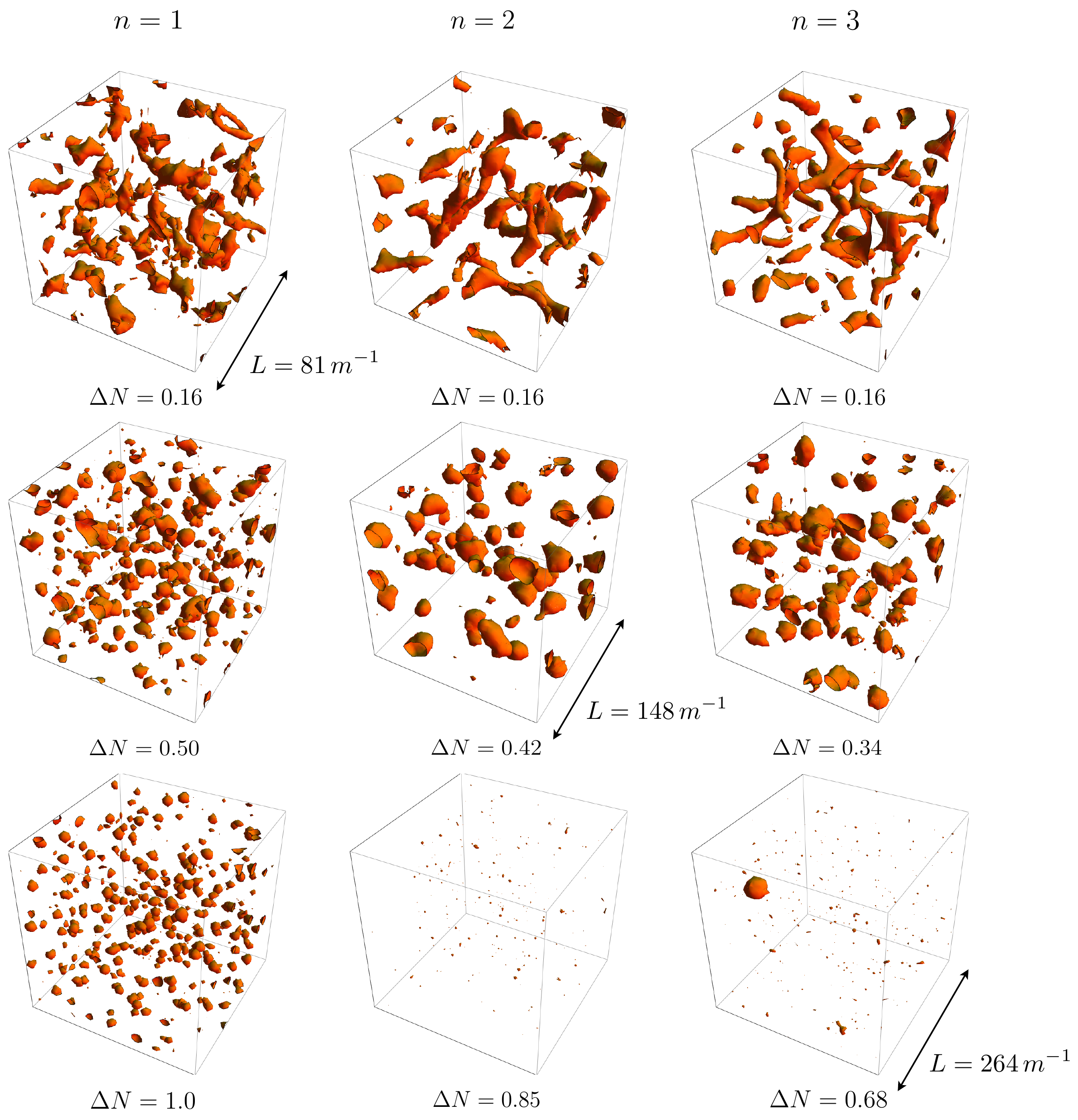} 
   \caption{The three columns show snapshots of density contours at $5\times$ the mean for $M\approx0.775\times10^{-2}\mpl$ for three different $n$ for the T-model. After the inflaton fragments it can form very stable objects (oscillons, $n=1$) lasting millions of oscillations or transient objects ($n>1$) lasting tens of oscillations. They are highly overdense regions, containing a substantial fraction of the energy of the universe for many {\it e}-folds of expansion ($n=1$) or for $\mathcal{O}(1)$ {\it e}-folds ($n>1$). The inflaton becomes virialized after transients decay. The physical size of the co-moving boxes is given in terms of the effective mass $m(\phi=M)$. The boxes are always subhozrion.}
   \label{fig:LatticeSnapshots}
\end{figure*}

\section{Nonlinear Dynamics}
\label{sec:NonLinDyn}

In this section we present our results from numerical simulations of the post-inflationary universe. For our simulations, we use \textit{LatticeEasy} \cite{Felder:2000hq}, a standard workhorse for calculating nonlinear field dynamics in an expanding FRW universe.

The inflaton evolution and the expansion of the universe are calculated according to eq. \eqref{eq:ActionEoM}:\footnote{We note that {\it LatticeEasy} actually uses a combination of the acceleration equation and the Friedman equation.} 
\Beq
&\partial_t^2\phi+3H\partial_t\phi-\frac{\nabla^2}{a^2}\phi+\partial_\phi V(\phi)=0\,,\\
&H^2=\frac{1}{3\mpl^2}\left\langle \rho\right\rangle_{\!\rm s}\,,\\
&\rho=\frac{\dot{\phi}^2}{2}+\frac{(\nabla\phi)^2}{2a^2}+V(\phi)\,,\\
\Eeq
where $\langle \hdots\rangle_{\rm s}$ stand for spatial averaging over the lattice, and $\rho$ is the energy density in the scalar field. Note that metric perturbations are ignored here. We will focus on subhorizon scales only since non-adiabatic resonant particle production happens predominantly on these scales. This allows us to plausibly ignore metric fluctuations for the duration of the simulation. We acknowledge that long term dynamics can be affected by gravitational clustering. A quantitative study of such gravitational clustering of $\phi$ is beyond the scope of this present work. However, we will {\it passively} calculate the metric fluctuations (Newtonian potentials and gravitational waves) generated by the fields in an future paper.

We adopt the standard initial conditions at the time of the beginning of preheating, i.e., a homogeneous inflaton condensate, $\bar{\phi}(t)$, with vacuum fluctuations, $\delta\phi(t,\bf x)$, on top of it. While the initial spectrum of fluctuations has a quantum mechanical origin, the evolution will be carried out classically because of our expectations that modes will become highly occupied. We initialize the simulations around the end of inflation, defined as the first instance when $\ddot{a}(t)=0$ (the results are insensitive to the exact time of initialization as long as it is near the end of inflation). 

Our typical simulation was carried out on a $N=256^3$ lattice. However, at times $N=512^3$ and even $N=1024^3$ became necessary to cover the required dynamical range, or to serve as a check on the lower resolution simulations. We will return to such numerical checks after discussing some of the results.

For future use, we define the spatially averaged equation of state parameter as 
\Beq
\label{eq:EoS}
w\equiv\frac{\langle p\rangle_{\rm s} }{\langle \rho\rangle_{\rm s}}=\frac{\langle\dot{\phi}^2/2-(\nabla\phi)^2/6a^2-V\rangle_{\rm s}}{\langle \dot{\phi}^2/2+(\nabla\phi)^2/2a^2+V\rangle_{\rm s} }\,,
\Eeq
where, $\rho$ and $p$ are the energy density and pressure of the inflaton field, respectively. The equation of state often rapidly oscillates compared to the characteristic expansion time scale -- a time average over many oscillations should be assumed when we refer to $w$ unless otherwise stated. 

As discussed in Section \ref{sec:LinAnal}, the interplay between parametric resonance and the Hubble expansion can be divided into two regimes: $M\sim\mpl$ and $M\ll\mpl$. Moreover, the relevance of the first narrow resonance band showed important differences between $n=1$ and $n>1$. Motivated by this analysis, we will consider four different regimes based on $\{n,M\}$.
\begin{figure*}[t!] 
   \centering
   \hspace{-0.0in}
   \includegraphics[height=2.30in]{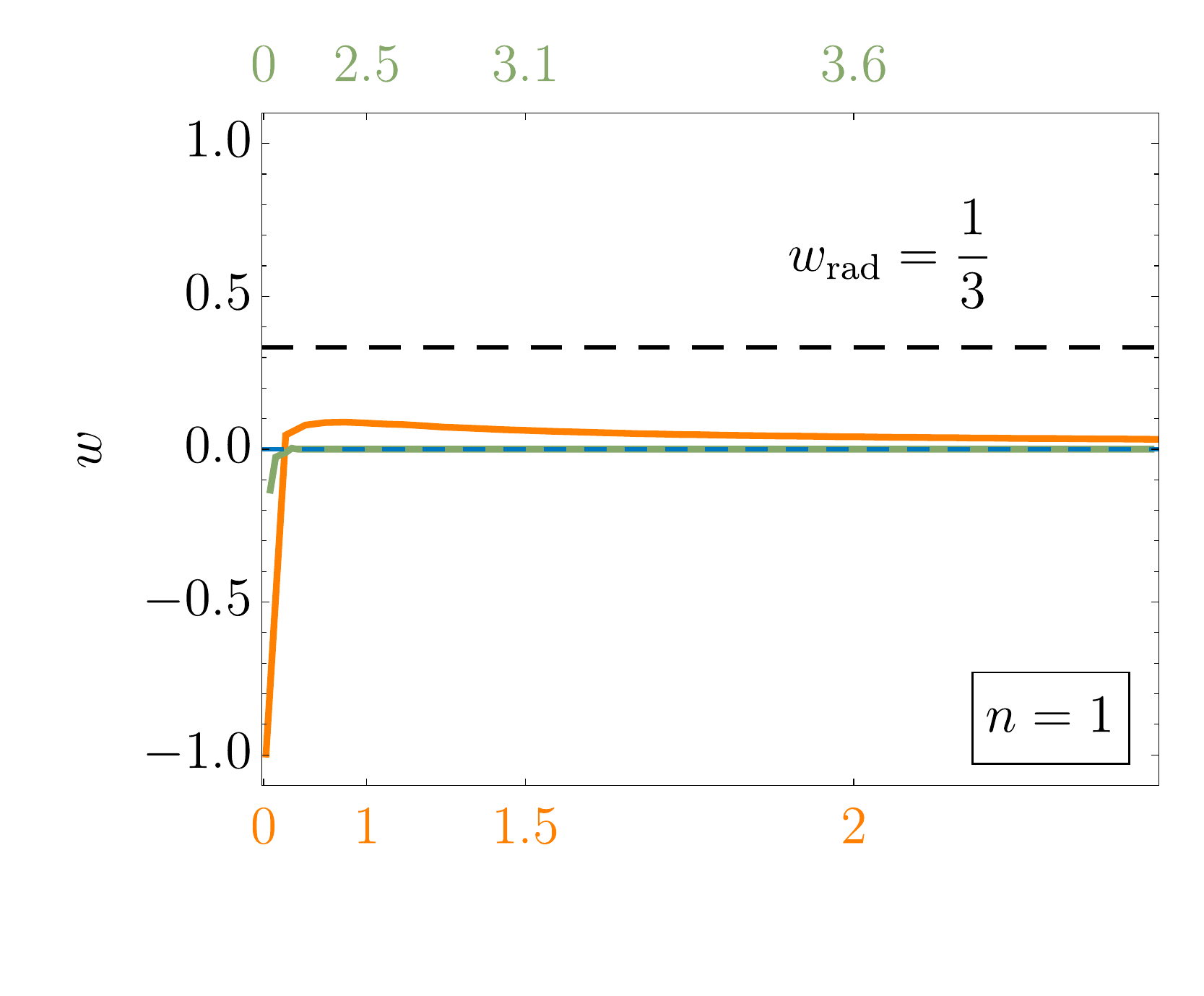} 
   \includegraphics[height=2.30in]{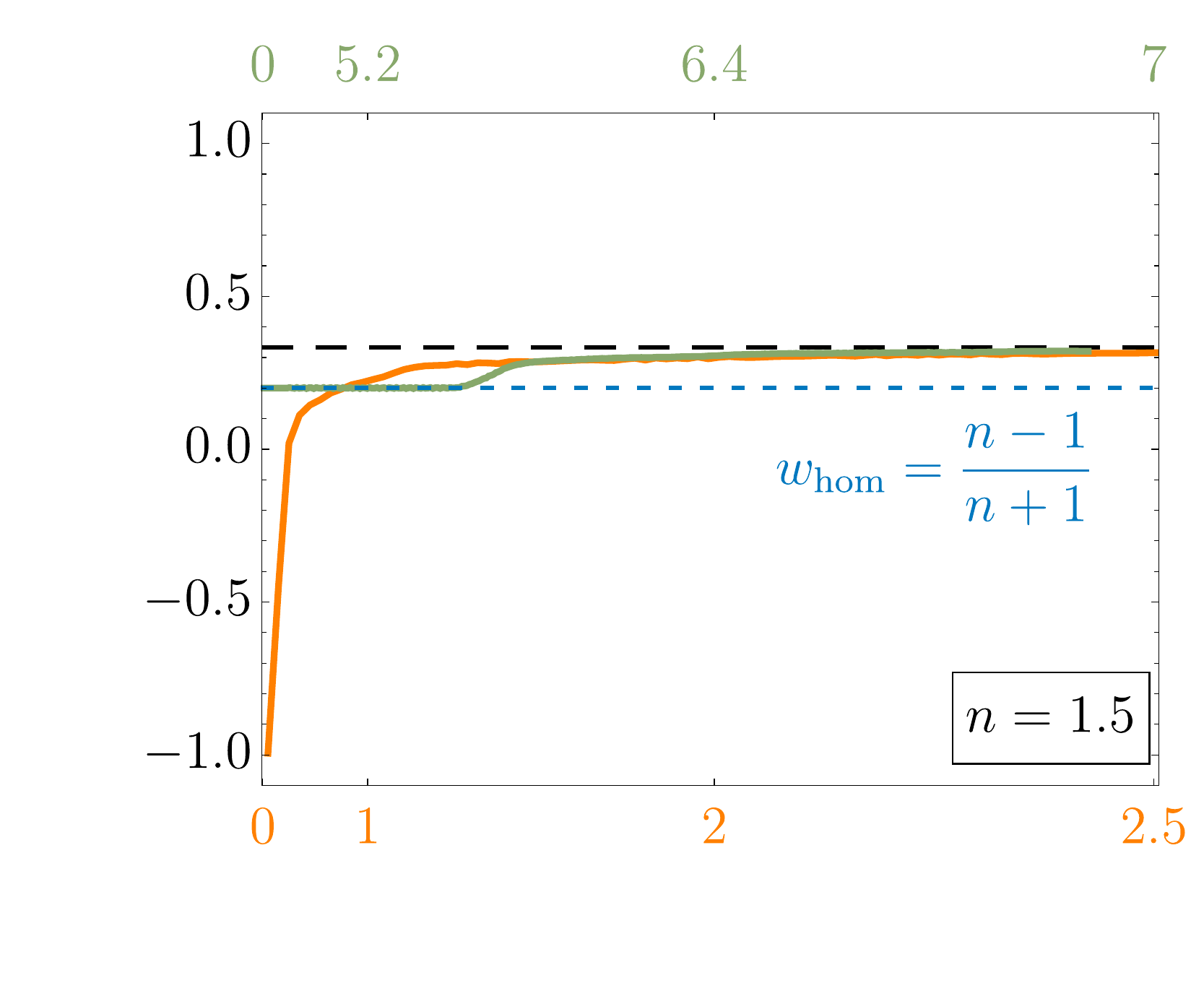}\\ 
   \includegraphics[height=2.30in]{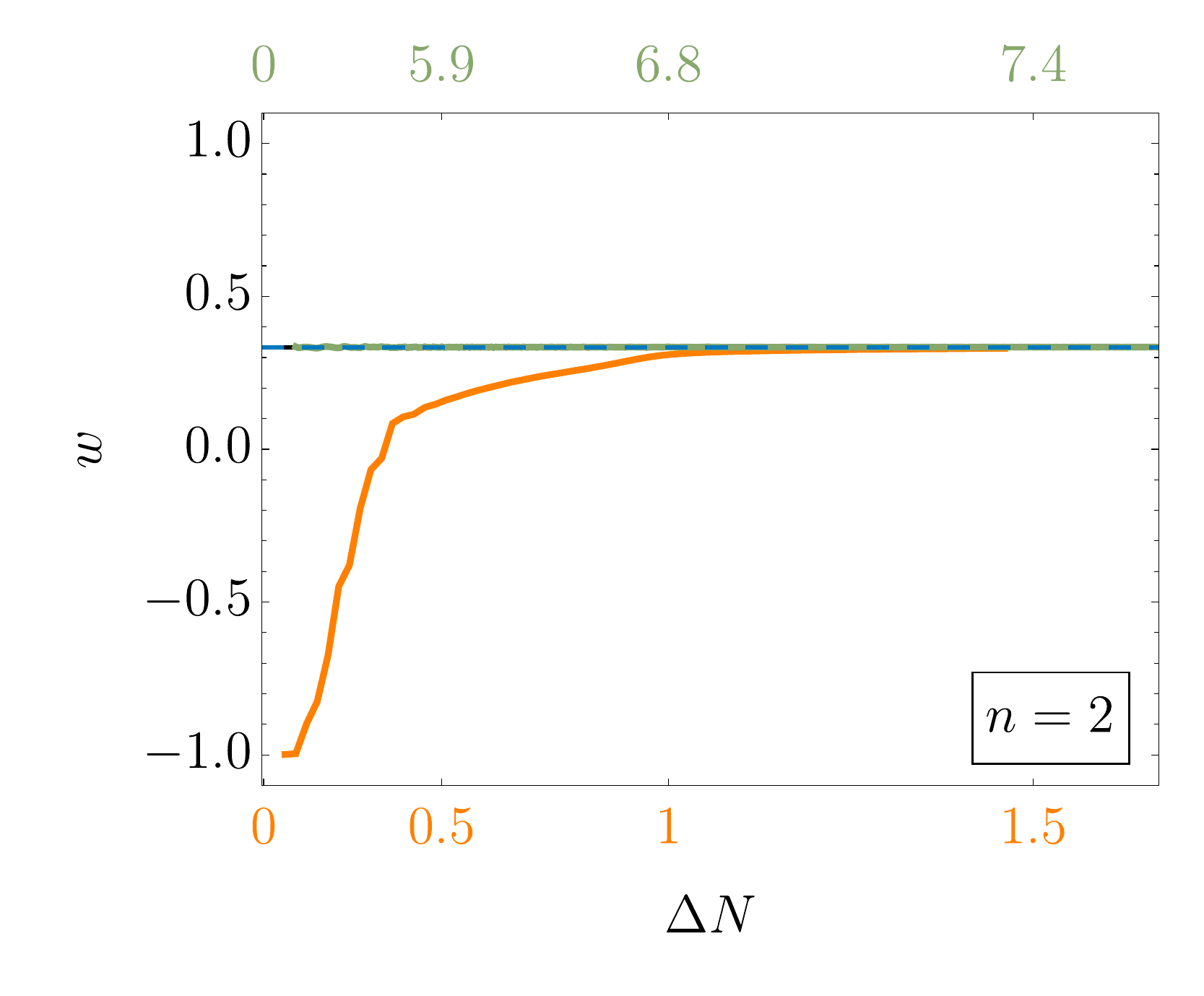} 
   \includegraphics[height=2.30in]{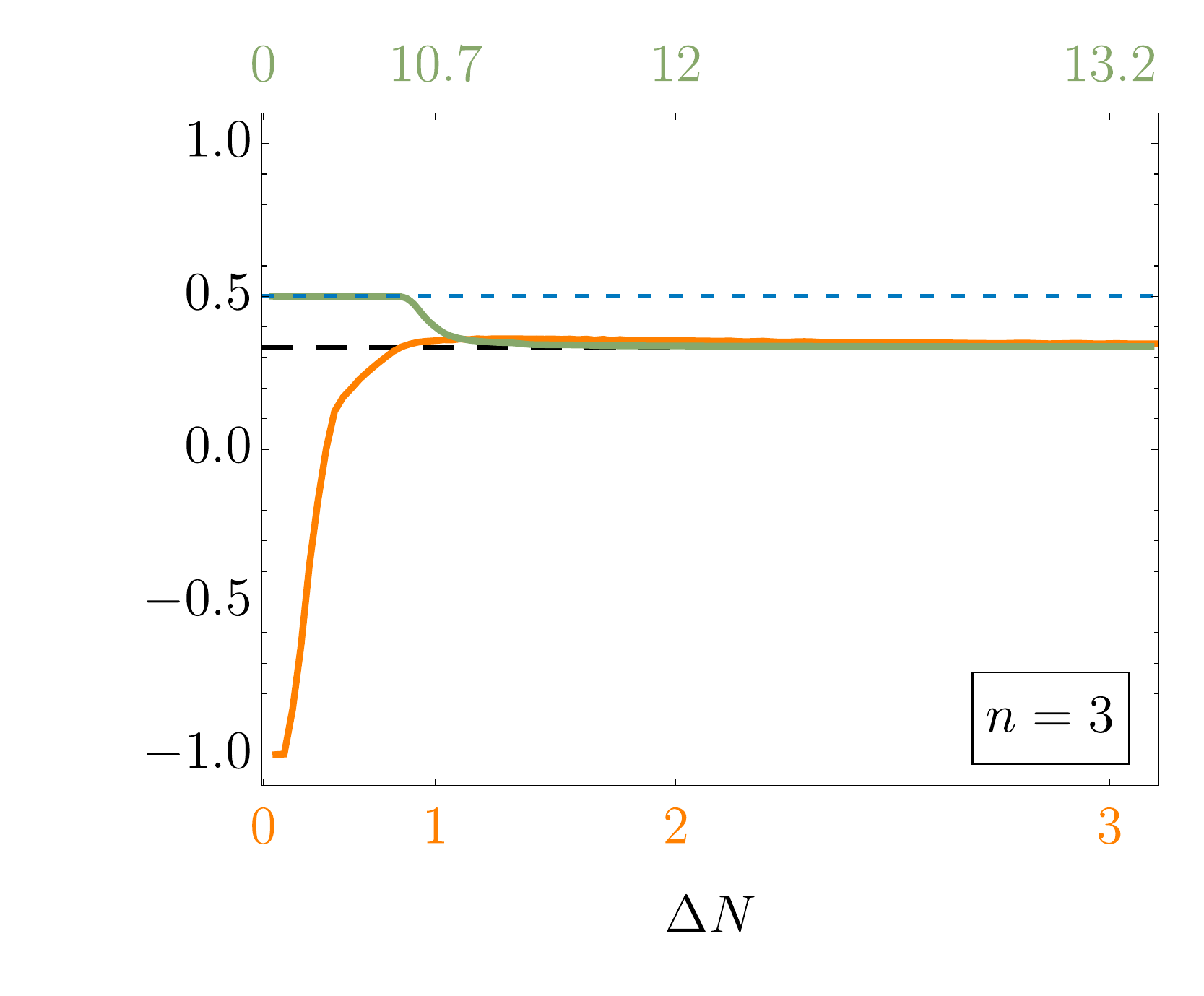}
   \caption{The evolution of the equation of state for different $M$ and $n$ in the T-models. On the vertical axes we have the equation of state and on the horizontal axes the number of {\it e}-folds of expansion after the end of inflation. We indicate the homogeneous equation of state (dotted blue line) and the equation of state for radiation domination (dashed black  line) for reference. The orange and green curves show the equation of state calculated from lattice simulations. For the orange curves ($M\approx7.75\times10^{-3}\mpl$), the resonance is efficient and leads to the complete fragmentation of the condensate within less than an {\it e}-fold of expansion. After fragmentation $w$ settles to $0$ for a quadratic minimum (since oscillons behave as pressureless dust) or $1/3$ for steeper power-laws (after the transient objects decay away). In the green cases ($M\approx2.45\mpl$), the resonance is inefficient and the condensate can oscillate for very long times. If $n=1$ the condensate never fragments due to self-resonance, whereas if $n>1$ backreaction eventually occurs at a predictable time, and the equation of state quickly settles to $1/3$ in a step-like manner.}
   \label{fig:EqOfState}
\end{figure*}
\subsection{Oscillons and Matter Domination}
\noindent\small{{\bf A1}: $\boldsymbol{n=1, M\ll \mpl$}}\\ \\
In this scenario, low-$k$ perturbations grow rapidly, which can be seen from the linear analysis result in the first column of Fig. \ref{fig:Floq}. The universe expands slowly with respect to the rate of oscillations of $\bar{\phi}(t)$. Our lattice simulations indeed confirm the expectation from the linear analysis. Initially, we observe the development of a broad low-$k$ peak in the spectrum of the inflaton. See left panel in Fig. \ref{fig:pspn1}. When the energy of the perturbations associated with this peak becomes comparable to that of the condensate, backreaction takes place. The backreaction process is quite efficient in the sense that the condensate fragments completely living a negligible amount of homogeneous condensate behind.
\subsubsection{Oscillons} 
The fragmentation of the condensate is followed by the formation of interesting nonlinear structures of fixed physical size. See the first column in Fig. \ref{fig:LatticeSnapshots}. These nonlinear structures survive for the entire duration of the simulation, and  explain the peak in the power spectrum in the left panel of Fig. \ref{fig:pspn1}. This peak never goes away and is only slowly shifted towards higher co-moving wavenumbers since the structures maintain a fixed physical size.

These localized, nonlinear objects are long-lived pseudo-solitonic objects called oscillons \cite{Bogolyubsky:1976yu,Gleiser:1993pt,Copeland:1995fq,Amin:2010jq,Amin:2013ika}. They normally form if $V$ is quadratic near the origin and flatter away from it, which is precisely the case we are considering. See for example \cite{Amin:2013ika} for more precise conditions. An oscillon profile is such that the nonlinear terms from $\partial_{\phi}V$ essentially cancel the dispersion term in the equation of motion, i.e., $\partial_t^2\phi_{\text{osc}}+\omega^2\phi_{\text{osc}}\approx0$ with $\omega^2=m^2(1-\hdots)\approx \rm{const.}$ For a more detailed explanation, see for example \cite{Amin:2013ika}. The oscillon field profile, $\phi_{\text{osc}}(r,t)$, is spherically symmetric, peaking at the centre of the oscillon and approaching  zero monotonically away from it. The profile is oscillating with time, hence the name. At leading order, all points of the profile oscillate in phase and at the same frequency, so one can write $\phi_{\text{osc}}(r,t)\approx R(r)T(t)$.  The energy contained in oscillons, however, is (approximately) constant with time. That is, inside an oscillon the energy density does not redshift. 

We note that $\phi_{\text{osc}}(r,t)$ is not actually an exact solution to the equations of motion, but only approximate. Oscillons have subleading multiple frequencies, and also decay non-perturbatively through classical \cite{Segur:1987mg} (or quantum \cite{Hertzberg:2010yz}) radiation eventually.  Typically oscillons last for millions of oscillations (corresponding to many Hubble times \cite{Salmi:2012ta}). For a discussion of their stability, see for example \cite{Amin:2013ika,Gleiser:2009ys,Amin:2010jq}.
\subsubsection{Equation of State}
In the top-left panel of Fig. \ref{fig:EqOfState},  the equation of state $w$ (see eq. \eqref{eq:EoS}) for $M\ll \mpl$, $n=1$ (orange curve) is shown as a function of the number of $e$-folds of expansion after inflation, $\Delta N$. Copious oscillon production is possible for $M\ll \mpl$, and collectively  the oscillons can lock a substantial fraction of the energy of the universe \cite{Amin:2011hj}. This gas of oscillons behaves as pressureless dust, thereby giving rise to $w=0$ and a matter dominated expansion of the universe.

Note that $w$ is not exactly $0$, but rather decays towards it. This is because there is some energy stored in relativistic modes outside the oscillons. However, since the energy in the oscillon gas redshifts as matter, $\rho\propto a^{-3}$, whereas in relativistic modes as $\propto a^{-4}$ it does not take long for $w\rightarrow0$. 

The approach to $w=0$ is rapid -- it takes less than an {\it e}-fold of expansion for the field to fragment and form oscillons and reach a matter-like equation of state in the T and E models. In these models the inflaton starts to oscillate in the broad band immediately after inflation, see Fig. \ref{fig:Floq}. We note that in the Monodromy models it can take up to $2$ {\it e}-folds for the amplitude of inflaton oscillations to be redshifted to the broad instability band peak near $\phi\sim M$.
\\ \\
\noindent{{\bf A2}: $\boldsymbol{n=1,M\sim\mpl}$}\\ \\
As expected from the linear analysis, there is no fragmentation of the condensate here. As a result, no oscillons form. The oscillating condensate still maintains the equation of state $w=0$, see top left panel of Fig. \ref{fig:EqOfState} (green curve). The universe remains matter dominated.

Thus in the case where $n=1$, for the universe to reach a radiation dominated state necessary for successful BBN, we need extra ingredients in the way of daughter fields coupled to the inflaton which eventually lead to the decay of the oscillons into relativistic matter. 
\begin{figure*}[t] 
   \centering
       \includegraphics[height=3.0in]{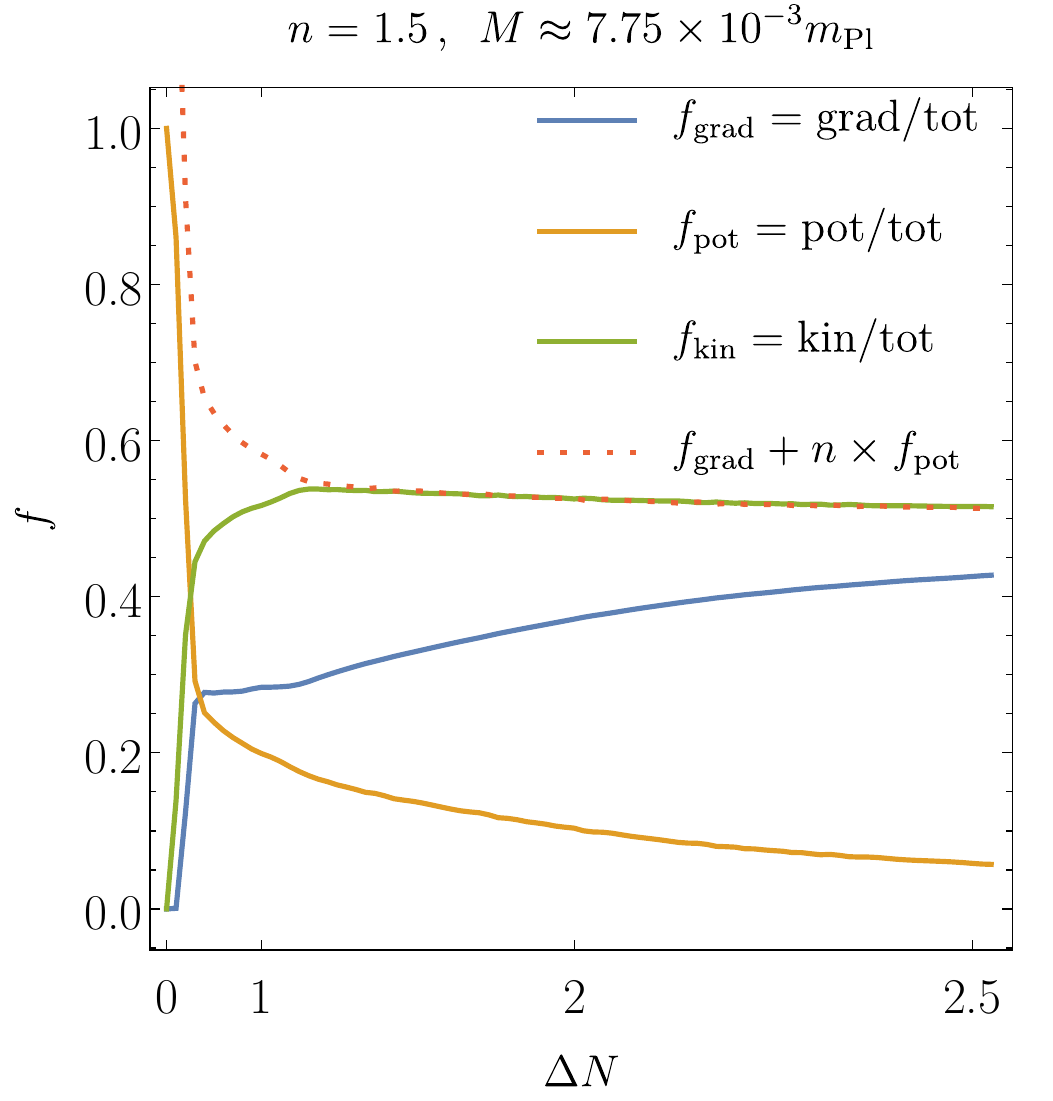}
       \includegraphics[height=3.0in]{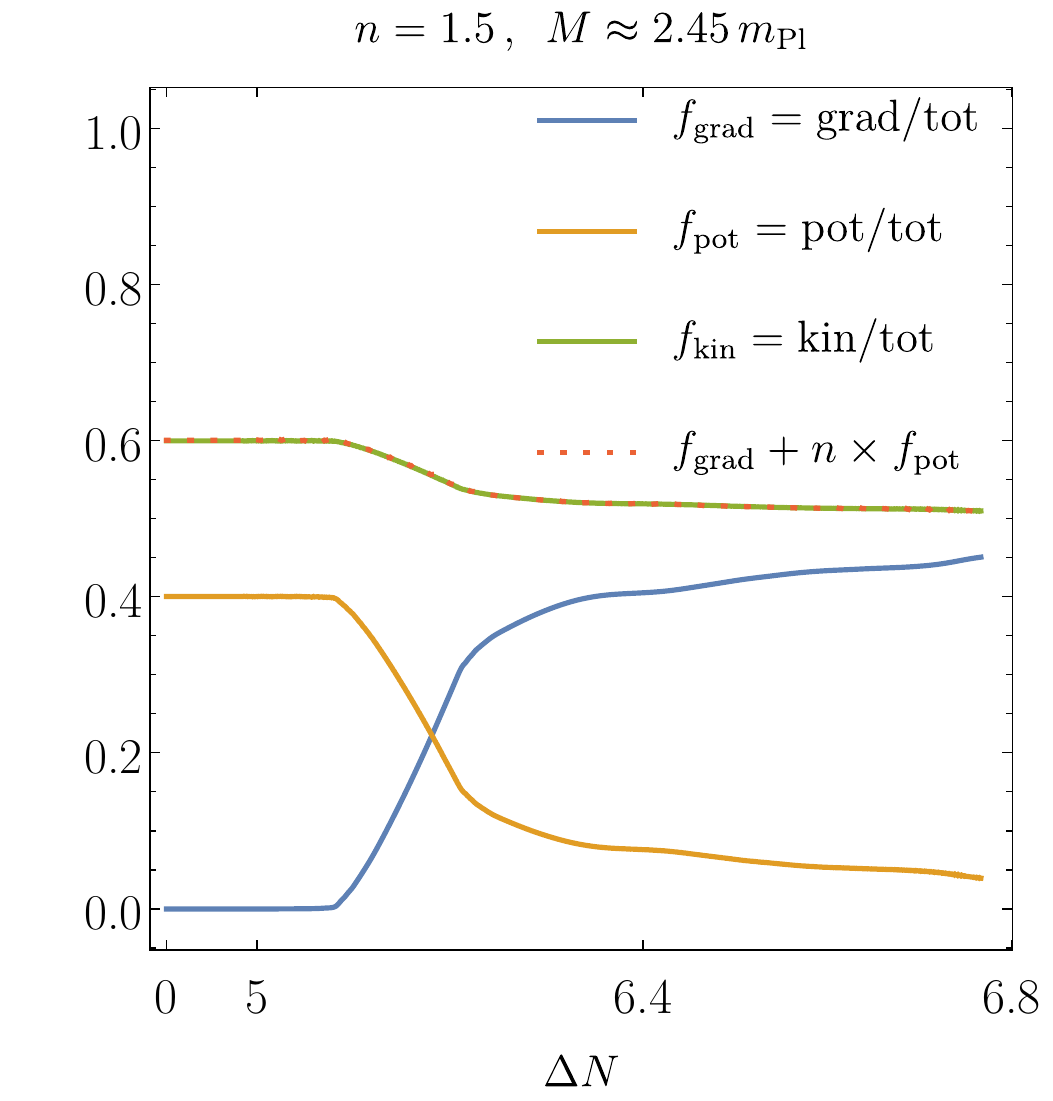}
   \caption{The evolution of the fraction of energy, $f$, stored in gradient (blue), potential (orange) and kinetic (green) terms. The red (dotted) line is the right-hand side of the virial expression in eq. \eqref{eq:Equipartition} divided by the total energy. All curves represent time averages over many oscillations and spatial averages over the simulation volume.  In the case on the left, the condensate fragments rapidly into transient objects, which survive for about an {\it e}-fold of expansion as evident from the plateau near $\Delta N=1$ in $f_{\rm{grad}}$. After that the transients decay away and the inflaton field becomes virialised. In the right panel, the first narrow instability band leads to slow but steady particle production. The condensate oscillates for over $5$ {\it e}-folds, as indicated by the initial plateaus in the three $f$s, before the excited modes backreact and the condensate fragments. Interestingly, the field remains completely virialised throughout its evolution. In both cases the self-interaction energy becomes increasingly subdominant with time.}
   \label{fig:virial}
\end{figure*}

\subsection{Transients and Radiation Domination}
\label{sec:Trans}

\noindent\small{{\bf B1}: $\boldsymbol {n>1,M\ll \mpl}$\\ \\
When $V(\phi)$ is steeper than quadratic near the minimum ($n>1$), and $M\ll \mpl$, we again observe a rapid growth of linear perturbations for $k_{\rm phys}\lesssim m$ (see Fig. \ref{fig:Floq}). 
\subsubsection{Transients}
Once backreaction kicks in, we observe the formation of nonlinear objects in a qualitatively similar manner to the formation of oscillons. We refer to the nonlinear objects formed in $n>1$ case as {\it transients} since they are much shorter-lived than oscillons. Transients survive for tens of oscillations only. Their formation and then subsequent decay act can be seen visually in the second and third columns in Fig. \ref{fig:LatticeSnapshots} and first column in Fig. \ref{fig:pspn1523}. Despite their short lives, they can dominate the energy budget of the universe for up to an {\it e}-fold of expansion and hence can be of cosmological relevance. 

Let us make a heuristic connection of transients with oscillons. If we assume that the dispersion term for transients is cancelled by all but the lowest order term from $\partial_{\phi}V$ in the equation of motion, then $\partial_t^2\phi_{\text{tr}}+m^2\phi_{\text{tr}}\approx0$. Note that here, $m^2\propto\phi_{\text{tr}}^{2n-2}\ne \textrm{const.}$ (distinct from the $n=1$ case). It is impossible to get an approximate solution of the form $\phi_{\text{tr}}(r,t)\approx R(r)T(t)$ and therefore the frequency of oscillations will be amplitude dependent, implying that different points of the profile will oscillate at different rates. In our simulations we indeed observe the formation of ripples on top of the oscillating $\phi_{\text{tr}}(r,t)$, which radiate away the energy of the objects, explaining their transient nature. 

We can in fact go a step further and generalize this statement and claim that any $V\propto|\phi|^{2n}$ ($n>1$) near the origin and flattening out away from it (hence leading to an attractive clustering of the field)  will lead to the formation, and temporary support of such transients. To the best of our knowledge this is the first time they have been reported in the literature.\footnote{It is interesting to think about the longevity of this objects if there is a mechanism which forces their profile to resemble a top-hat (akin to the flat-top oscillons \cite{Amin:2010jq}). The constancy of the amplitude within an extended region should suppress the formation of ripples and prolong the lifetime of the transients. We leave this investigation for a future work.}

The rapid appearance, approximately adiabatic evolution for an $\lesssim e$-fold, and sudden disappearance can lead to interesting gravitational wave signatures. We will explore the gravitational waves generated from the formation and eventual dispersion of these transients in a future paper \cite{Lozanov:2017b}.
\begin{figure*}[t] 
   \centering
   \includegraphics[clip, trim=0cm 1cm 0cm 0cm, height=1.75in]{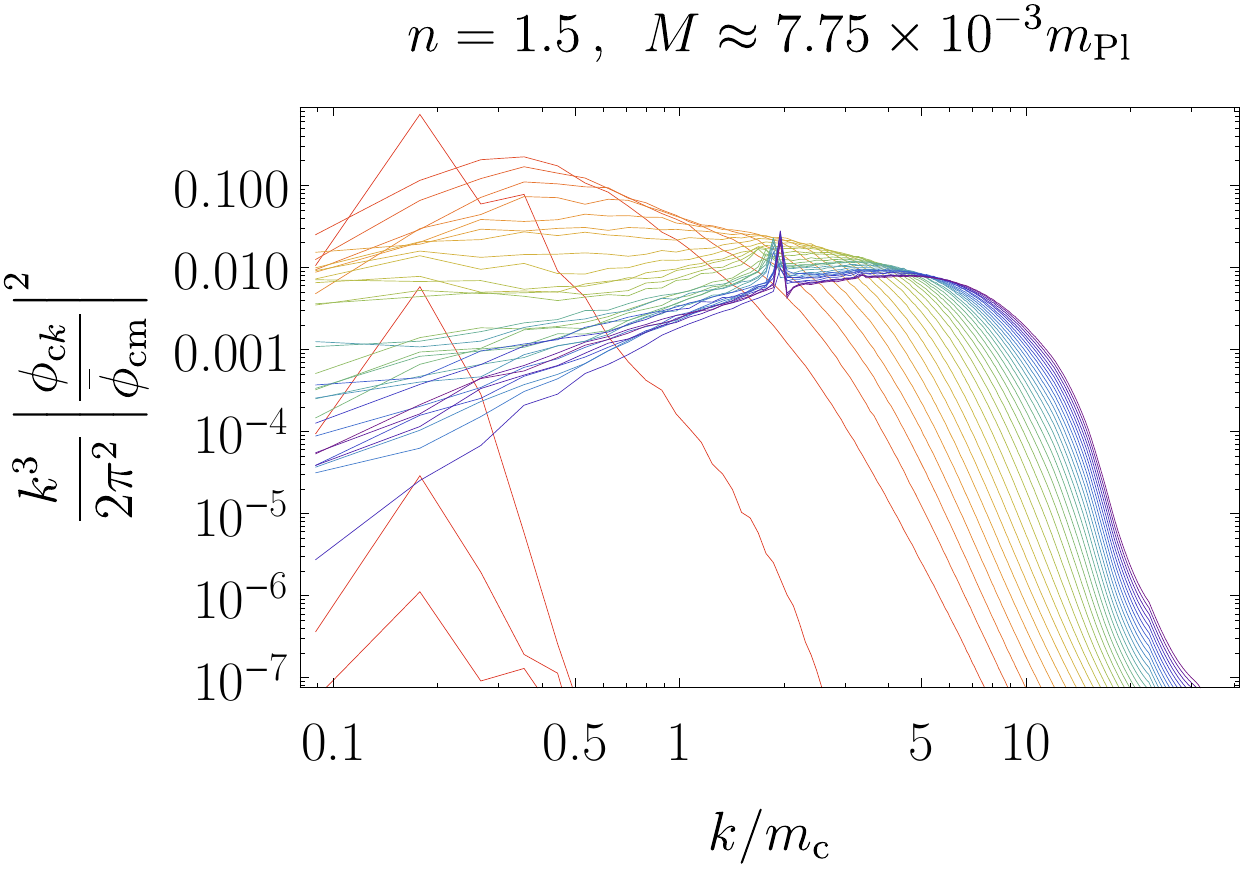}
   \hspace{0.2in} 
   \includegraphics[clip, trim=1.6cm 1cm 0cm 0cm, height=1.75in]{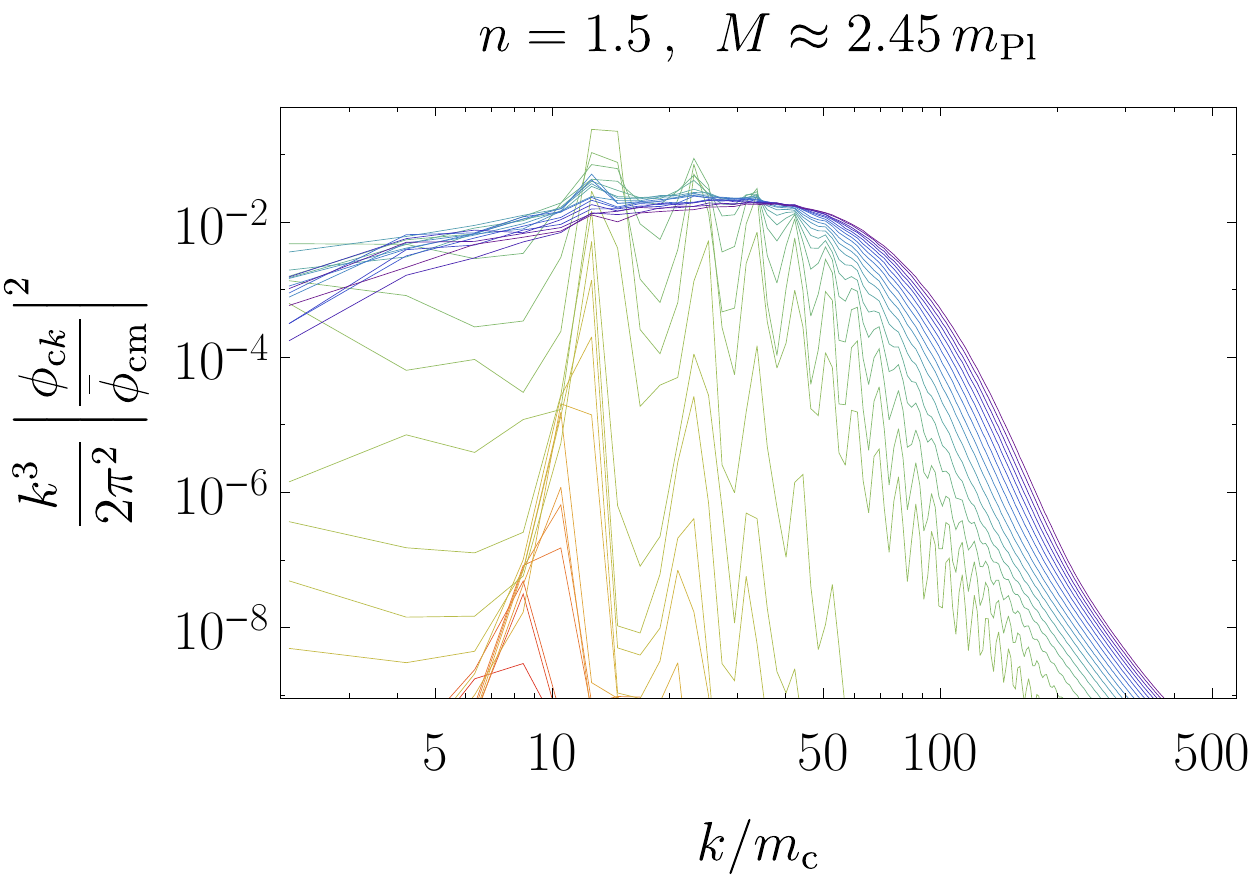}\\
   \vspace{0.2in}
   \includegraphics[clip, trim=0cm 1cm 0cm 0cm, height=1.75in]{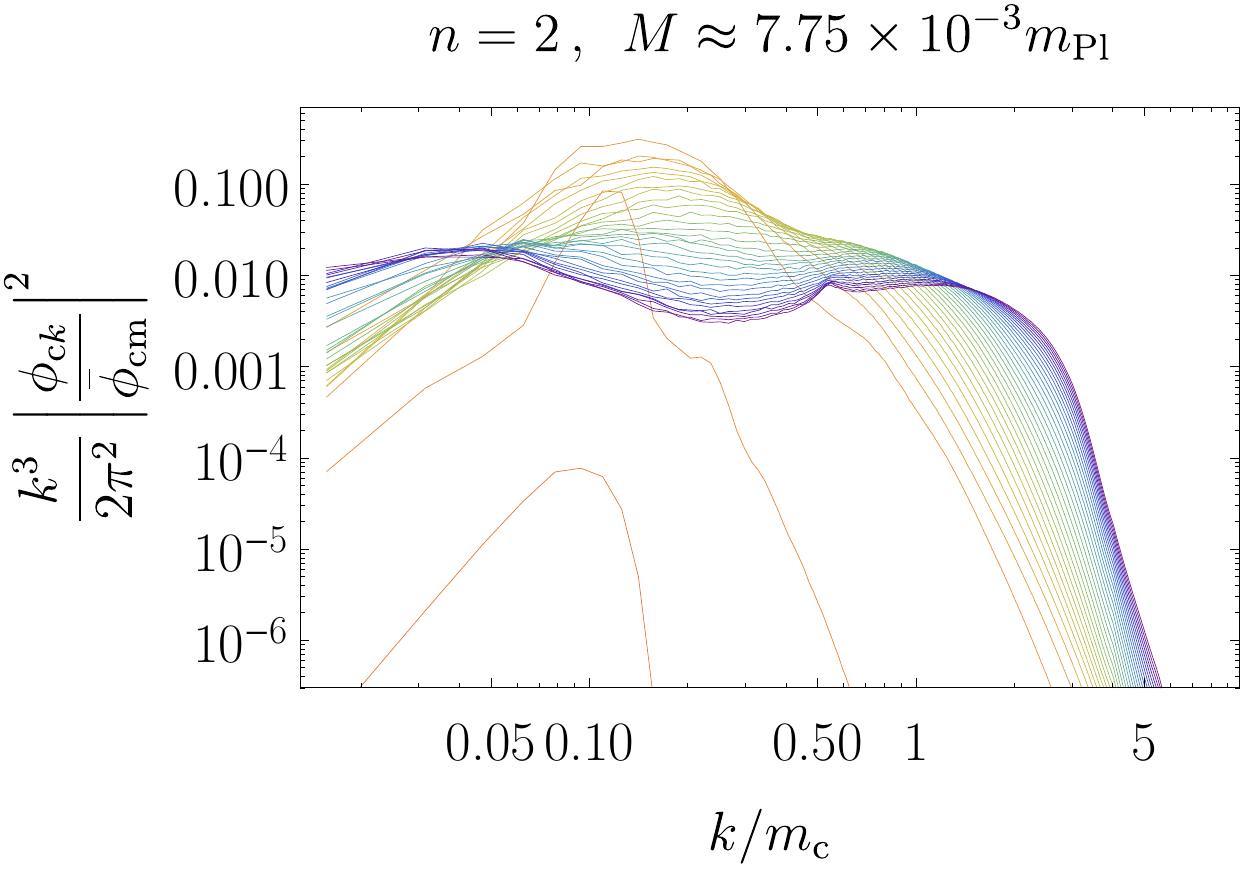}
   \hspace{0.2in} 
   \includegraphics[clip, trim=1.6cm 1cm 0cm 0cm, height=1.75in]{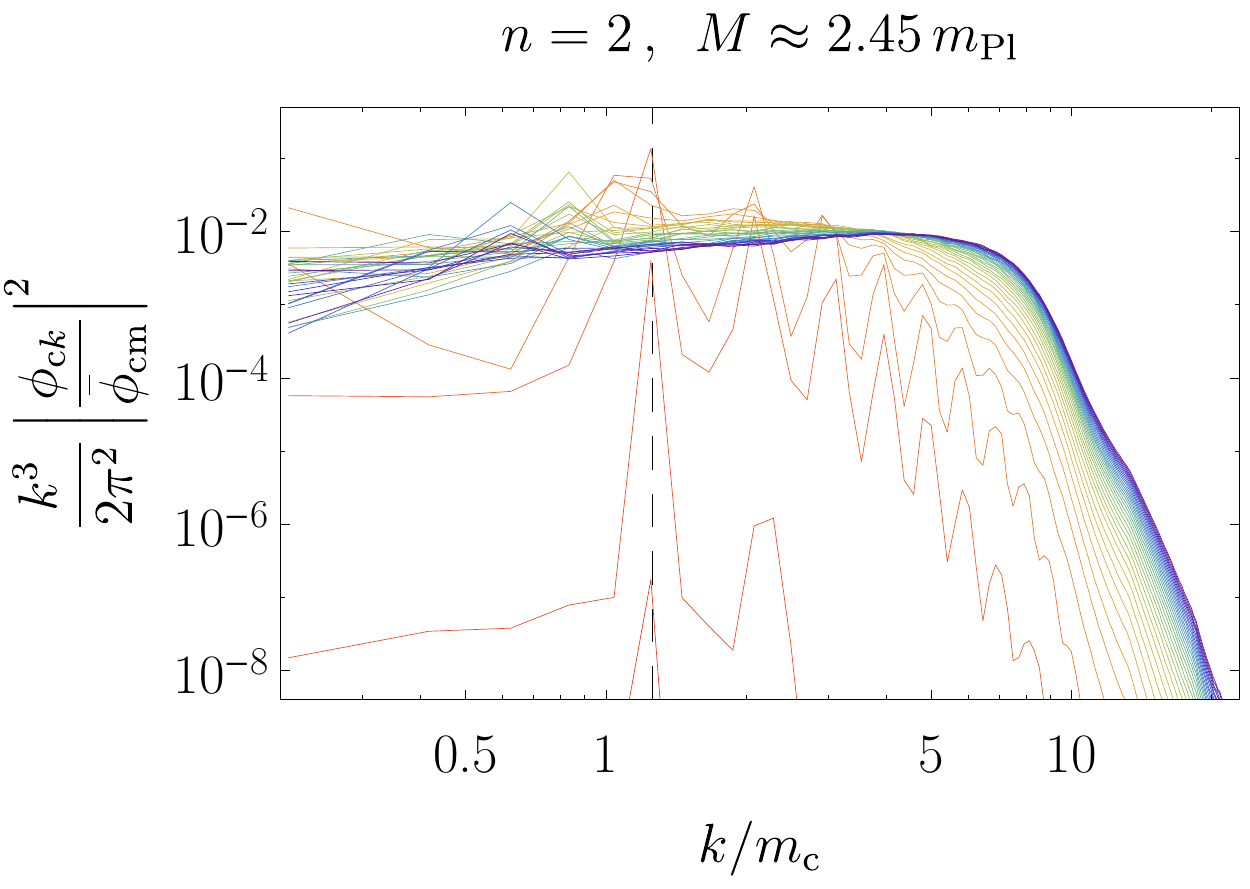}\\
   \vspace{0.2in}
   \includegraphics[height=1.99in]{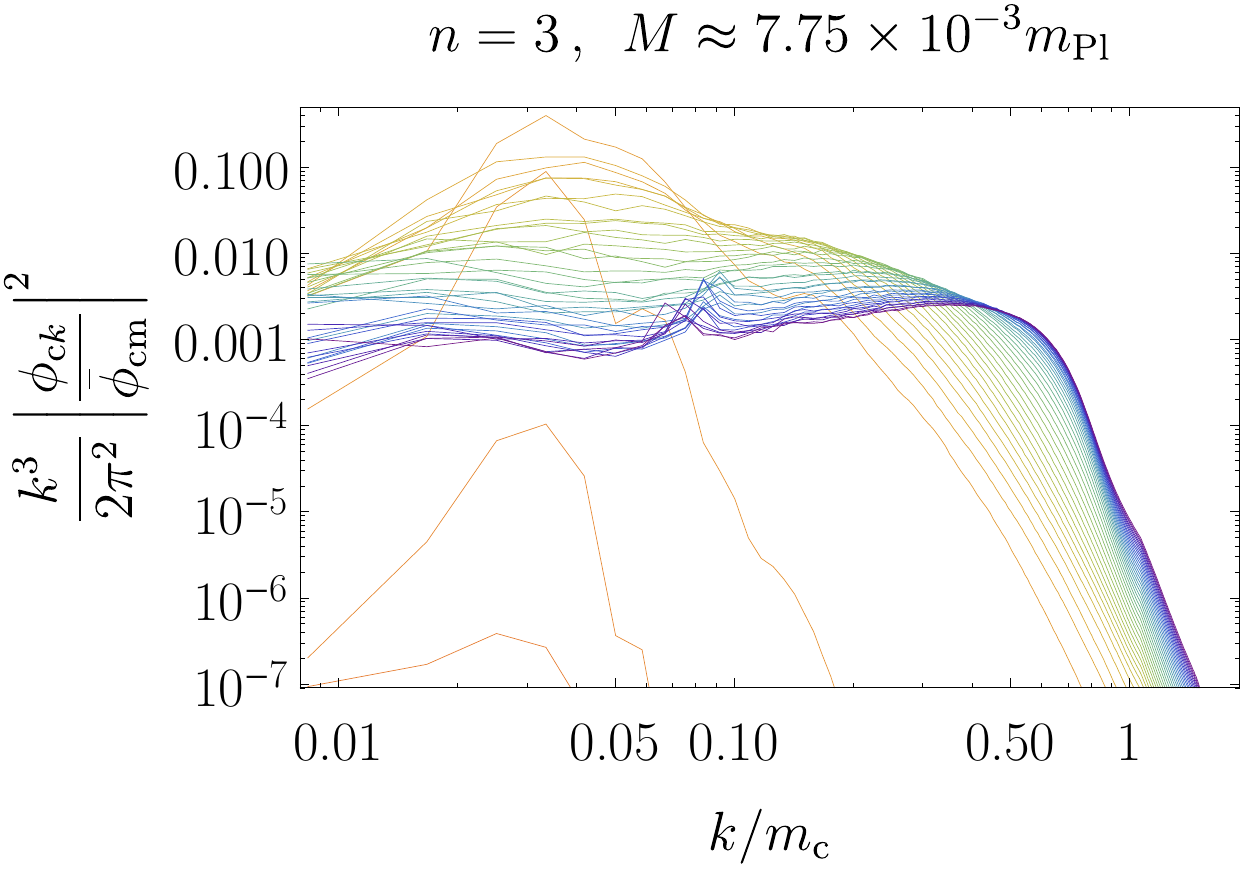}
   \hspace{0.2in} 
   \includegraphics[clip, trim=1.6cm 0cm 0cm 0cm, height=1.99in]{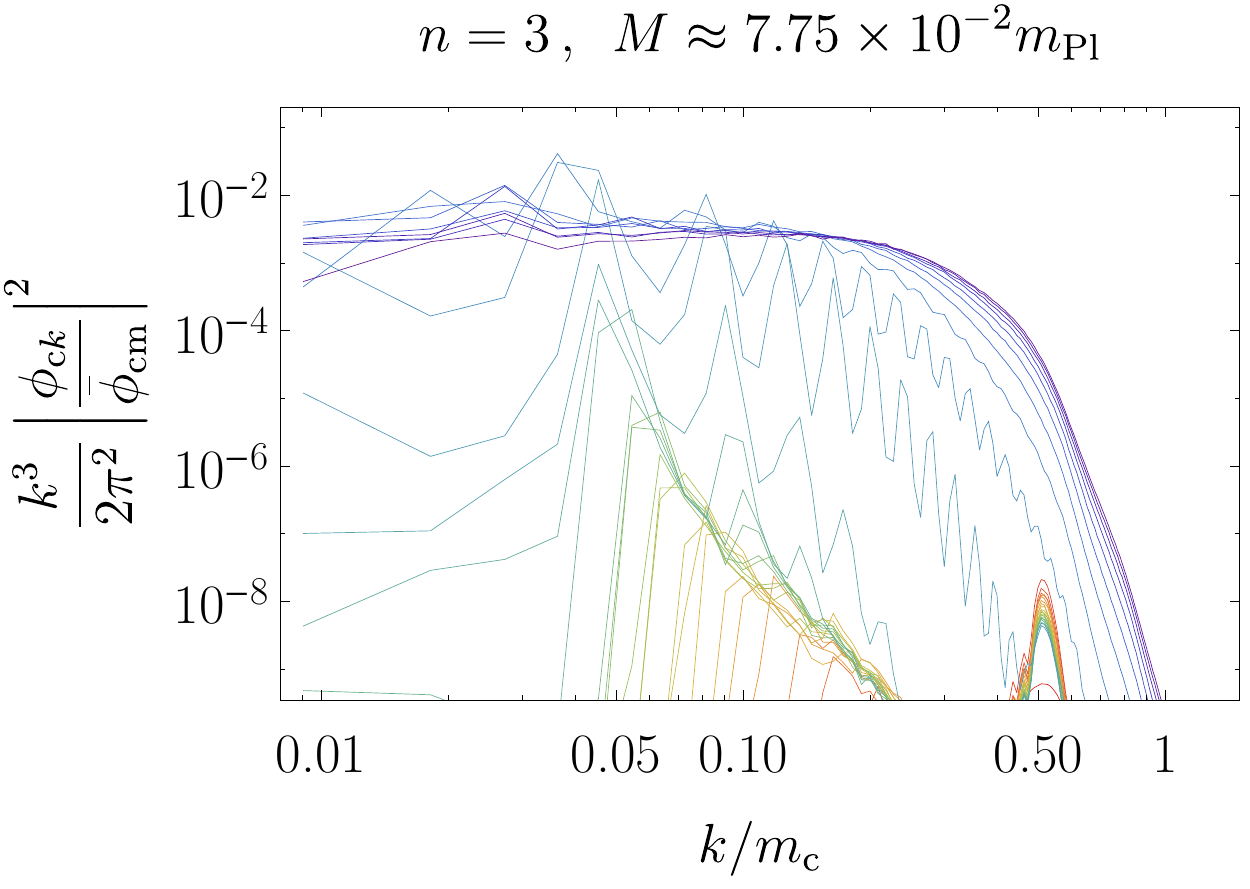}
   \caption{Representative power spectra of inflaton fluctuations for $n>1$. The left column is for sufficiently small $M$, allowing for the broad instability band to fragment the condensate and form transients. As the transient objects decay, the broad peaks in the power spectra disappear, shifting power to the  UV modes. The right column is for larger $M$, for which the first narrow instability band leads to slow, but steady particle production in a narrow co-moving band. The peak of this band shifts with time towards higher ($n<2$), lower ($n>2$) co-moving modes or stays fixed ($n=2$) at $k\approx1.27m_{\rm{c}}$. The generation of multiple re-scattering peaks is also evident in the second column. The growth is eventually shut off by backreaction and fragmentation without the formation of any transient nonlinear objects. In all six panels, power cascades slowly towards the UV at late times. Since there is a subdominant remnant oscillating condensate, some particle production from the first narrow instability band occurs at late times (clearly visible in the first column). The notation is the same as in Fig. \ref{fig:pspn1}.}
   \label{fig:pspn1523}
\end{figure*}
\subsubsection{Equation of State}
In Fig. \ref{fig:EqOfState}, we can qualitatively see the  effect of transients on the equation of state (orange curves, $n=1.5, 2,3$). After the transients form, and before they decay, $w_{\rm trans}$ is expected to be matter-like since transients, like oscillons, behave collectively as pressureless matter. Note that because of their short lived nature, this stage is hard to see cleanly in the behavior of the equation of state.

After the transients decay away we get $w=1/3$, for {\it all} $n>1$. This can be understood by looking at the evolution of the fraction of energy stored in the form of kinetic, gradient and potential energies, as shown in Fig. \ref{fig:virial}. Numerically, we find that after the transients decay, they leave a completely virialized \cite{Boyanovsky:2003tc} inflaton:
\Beq
\label{eq:Equipartition}
\frac{1}{2}\braket{\dot{\phi}^2}_{\rm{s,t}}=\frac{1}{2}\braket{(\nabla\phi/a)^2}_{\rm{s,t}}+n\braket{V}_{\rm{s,t}}\,.
\Eeq
\begin{figure*}[t] 
   \centering
   \includegraphics[width=2.9in]{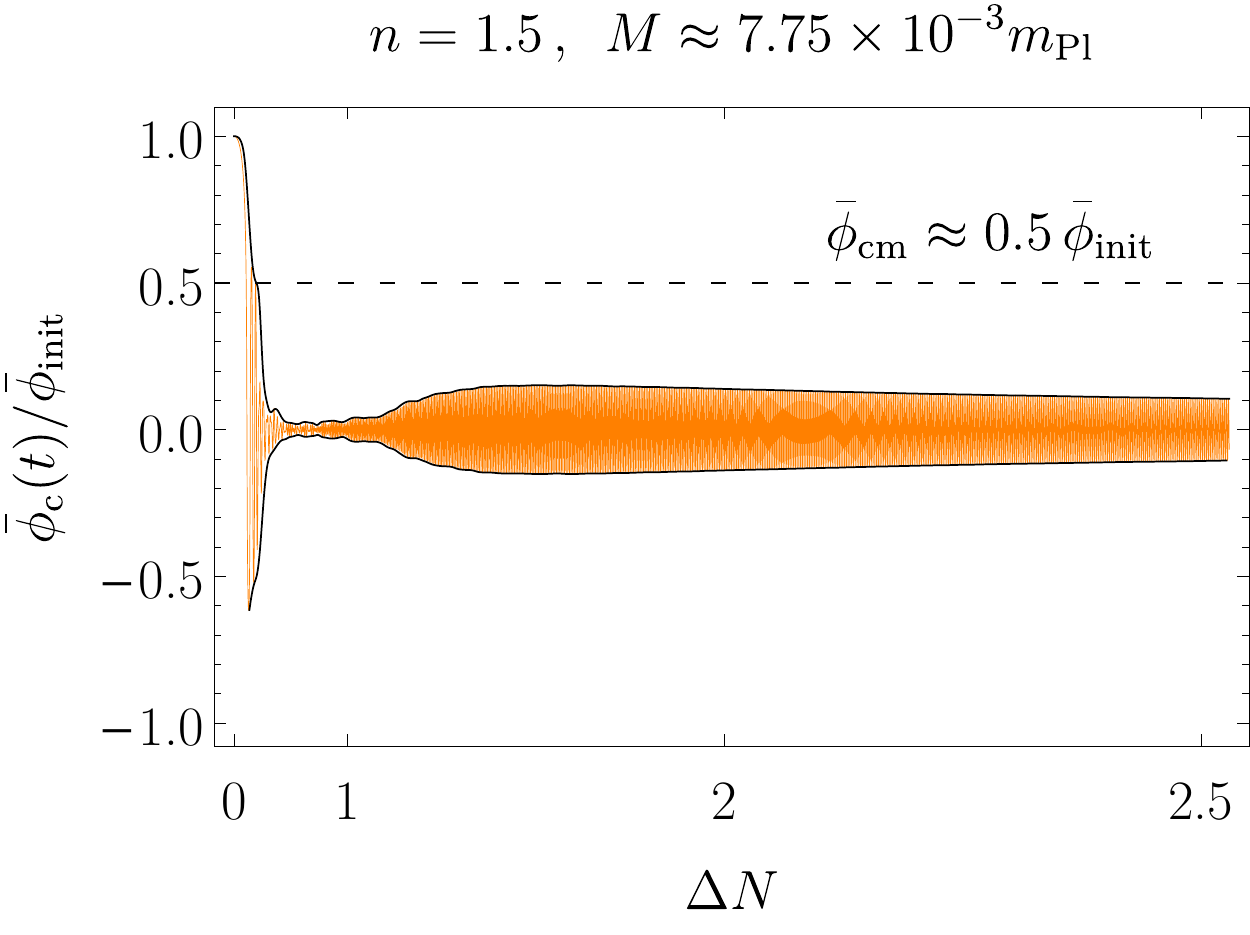}
   \hspace{0.2in} 
   \includegraphics[width=3.0in]{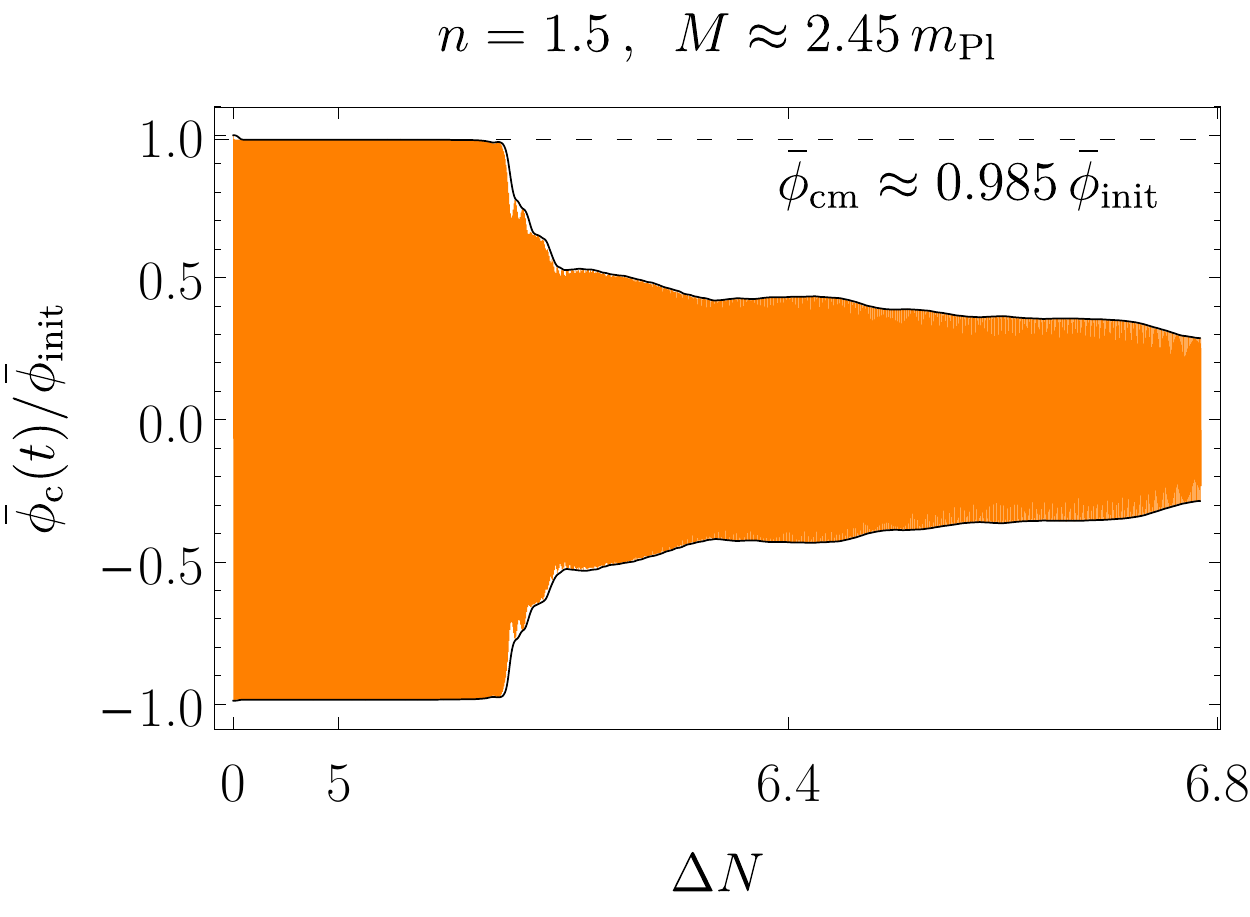}
   \caption{The evolution of the spatially averaged inflaton field (for the T-models). The initial value of the field is $\bar{\phi}_{\rm{init}}$ (defined at the end of inflation). We plot the rescaled `conformal' value $\bar{\phi}_{\rm{c}}(t)\equiv (a/a_{\rm{init}})^{3/(n+1)}\bar{\phi}(t)$ to compensate for the redshifting of the amplitude of inflaton oscillations. The left panel is for a case when the mass $M$ is small enough for the broad instability band to play a major role and lead to the formation of transients. Shortly after initialization, the condensate undergoes several oscillations with a nearly constant conformal amplitude, $\bar{\phi}_{\rm{c,m}}\approx0.5\bar{\phi}_{\rm{init}}$ (for Monodromy models, it can undergo $10$s of oscillations). During this period, the broad instability band excites multiple modes, eventually causing backreaction. The oscillating condensate disappears for about an {\it e}-fold of expansion, reflecting its complete fragmentation and the formation of localized transients. As the objects decay away, a part of the condensate reappears, because a non-trivial dynamical equilibrium is established \cite{Khlebnikov:1996mc,Micha:2004bv,Micha:2002ey}. The new condensate, however, is increasingly subdominant in energy. The right panel is for a case when $M$ is so large, that only the first narrow instability band plays an important role. The condensate undergoes many oscillations with conformal amplitude very close to $\bar{\phi}_{\rm{init}}$ (this is typical for $M\gtrsim\mpl$). As the slow, but steady particle production due to the first narrow band causes backreaction, the condensate fragments partially. Nonlinear transient objects do not form in this case, but just like after their decay in the other case, the remnant condensate becomes energetically less important with time. }
   \label{fig:means}
\end{figure*}
Using this in eq. \eqref{eq:EoS} we have\footnote{If we assume that $\left<\frac{\braket{...}_{\rm{s}}}{\braket{...}_{\rm{s}}}\right>_{\rm{t}}=\frac{\braket{...}_{\rm{s,t}}}{\braket{...}_{\rm{s,t}}}$, which turns out to be an excellent approximation at late times.}
\Beq
w=\frac{1}{3}+\frac{2}{3}\frac{(n-2)}{\left(n+1\right)+\frac{\braket{(\nabla\phi/a)^2}_{\rm{s,t}}}{\braket{V}_{\rm{s,t}}}}\rightarrow \frac{1}{3}+\hdots\,
\Eeq
The last implication, can be understood from Fig. \ref{fig:virial}, were we can see $\braket{V}_{\rm{s,t}}\ll \braket{\dot{\phi}^2}_{\rm{s,t}},\braket{(\nabla\phi/a)^2}_{\rm{s,t}}$.  

The fact that $w\rightarrow 1/3$ is somewhat unexpected result (at least for $n<2$). The reason why it is surprising is the following. Recall that the density of a condensate oscillating in $V\propto|\phi|^{2n}$ redshifts as $\rho\propto a^{-6n/(n+1)}$,\footnote{For a rapidly oscillating scalar condensate in an expanding universe $0=\braket{\phi(\ddot{\phi}+\partial_{\phi}V)}_{\rm t}=-\braket{\dot{\phi}^2}_{\rm t}+\braket{\phi\partial_{\phi}V}_{\rm t}$, from which follows $\braket{\dot{\phi}^2}_{\rm t}=2n\braket{V}_{\rm t}$. After substitution in eq. \eqref{eq:EoS} we obtain eq. \eqref{eq:EoShom}, whereas using energy conservation $\dot{\rho}+3H(\rho+p)=0$, we arrive at $\rho\propto a^{-6n/(n+1)}$.\label{refnote}} i.e., slower than radiation for $n<2$. Hence, for such $n$, whatever condensate (coherent low-$k$ modes) is left after the decay of the transients, its energy should redshift more slowly than the energy stored in the relativistic modes ($\rho_{\rm rel}\propto a^{-4}$) and eventually become the dominant component, yielding the homogeneous equation of state \cite{Turner:1983he,Johnson:2008se}
\Beq
\label{eq:EoShom}
w_{\text{hom}}=\frac{n-1}{n+1}\,.
\Eeq 
Instead, numerically we find that power cascades slowly towards the UV akin to the turbulent evolution described in \cite{Micha:2004bv,Micha:2002ey} and that the energy stored in the relativistic modes always dominates over the remnant condensate, leading to $w\rightarrow1/3$, for all $n>1$. This is a purely nonlinear effect.
\\ \\
\noindent\small{{\bf B2}: $\boldsymbol {n>1,M\sim \mpl}$
\setcounter{subsubsection}{0}
\subsubsection{Slow particle production}
\label{sec:RadDom}
In this case the expansion of the universe immediately after inflation is more important than the particle production from the broad resonance band. For all $n$ the amplitude of $\bar{\phi}$ decays rapidly, and it does not undergo a significant number of oscillations while the low-$k$ modes lie in the broad instability band (see Fig. \ref{fig:Floq}, $n>1$ cases). Our simulations indeed reveal that none of the $\delta\phi_{\bk}$ grow much initially. They experience brief excitations due to the crossing of multiple instability bands, but not large enough to back-react on the condensate. $\bar{\phi}(t)$ continues to oscillate around the bottom of $V\propto|\phi|^{2n}$.

However, as Fig. \ref{fig:Floq} suggests, despite the fast expansion of the universe compared to the growth rate of the perturbations, narrow resonance effects play an important role at late times for $n>1$. Co-moving modes within the narrow bands remain unstable as Hubble expansion drives $\bar{\phi}\rightarrow0$ (see the discussion in Sec. \ref{sssec:NKB}). 

In the right column in Fig. \ref{fig:pspn1523}, we show the inflaton spectra obtained from lattice simulations at different times. After a short period of excitation of low-$k$ modes, $\bar{\phi}$ has decayed significantly (due to expansion). Thereafter, it is the narrow instability bands that become important. The modes within the first narrow resonance band grow ever faster when compared to the expansion time scale, since $|\Re(\mu_{k})|\propto m$ and $H\sim m\bar{\phi}/\mpl$, i.e., $|\Re(\mu_{k})|/H\propto \mpl/\bar{\phi}$, developing a prominent narrow \textit{primary} peak. This peak is shifted towards higher co-moving wavenumbers  for $n<2$, and towards lower co-moving wavenumbers $n>2$. For $n=2$, it is fixed in co-moving space at $k_1=\kappa_1a(t)m(t)=\rm{const}$, $\kappa_1\approx1.27$.

Interestingly, before the deposited energy in the primary peak becomes comparable to that of the condensate, a series of \textit{secondary} peaks develops. Initially at low $k$ and then at ever higher $k$. We call them secondary because they do not follow from the linear analysis (the linear analysis yields a much slower growth near the secondary peaks). They result from rescattering processes (we confirmed their rescattering origin by removing the initial fluctuations above a certain cut-off, e.g. $k>1.2k_1$ for $n=2$). The $k$ close to $0$ appears first as a consequence of the strongest rescattering -- between particles from the primary peak and the condensate. The higher $k$ peaks then follow from `primary-primary' and `primary-secondary' rescattering processes. 

The growth of perturbations (seen in both primary and secondary peaks) is eventually shut off by the backreaction on the condensate. All peaks smear out and again (just like in the transients case) the field is virialized, see Fig. \ref{fig:virial}.  The power cascades slowly towards the UV and the energy stored in the relativistic modes ends up dominating over the remnant condensate, leading to $w\rightarrow1/3$. We re-iterate, that this is a purely nonlinear process. It has been observed in pure $\lambda\phi^4$ theory \cite{Khlebnikov:1996mc,Micha:2004bv,Micha:2002ey}. Here we see it for general $n>1$.

Note that $\bar{\phi}$ never disappears completely (see Fig \ref{fig:means}) -- it is in a non-trivial dynamical equilibrium with the highly occupied modes and if it is removed artificially it reappears due to Bose condensation (see also \cite{Khlebnikov:1996mc}).
\subsubsection{Duration to Radiation Domination}
\begin{figure}[t] 
   \centering
   \includegraphics[height=2.23in]{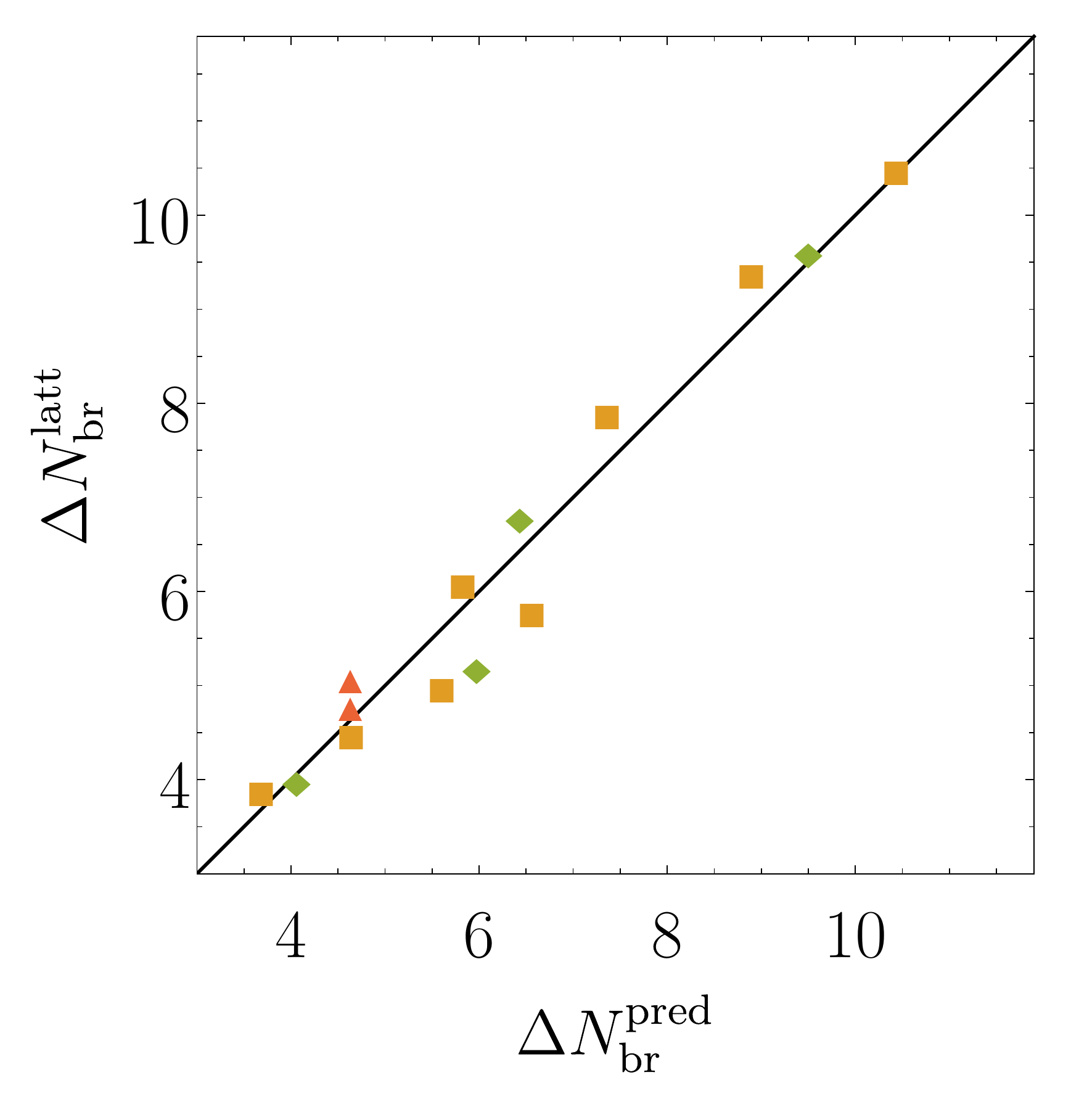}
   \hspace{1.0in}
   \includegraphics[height=2.23in]{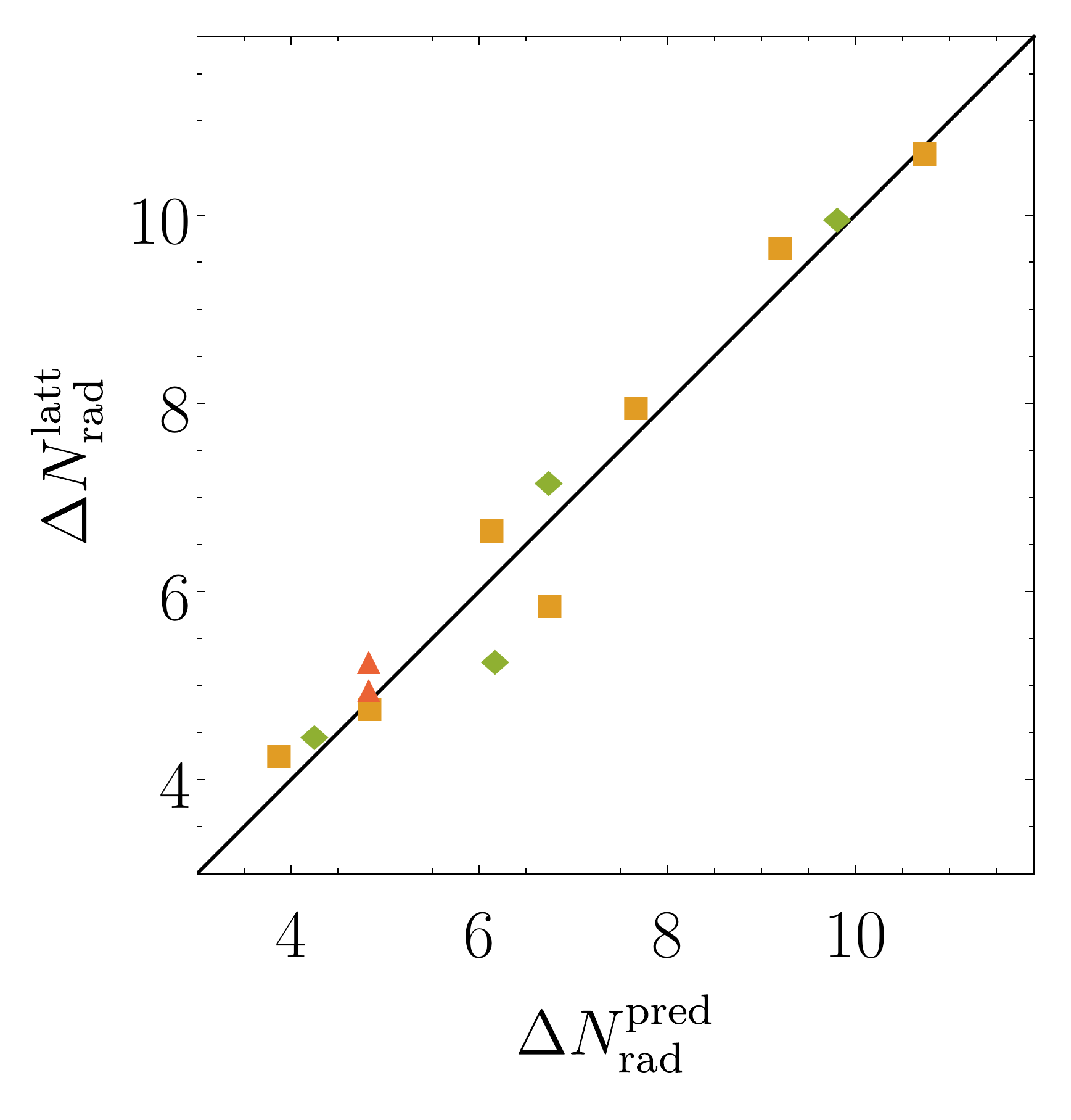}
   \caption{On the top we show the number of {\it e}-folds of expansion after the end of inflation when backreaction takes place due to particle production from the first narrow instability band. On the horizontal axis we plot the predicted values (eq. \eqref{eq:DeltaNbr} for $1<n\neq2$ and eq. \eqref{eq:DeltaNbrQuart} for $n=2$) and on the vertical axis the measured values from lattice simulations, for different $M$ and $n$. The orange squares are for the T-models, the green rhombi are for the E-models and the red triangles are for the Monodromy models for $q=0.5$, $1$. We find that for all models and parameters, the data fits the $45^{\rm o}$ degree line for $\delta\approx0.126$. We also found that changing $\delta$ to $0.100$ describes well the time the equation of state approaches $w_{\rm{rad}}=1/3$, as shown on the bottom. }
   \label{fig:DeltaN}
\end{figure}
After fragmentation, the inflaton field is again virialized with kinetic and gradient energies much greater than the potential energy and $w$ asymptoting to $1/3$.  We find that the fragmented inflaton almost immediately reaches a radiation dominated state of expansion, i.e., $\Delta N_{\text{rad}}\gtrsim\Delta N_{\text{br}}$. The expected time for backreaction from the linear analysis in Section \ref{sec:LinAnal}, see eqs. (\ref{eq:DeltaNbr},\ref{eq:DeltaNbrQuart}), agrees well with the lattice simulations (Fig. \ref{fig:DeltaN}) when the first narrow instability band plays an important role. The moment of backreaction obtained from our lattice simulations fits the linear analysis predictions for $\delta\approx0.126$, for all four models (T, E and Monodromy $q=0.5$, $1$). We also find that decreasing the small parameter to $0.100$ describes well the data for the time when $w$ settles to $1/3$. When the broad instability band causes the fragmentation of the condensate into transient objects we find that $\Delta N_{\text{rad}}\gtrsim\Delta N_{\text{br}}\sim1\,(2.5)$ for the T and E (Monodromy) models.

We note that in terms of actual values of $w$ from simulations, by radiation domination we mean the moment when the equation of state approaches, $w_{\rm rad} = 1/3 \pm 0.03$. The $\pm10\%$ width makes the effects in inflationary observables (see the next section) due to numerical uncertainties $< 1\%$.

\section{Observational Implications for Inflationary Observables}
\label{sec:ObsImpl}
The numerical studies presented in the previous section give us a new look on the expansion history of our universe. We have shown that for all potentials that are steeper than quadratic near the origin, $n>1$, and any value of $M$, the oscillating inflaton condensate fragments due to self-resonance. The equation of state approaches that of a radiation dominated universe at sufficiently late times (see Fig. \ref{fig:EqOfState}). For $M\ll \mpl$, this duration to radiation domination in $e$-folds: $\Delta N_{\rm rad}\lesssim 1$. For $M\sim\mpl$, $\Delta N_{\rm rad}$ is given by eqs. \eqref{eq:DeltaNbr} and \eqref{eq:DeltaNbrQuart}. Note that with $\delta \approx 0.126\rightarrow 0.1$, $\Delta N_{\rm br}\rightarrow \Delta N_{\rm rad}$.

The duration to radiation domination we have calculated can be used as an upper bound if we include perturbative decay to other light species. This upper bound arises because if the interactions with other relativistic species are stronger, then the production of relativistic daughter particles is even more effective than due to self-resonance, and the transition is faster. Note that if the daughter fields are massive, or there are non-perturbative dynamics, our statement about the upper bound does not hold.

These insights can help in reducing the uncertainties in the predictions of individual models of inflation without the need for a specific reheating scenario. In the remainder of the section we derive the improved predictions for two cosmological observables -- the scalar spectral index, $n_{\rm{s}}$, and the tensor-to-scalar ratio, $r$, and compare them to the most recent constraints from measurements of the CMB.
\subsection{Slow-Roll Parameters}
\label{sssec:SRP}
Slow-roll inflation yields a nearly scale-invariant power spectrum of the curvature and tensor perturbations,
\Beq
\label{eq:DR}
\Delta^2_{\mathcal{R}}(k)=A_{\rm{s}}\left(\frac{k}{k_{\star}}\right)^{n_{\rm{s}}-1}\,,\qquad\Delta^2_{\rm{t}}(k)=A_{\rm{t}}\left(\frac{k}{k_{\star}}\right)^{n_{\rm{t}}}\,,
\Eeq
where $A_{\rm s}$ and $A_{\rm t}$ characterize the amplitude, whereas $n_{\rm s}$ and $n_{\rm t}$ are the spectral indices of the scalar and tensor perturbations.
Observations have shown $A_{\rm{s}}\approx 2.2\times10^{-9}$ (with $k_{\star}=0.05\,\rm{Mpc}^{-1}$) \cite{Ade:2015lrj}. While tensors have yet to be detected, the tensor-to-scalar ratio
\Beq
r=\frac{A_{\rm{t}}}{A_{\rm{s}}}\,,
\Eeq 
is constrained to be $\lesssim 0.1$ (with $k_{\star}=0.002\,\rm{Mpc}^{-1}$)\cite{Ade:2015lrj}. The constraints on $r$ and $n_{\rm s}$ are shown in Fig. \ref{fig:Planck}. Note that while $(n_{\rm s},A_{\rm s})$ and $r$ use different $k_\star$, this is of little consequence for what follows, primarily because there is no detectable running of the scalar spectral index. Henceforth, we will use $0.002\,\rm{Mpc}^{-1}$ as the value of the pivot scale when calculating predictions for $n_{\rm{s}}$ and $r$.\footnote{For the calculation of the inflaton potential parameters in Section \ref{sec:NonLinDyn}, we have used the value of $A_{\rm{s}}$ measured at $k_{\star}=0.05\,\rm{Mpc}^{-1}$. We have checked that our results do not change with the variation of the model parameters for pivot scales in the range $0.002\,{\rm{Mpc}}^{-1}\leq k_{\star}\leq 0.05\,\rm{Mpc}^{-1}$, assuming eq. \eqref{eq:DR}.}

Connecting $A_{\rm s}$, $n_{\rm s}$ and $r$ to the inflationary potential in the slow-roll approximation, we have
\Beq
\label{eq:nrVstar}
A_{\text{s}}&=\frac{1}{12\pi^2}\frac{V_\star^3}{\mpl^6V_\star'{}^2}\,,\\
n_{\rm{s}}&=1-3\mpl^2\left(\frac{V_{\star}'}{V_{\star}}\right)^2+2\mpl^2\frac{V_{\star}''}{V_{\star}}\,,\\
r&=8\mpl^2\left(\frac{V_{\star}'}{V_{\star}}\right)^2\,,
\Eeq
where $V_{\star}\equiv V(\phi_{\star})$, etc. and $\phi_{\star}$ is the value of the inflaton field at the time when the co-moving pivot scale crossed the Hubble radius $k=k_{\star}=a_{\star}H_{\star}$. 
\subsection{$e$-folds of Inflation}
The number of $e$-folds of expansion before the end of inflation, $N_{\star}$, when the pivot scale exited the horizon is
\Beq
\label{eq:NstDef}
N_{\star}\approx \left|\int_{\phi_{\star}}^{\phi_{\rm{end}}}\frac{V}{V'}\frac{d\phi}{\mpl^2}\right|\,.
\Eeq
Here $\phi_{\rm{end}}$ is the value of the inflaton field at the end of inflation, $\ddot{a}=0$.\footnote{In principle, one can calculate $\phi_{\rm{end}}$ from the inflaton dynamics during inflation. Nevertheless, a good estimate, sufficient for our purposes, can be obtained from setting the potential slow-roll parameter, $\epsilon_{\,\!_{\rm{V}}}=\mpl^2\left(V'/V\right)^2/2$, to $1$.}

In conventional studies of reheating $N_\star$ is effectively treated as a free parameter in the range $50<N_{\star}<60$ due to uncertainties related to the post-inflationary expansion history. 
With the current study, we have a better understanding of $w(\Delta N)$ and an upper bound on $\Delta N_{\rm{rad}}$ for the models under consideration. Thus, we can treat $N_\star$ as a known quantity (with a known variation) and use it to constrain inflaton potential parameters. To this end we need an expression for $N_\star$ from the usual mapping of modes between horizon crossing, $k_\star=a_\star H_\star$, during inflation and re-entry, $k_\star=a_0 H_0$, at late times. We start from \cite{Dodelson:2003vq,Liddle:2003as}
\Beq
\label{eq:kstar}
\frac{k_\star}{a_0 H_0}&=\frac{a_\star H_\star}{a_0 H_0}\,,\\
&=\frac{a_\star}{a_{\rm{end}}}\frac{a_{\rm{end}}}{a_{\rm{rad}}}\frac{a_{\rm{rad}}}{a_0}\frac{\rho_{\rm{rad}}^{1/4}}{H_0}\frac{\rho_{\rm{end}}^{1/4}}{\rho_{\rm{rad}}^{1/4}}\frac{H_\star}{\rho_{\rm{end}}^{1/4}}\,,
\Eeq
where $a_{\rm end}$ and $\rho_{\rm end}$ are the scale factor and energy density at the end of inflation, $N_\star\equiv\ln(a_{\rm{end}}/a_\star)$ and $\Delta N_{\rm{rad}}\equiv\ln(a_{\rm{rad}}/a_{\rm{end}})$. Note that $\rho_{\rm{rad}}$ is the mean energy density at the beginning of radiation domination, $w\rightarrow w_{\rm{rad}}=1/3$, which is captured by our lattice simulations. The universe need not be in thermal equilibrium at that time. In fact, thermal equilibrium could be reached much later when
\Beq
\label{eq:rhoth}
\rho_{\rm{rad}}^{1/4}a_{\rm{rad}}=\rho_{\rm{th}}^{1/4}a_{\rm{th}}\,,
\Eeq
where we assume that the universe is dominated by relativistic degrees of freedom while $a_{\rm{rad}}<a<a_{\rm{th}}$. Taking the logarithm of eq. \eqref{eq:kstar} after plugging-in eq. \eqref{eq:rhoth} and $H_{\star}^2\approx V_\star/(3\mpl^2)$ yields
\Beq
\label{eq:Nst1}
N_\star\approx&\,\ln\left(\frac{\rho_{\rm{th}}^{1/4}}{\sqrt{3}H_0}\frac{a_{\rm{th}}}{a_0}\right)+\frac{1}{4}\ln\left(\frac{V_\star^2}{\mpl^4\rho_{\rm{end}}}\right)\\
        &-\ln\left(\frac{k_\star}{a_0 H_0}\right)-\Delta N_{\rm{rad}}+\frac{1}{4}\ln\left(\frac{\rho_{\rm{rad}}}{\rho_{\rm{end}}}\right)\,.
\Eeq
The first term evaluates to $\approx 66.89-(1/12)\ln g_{\rm th}$ where $g_{\rm th}$ is the number of bosonic degrees of freedom in the early universe, and the number $66.89$ is determined from late universe cosmological observations which do not rely on the inflationary potential (see Section \ref{ssec:TRD} for a derivation). The second term depends only on the inflaton potential parameters and $\phi_\star$. The third term is a function of the pivot scale which we have already fixed. The fourth and fifth terms is where we make contact with our present work.
\begin{figure*}[t] 
   \centering
   \includegraphics[width=2.8in]{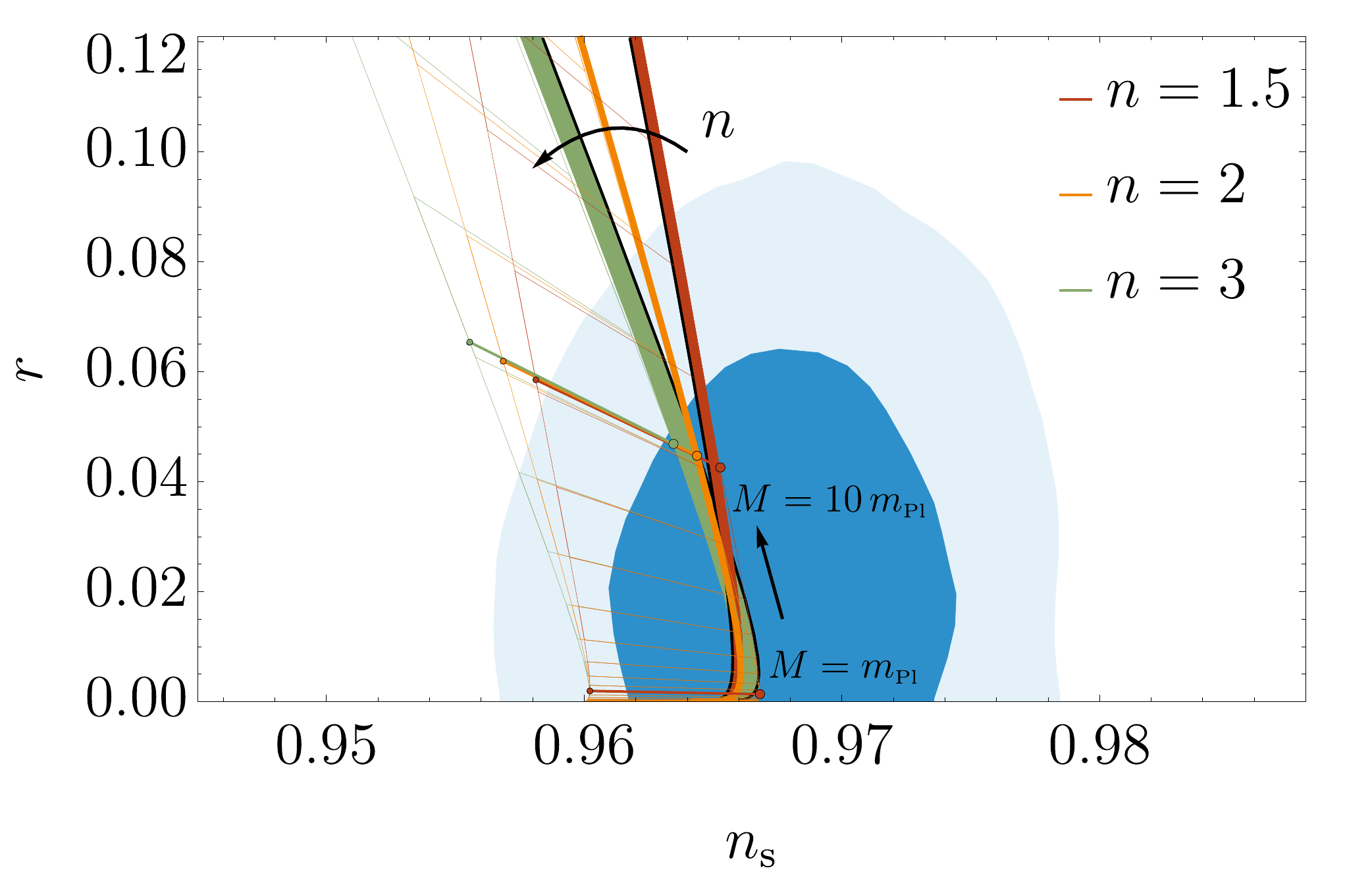} 
    \hspace{0.2in}
   \includegraphics[width=2.8in]{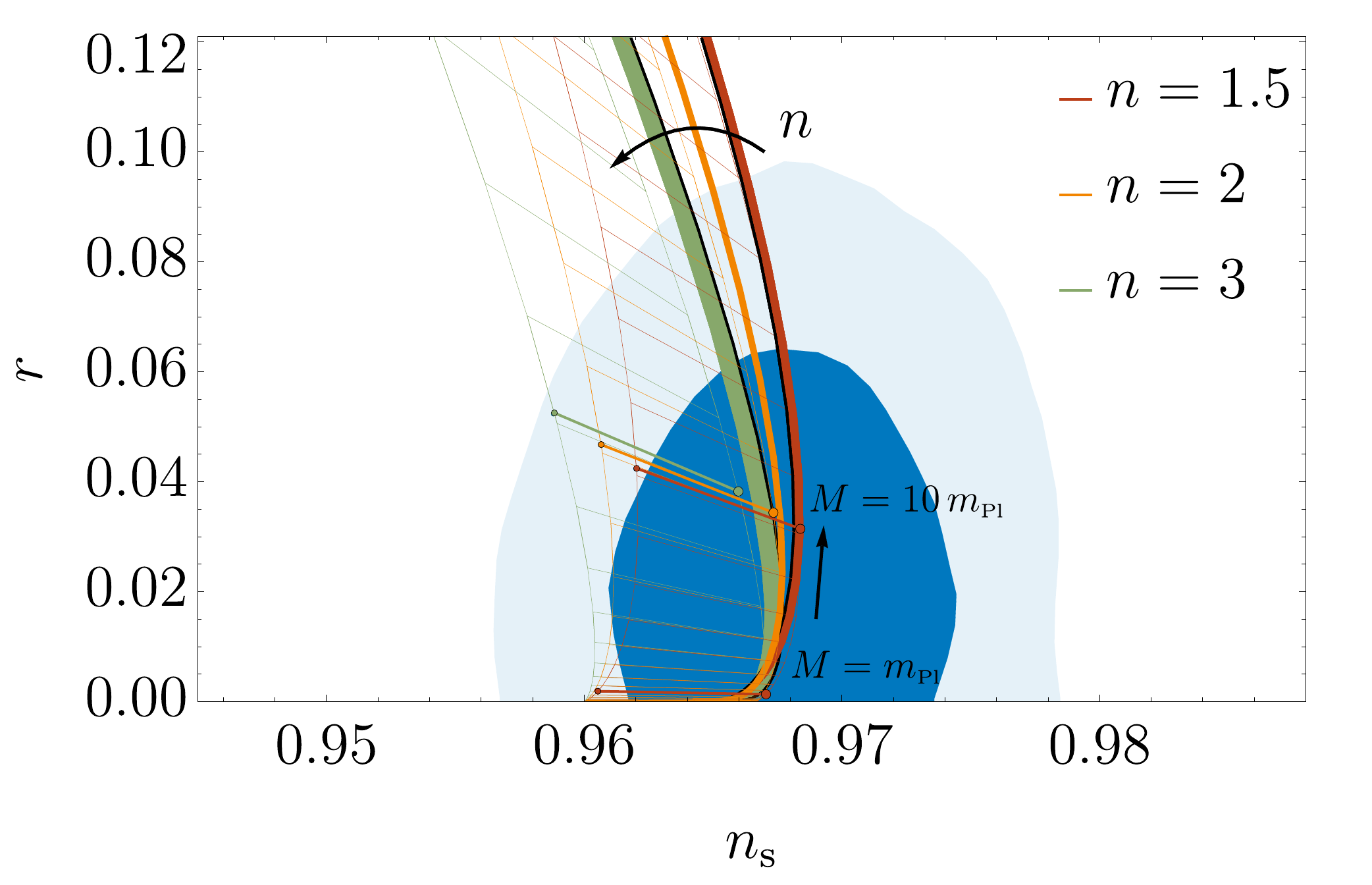} \\ \vspace{0.2in}
   
   \includegraphics[width=2.8in]{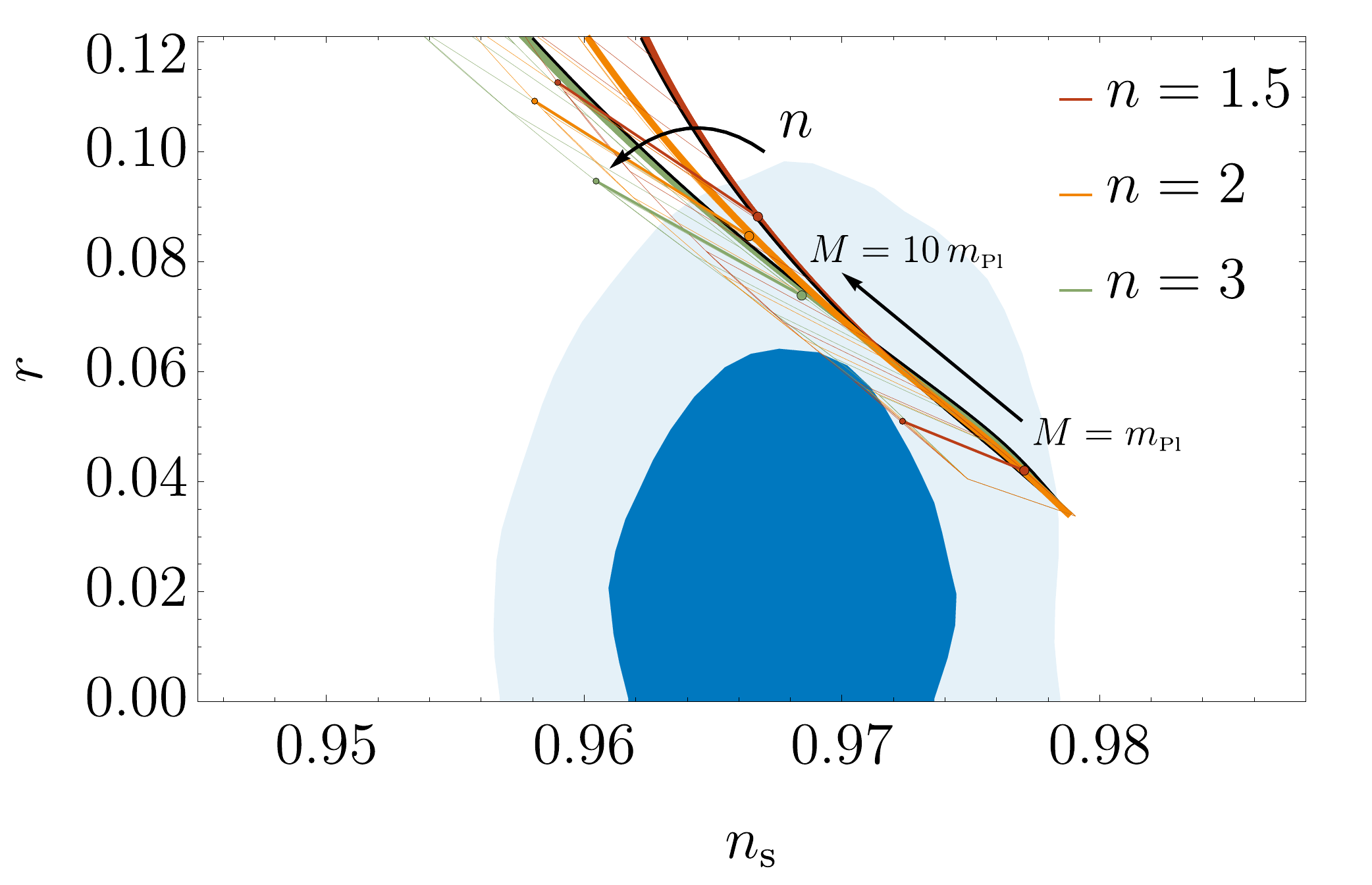} 
   \hspace{0.2in}
   \includegraphics[width=2.8in]{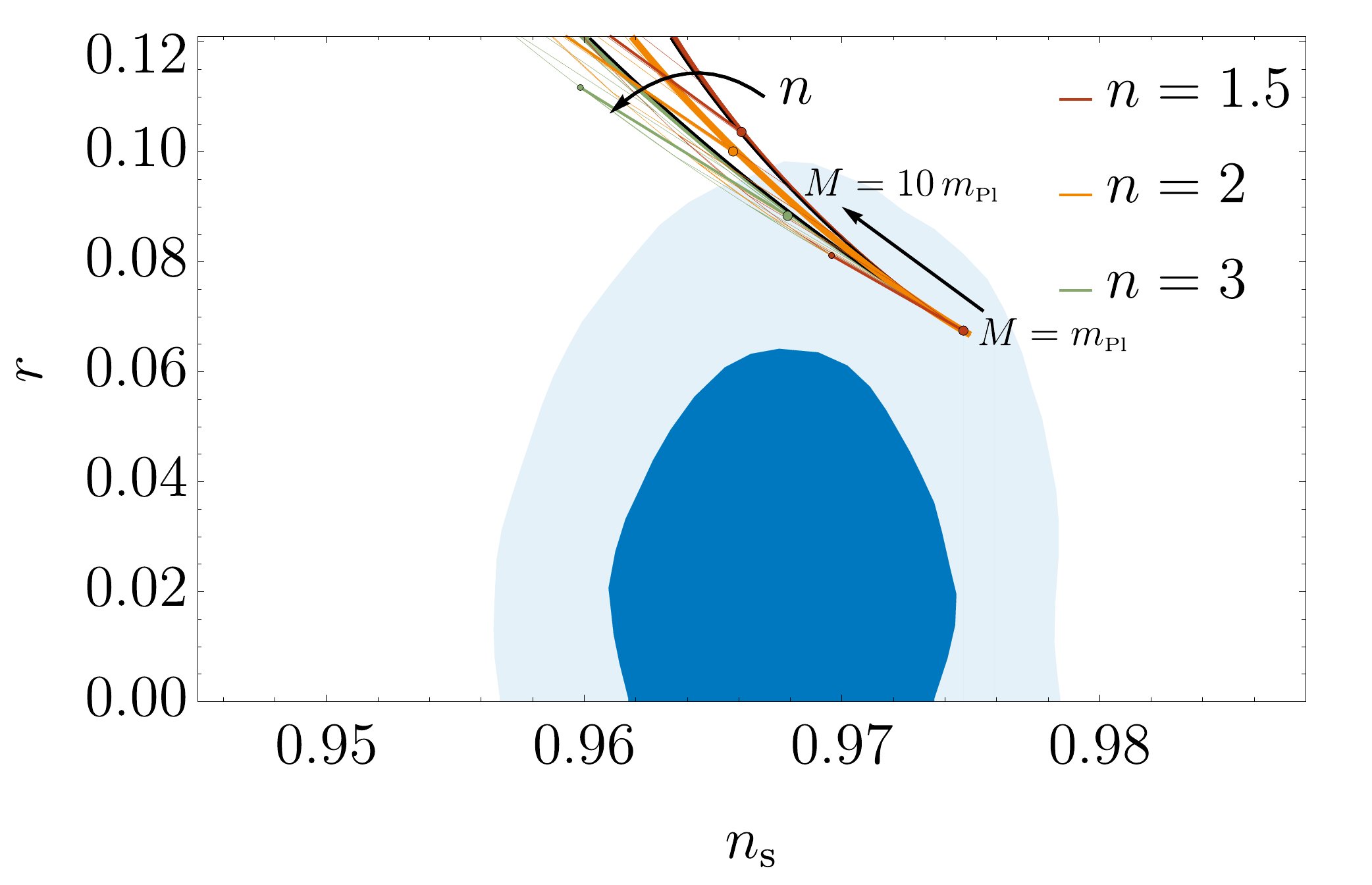} 
   \caption{Top row from left to right: $\alpha$-attractor T and E models. Bottom row from left to right: Monodromy models, $q=0.5,\,1$. The recently reported constraints on $r$ and $n_{\text{s}}$ by the Planck Collaboration \cite{Ade:2015lrj} are shown with the blue shaded regions. Our predictions for $n=1.5,\,2,\,3$ are given by the thick green, orange and red lines for different values of $M$. As the straight arrows indicate, as we increase $M$ we move up the lines. The width of each thick line reflects the uncertainty from our analysis -- $\Delta N_{\rm{rad}}\leq\Delta N_{\rm{rad}}^{\rm{latt}}$, i.e., the uncertainty from coupling the inflaton to other light fields. The black edges are for the upper bound. On the other hand the broad green, orange and red striped bands give the standard predictions for the same $M$ and $n$. The width of each band accounts for the standard reheating-related uncertainties -- $50<N_\star<60$. Our bounds on the expansion history after inflation, reduce the uncertainties in $N_\star$ significantly and hence in the predictions for the two CMB observables, as can be seen in the figures.}
   \label{fig:Planck}
\end{figure*}

We have shown in Section \ref{sec:Trans} that even in the absence of couplings to other fields, for generic, observationally consistent potentials with $n>1$, we reach $w\rightarrow 1/3$. Thus, $\Delta N_{\rm rad}$ and $\rho_{\rm end}/\rho_{\rm rad}$ are bounded from above and these bounds can be calculated from the parameters of the inflationary potential. Even if additional relativistic species are coupled to the inflaton,\footnote{As long the decay is perturbative to avoid metastable, nonlinear states.} our calculated $\Delta N_{\rm rad}$ and $\rho_{\rm end}/\rho_{\rm rad}$ are robust upper bounds. The lower bound is of course $\Delta N_{\rm rad}=0$ and $\rho_{\rm end}/\rho_{\rm rad}=1$. Thus the fourth and fifth terms together are bounded, and calculable from the inflationary potential parameters, which reduces the uncertainty in $N_\star$ significantly. 

As an example, note that for $n>1,M\sim \mpl$, the expansion history takes a very simple form since the transition from the homogeneous equation of state $w_{\rm hom}=(n-1)/(n+1)$ to $w_{\rm rad}=1/3$ happens in less than an $e$-fold (see the green curves in Fig. \ref{fig:EqOfState}). Hence 
\Beq
\label{eq:rhoend}
\rho_{\rm{end}}a_{\rm{end}}^{6n/(n+1)}\approx \rho_{\rm{rad}}a_{\rm{rad}}^{6n/(n+1)}\,,
\Eeq
finally leading to
\Beq
\label{eq:NstarFinal}
N_\star\approx &\,66.89-\frac{1}{12}\ln\left(g_{\rm{th}}\right)+\frac{1}{4}\ln\left(\frac{V_\star^2}{\mpl^4\rho_{\rm{end}}}\right)\\
        &-\ln\left(\frac{k_\star}{a_0 H_0}\right)+\frac{n-2}{2(n+1)}\Delta N_{\rm{rad}}\,.
\Eeq
If $\Delta N_{\rm rad}$ is given by eq. \eqref{eq:DeltaNbr}, this provides an upper bound on $N_\star$. While we considered $n>1,M\sim \mpl$ case as an example, a similar analysis can be done for $M\ll \mpl$. 

Apart from the logarithmic dependence on the unknown $g_{\rm th}$, the upper bound on $N_{\star}$ only depends on $\phi_\star$ and the parameters in the inflationary potential $(n,M)$, also ($q$ for Monodromy models). We can now solve eqs. \eqref{eq:NstDef}, \eqref{eq:NstarFinal} and the $A_{\rm s}$ constraint in \eqref{eq:nrVstar} simultaneously for $N_{\star}$, $\Lambda$ and $\phi_\star$, for given values of $M$, $n$ (and $q$) and substitute the results in the expressions for $n_{\rm s}$ and $r$ in eq. \eqref{eq:nrVstar}. 

The resulting $n_{\text{s}}(M,n)$ and $r(M,n)$ are shown in Fig. \ref{fig:Planck} as thick green, orange and red lines for $n=1.5$, $2$, $3$, respectively.\footnote{We have used $g_{\rm{th}}=10^3$, however, letting it vary in the range $1-10^5$, does not change the location and the thickness of the lines in any visible way.} The width of each line reflects the uncertainty from coupling to other light fields, i.e., $\Delta N_{\text{rad}}\leq\Delta N_{\text{rad}}^{\text{latt}}(M,n)$. This width  reflects the uncertainty in $N_\star$ in this model. For comparison we also give the predictions for the same $M$, $n$ (and $q$), assuming the standard reheating related uncertainties $50<N_\star<60$, with the slanted thin lines. The figures clearly indicate that our analysis significantly reduces the theoretical uncertainties in the predictions for the CMB observables.
\subsection{Thermalization and Radiation Domination}
\label{ssec:TRD}
We end this subsection by discussing the distinct roles of thermalization and radiation domination, and a derivation of $\approx 66.89-(1/12)\ln g_{\rm th}$ used for the first term in eq. \eqref{eq:Nst1}.

Note that the ratios $a_{\rm{th}}/a_{\rm{rad}}$ and $\rho_{\rm{th}}/\rho_{\rm{rad}}$ do not appear in eq. \eqref{eq:Nst1}. Their absence is a manifestation of the importance of the expansion history over the thermal history of the universe in the context of mapping cosmological perturbation modes to early times. It is sufficient to know the evolution of the scale factor and the moment when $w\rightarrow1/3$. The value of the redshift at which the universe reached local thermal equilibrium has no effect on the mapping of the modes as long as $w=1/3$, while $a_{\rm{rad}}<a<a_{\rm{th}}$. Thus, one can in principle employ classical lattice simulations to calculate the expansion history up to $a=a_{\rm{rad}}$ and, thereby, connect inflationary predictions with observations without having to worry about thermalization, $T_{\rm{th}}$ and the end of reheating as a whole.\footnote{It is worth mentioning that re-emergence of moduli domination \cite{Kane:2015jia} after the initial radiation domination obviously complicates our analysis.}

To show that this is indeed the case, let us calculate the first term in eq. \eqref{eq:Nst1}. It is reasonable to assume that entropy, $s\sim g T^3$, ($g$ being the number of effective bosonic degrees of freedom in thermal equilibrium) is conserved between the end of reheating and today
\Beq
s_{\rm{th}}a_{\rm{th}}^3=s_{\rm{0}}a_{\rm{0}}^3\,,
\Eeq
whence, 
\Beq
\label{eq:firstterm}
\frac{\rho_{\rm{th}}^{1/4}}{\sqrt{3}H_0}\frac{a_{\rm{th}}}{a_0}&=\frac{\left(\pi^2g_{\rm{th}}T_{\rm{th}}^4/30\right)^{1/4}}{\sqrt{3}H_0}\left(\frac{g_{\rm{0}}T_{\rm{0}}^3}{g_{\rm{th}}T_{\rm{th}}^3}\right)^{1/3}\,,\\
                                                               &=\left(\frac{\pi^2}{30}\right)^{1/4}\frac{g_0^{1/3}g_{\rm{th}}^{-1/12}}{\sqrt{3}}\frac{T_0}{H_0}\,,\\
                                                               &\approx e^{66.89}g_{\rm{th}}^{-1/12}\,,
\Eeq
where $T_0=0.235\,\rm{meV}$ is the CMB temperature today, $g_0=43/11$ and $H_0=67.6\,\rm{km}\,\rm{s}^{-1}\,\rm{Mpc}^{-1}$.  The effective bosonic d.o.f, $g_{\rm{th}}$, is largely unknown. At the time of big bang nucleosynthesis, $g_{\rm{BBN}}=106.75$. In our analysis, we let $g_{\rm{th}}=1-10^5$. Luckily, due to the small power it is raised by in eq. \eqref{eq:firstterm} and the fact that it is inside a logarithm in the equation for $N_\star$, $g_{\rm{th}}$ does not affect the predictions for $n_{\rm{s}}$ and $r$ significantly.

We end this section by noting that with additional assumptions, it is possible to connect reheating temperature (temperature at thermalization) to the parameters of the inflationary potential. One possible assumption is to make $a_{\rm rad}\approx a_{\rm th}$, and hence $\rho_{\rm th}\approx \rho_{\rm rad}$. Since we can calculate $\rho_{\rm rad}$ from the parameters in the inflationary potential, we can use $\rho_{\rm rad}\approx\rho_{\rm th}=\pi^2g_{\rm th}T_{\rm th}^4 /30$ to determine $T_{\rm th}$ (with a dependence on $g_{\rm th}$). If observational constraints (such as $T_{\rm th}>T_{\rm BBN}$) or theoretical prejudice constrains $T_{\rm th}$, one can use these to either constrain inflationary parameters, or the strength of couplings of the inflaton to other fields. A more detailed discussion is delegated to an Appendix.

\section{Discussion of Numerics and Assumptions}
\label{sec:Discussion}
\begin{figure}[t] 
   \centering
   \includegraphics[width=3.0in]{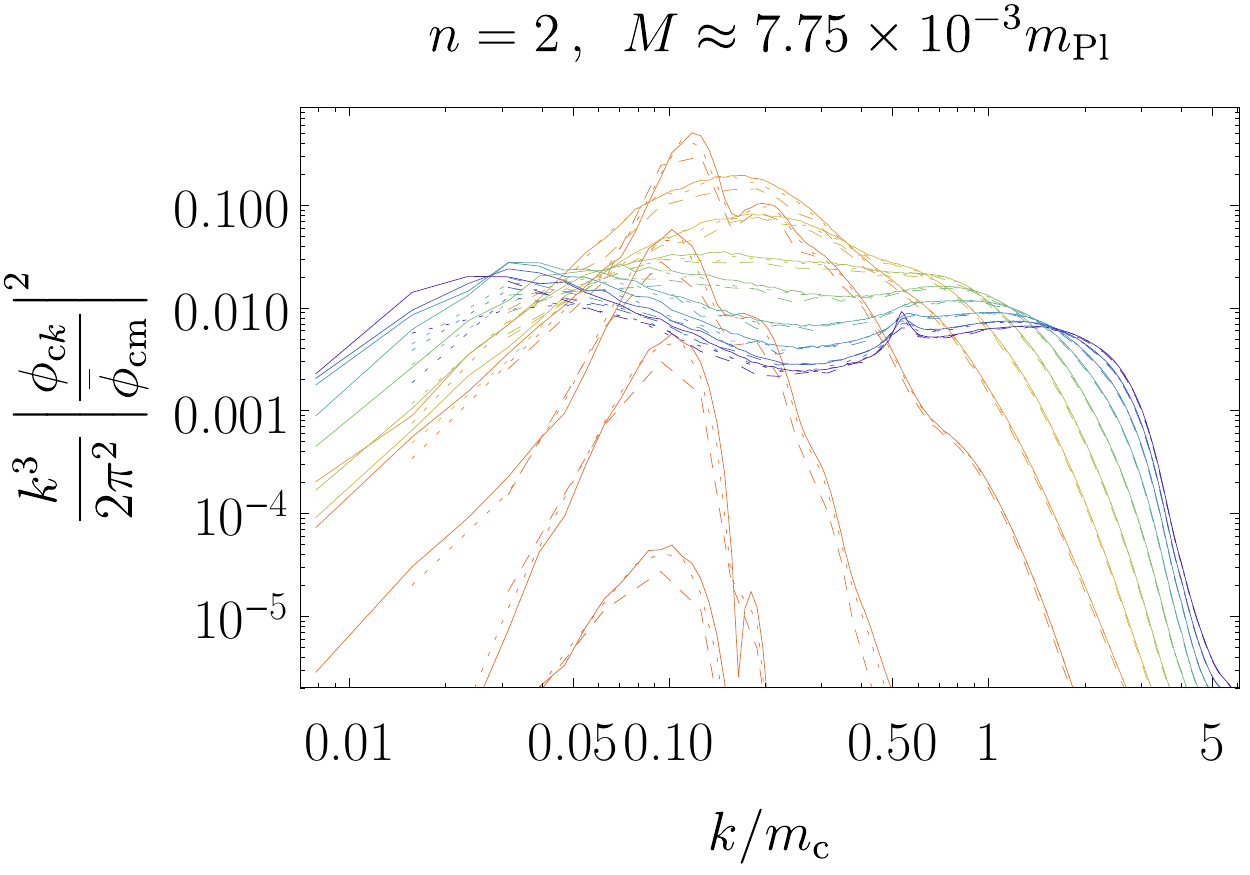}
   \caption{The evolution of the power spectrum for a case when transients form and then decay away. The data is from three simulations, differing in their IR resolution. The solid lines are for a $1024^3$ run, the dotted for a $512^3$ run and the dashed for a $256^3$ run. After the initial particle production in a broad co-moving band, the condensate fragments and forms transient objects, as indicated by the broad peak at intermediate times. As the transients decay, the broad peak goes away, power cascades towards the UV, with low-energy long wavelength modes also being excited. Importantly, the IR cut-off does not affect the evolution of the intermediate and short wavelength modes which dominate the energy budget.}
   \label{fig:pspn2alphaemins5Ncheck}
\end{figure}
\begin{figure}[t] 
   \centering
   \includegraphics[width=3.0in]{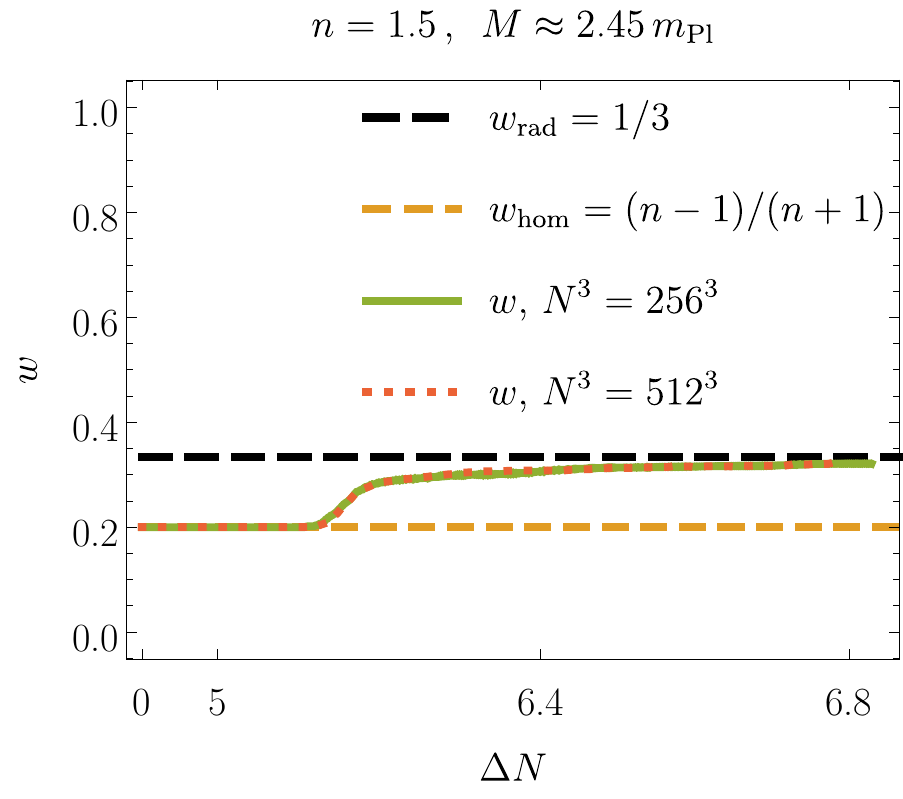}
   \caption{The evolution of the equation of state for a case when $M$ is large enough so that it particle production from the first narrow instability band that causes backreaction and fragmentation. The green solid and the red dotted lines are for two runs with the same IR cut-off, but different UV resolution. The inflaton dynamics and $w$ are not affected by the UV cut-off for the duration of the simulations.}
   \label{fig:wtanh15alpha0check}
\end{figure}
\subsection{Numerics}
\label{ssec:Numerics}
We have carried out various numerical checks to test the robustness of our results against changes in the Infrared (IR -- size of simulation box) and Ultraviolet (UV -- lattice spacing) resolutions of our simulations. While often a $256^3$ lattice turned out to be large enough to cover the relevant dynamical range, we typically ran $512^3$, sometimes going up to $1024^3$ to make sure that finite resolution effects do not lead to spurious results.
\\ \\
\noindent{(a) \it $n=1$ cases}:
\\ \\
For the case when $n=1$, but $M\ll\mpl$ we have copious oscillon production. The numerical challenge is two-fold. (i) We must make sure that the full width of the initial broad low-momentum instability band is captured (setting the IR cut-off). (ii) At the same time we have a large enough UV cut-off that allows for the resolution of the small (with respect to the horizon) objects. The second condition is especially relevant for oscillons since the lattice is fixed in co-moving space, i.e., expands along with the universe, and oscillons have fixed physical size. We inevitably run out of resolution on small sales. A $512^3$ run with the product of the physical length of the edge of the lattice and the Hubble parameter at the time of backreaction $(LH)_{\rm{br}}\sim0.1$, allows to keep track of the oscillons for about 1-2 {\it e}-folds of expansion after their formation, while also resolving the broad resonance band at earlier times. 

For $n=1$, $M\sim \mpl$, the simulations are relatively straightforward. In this case, very little energy is transfered from the homogeneous inflaton to relativistic modes, backreaction never takes place due to self-resonance, the condensate remains intact and keeps oscillating (however, if gravity is included at first order, the condensate must eventually fragment \cite{Easther:2010mr}; this is beyond the scope of our paper). 
\\ \\
\noindent{(b) \it $n>1$ cases}:
\\ \\
When $n>1, M\ll \mpl$ and transients form, we are faced with the same challenges as in the oscillons case, with some differences. Let us first focus on the IR-cutoff set by the size of our simulation box. As transient objects decay, most of the power is transferred to UV modes with a smaller fraction of the energy going to the IR modes. We have verified that the IR cut-off does not affect the dynamics on small scales or the features on intermediate scales (see Fig. \ref{fig:pspn2alphaemins5Ncheck}), primarily because the IR modes tend to be energetically subdominant at late times. In particular, the IR cut-off does not affect the evolution of the equation of state, which in this case is determined by the high energy modes.

 We note that the low $k$-modes, while not relevant for the equation of state, are affected by the IR-cutoff {\it and} UV dynamics because of non-trivial dynamical equilibrium \cite{Khlebnikov:1996mc,Micha:2004bv,Micha:2002ey} between the energetic (i.e., UV) and subdominant (i.e., intermediate and IR) modes. The quantity that we find to be affected by the IR-cutoff is the spatial average of the inflaton, $\bar{\phi}(t)$, after transients decay (see Fig. \ref{fig:means}). The amplitude of the oscillations of the condensate that forms after transients go away turns out to vary weakly (normally decrease) as we improve the IR resolution. This is reasonable, since we capture more long wave modes (the non-trivial dynamical equilibrium also implies in general some variation in the occupation number of the IR modes). But again, since the IR modes and the condensate in particular are subdominant in energy at late times, their sensitivity on the IR resolution has a negligible effect on the distribution of energy between kinetic, gradient and potential energies. 

\begin{figure*}[t] 
   \centering
   \includegraphics[width=4.5in]{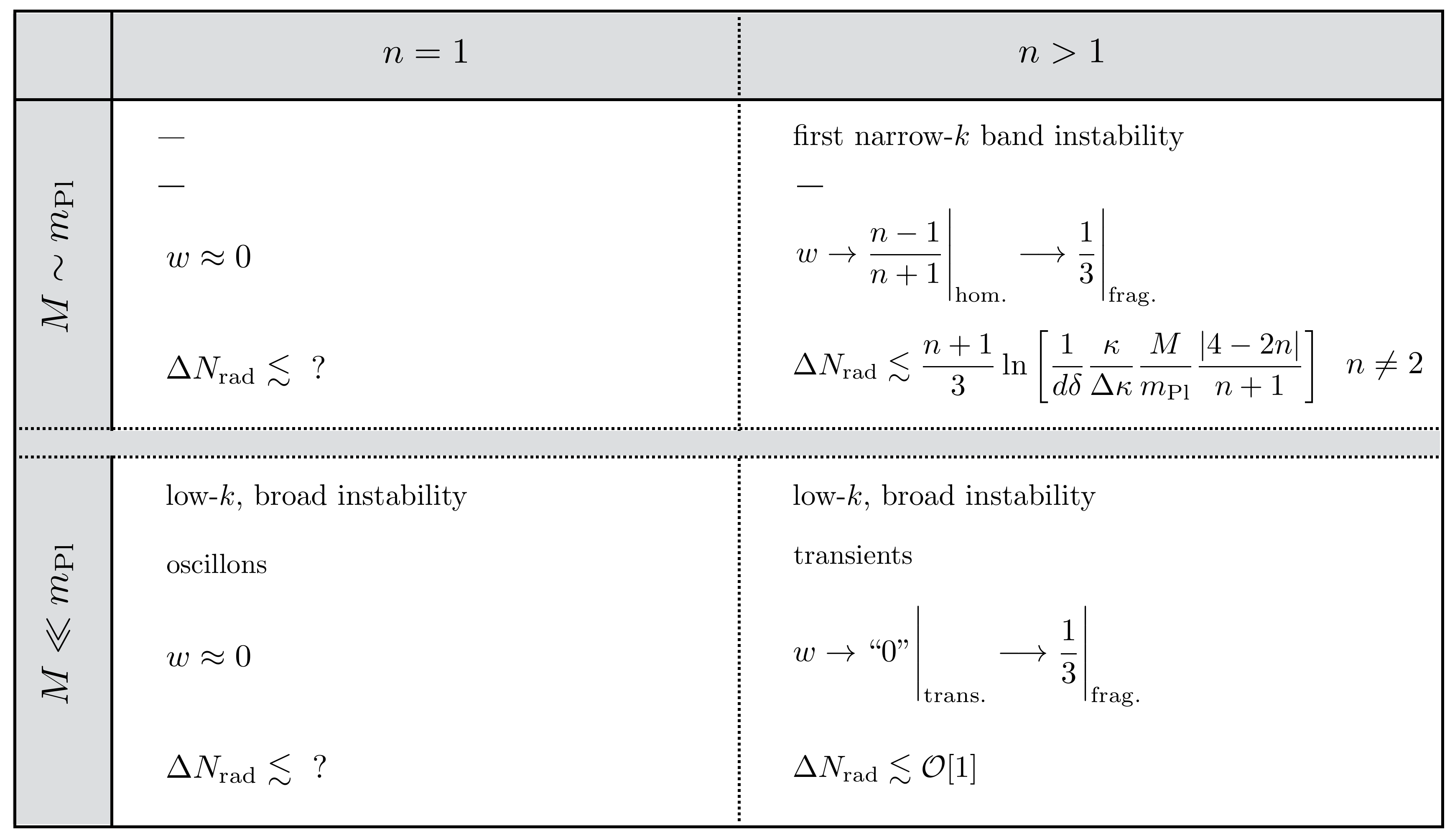}
   \hspace{.0in}
  \raisebox{.8\height}{\includegraphics[width=1.9in]{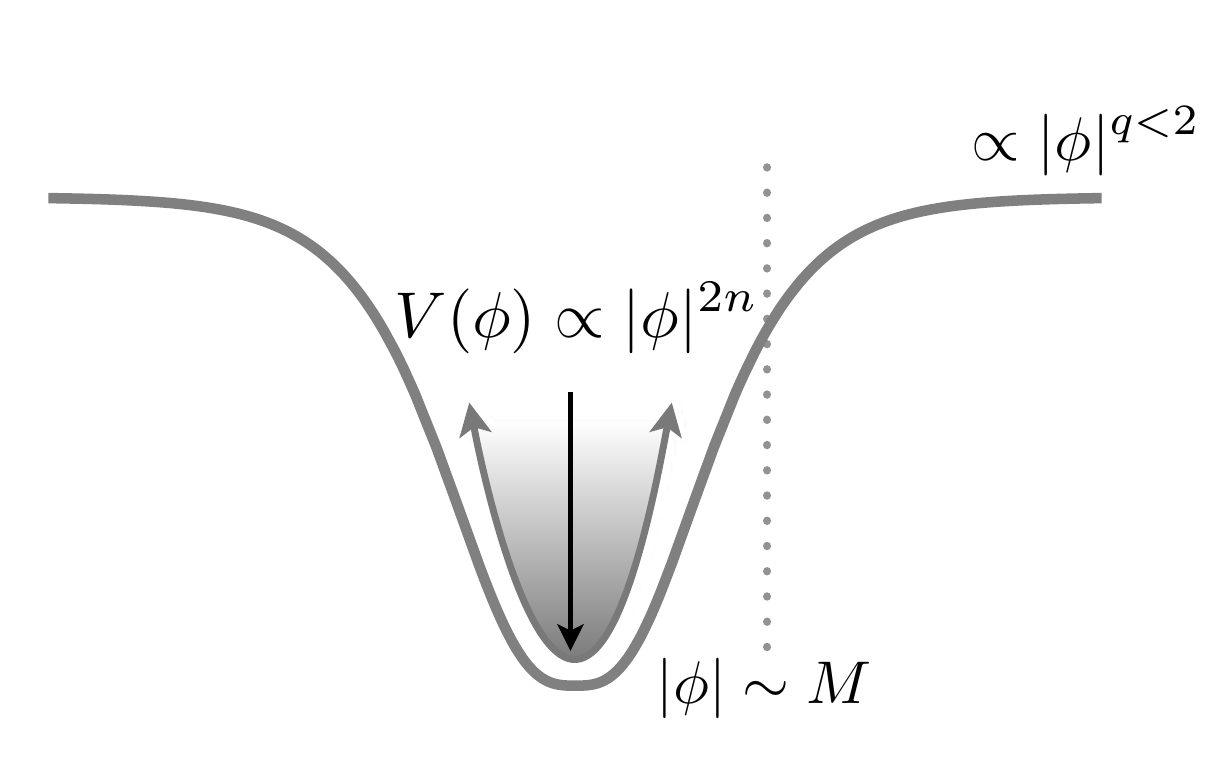}}
   \caption{A summary of the dynamics of the inflaton, the equation of state and the duration to radiation domination for different values of $n$ and $M$. Note that $n$ determines the shape of the minimum of the potential ($n=1$ is quadratic). Whereas $M$ determines the scales where the potential changes from $n\ge 1$ ``bowl" to flatter ``wings" . The field dynamics at the end of inflation are fast compared to Hubble when $M\ll \mpl$. The parameter $\delta\approx 0.126$, whereas $d$ and $\kappa/\Delta\kappa$ are calculable in terms of the parameters in the potential (see Fig. \ref{fig:1stband}). The above summary table is a coarse version of the results, see text for details and caveats.}
   \label{fig:ST}
\end{figure*}

We now turn to the UV-cutoff. Since in the cases of radiation domination, energy always cascades slowly towards high-$k$ modes it was also important to verify that the UV cut-off does not affect the inflaton dynamics and the evolution of the equation of state, see Fig. \eqref{fig:wtanh15alpha0check}. We found that with $256^3$ and $512^3$ boxes with the same IR cut-off, effects due to the finite UV resolution become important after $2$ {\it e}-folds after backreaction for $n\lesssim2$ and much later for $n>2$, i.e., always long after radiation domination is established. We always stop our simulations before reflections from the UV-cutoff become important.

In the case when $n>1$, $M\sim \mpl$, the UV-cutoff considerations at late times are identical to the $M\ll \mpl$ case. The rest are less stringent since no transients are formed.

Finally, we note that in terms of actual values of $w$ from simulations, by radiation domination we mean the moment when the equation of state approaches, $w_{\rm rad} = 1/3 \pm 0.03$. The $\pm10\%$ width makes the effects in inflationary observables due to numerical uncertainties $< 1\%$.
\subsection{Assumptions and Caveats}
Beyond the discussion of numerical considerations, for convenience, we collect some of the important theoretical assumptions underlying our work below.
\subsubsection{Metric Perturbations}
We included expansion of the universe self-consistently, but ignored perturbations in the metric.  While the early time dynamics are fast and driven by self-interactions of the field, we acknowledge that very long term dynamics can be affected by gravitational clustering, especially in the $n=1$ case. In the $n>1$ case (relevant for duration to radiation domination), the domination of the gradient and kinetic energies indicates that gravitational clustering would be difficult and unlikely. Moreover, for calculations of spatially averaged quantities like the equation of state and expansion history, metric potentials are unlikely to play a dominant role.

One might wonder whether the dense, localized oscillons and transients lead to gravitational potentials that might get large. In a future paper \cite{Lozanov:2017b} we {\it passively} calculate the metric fluctuations (Newtonian potentials) generated by these objects. We find find that the Newtonian potential remains quite small compared to unity. We re-iterate that the smallness of metric potentials just means we can ignore general relativity. Newtonian potentials can still change the detailed field dynamics on long time-scales (for $n=1$), but less so for $n>1$ case.

In the same future paper \cite{Lozanov:2017b}, we will also calculate gravitational waves generated by the fragmentation of the field.

\subsubsection{Inflaton Potential}
We assumed a rather generic shape of the inflaton potential. While a quadratic minimum ($n=1$) is the norm,  $n>1$ case (non-quadratic minimum) is unusual, especially for $n\ne 2$ (non-quartic case). One might also expect a mass to be generated by quantum effects, possibly $\sim H^2$. However, since the effective mass $m^2$ (see eq. \ref{eq:EffMass}) is larger than $H^2$ at the end of inflation, and redshifts slower than $H^2$, one might expect our results to apply for a long time.
\subsubsection{Couplings to other fields}
For most of the paper we ignore couplings to other fields. This is of course an approximation, couplings to other fields must exist so that the universe can be eventually populated by Standard Model fields. Given a theoretical prejudice of  $T_{\rm BBN}\ll E_{\rm inf}$, the couplings to other fields can be very, very small and still be consistent with all observations. As we have discussed, even if the couplings are larger, our calculated $\Delta N_{\rm rad}$ serves as an upper bound. 

We made an assumption that the fields coupled to the inflaton are relativistic. This is reasonable given the expected energy scale at the end of inflation. Nevertheless, long-lived massive fields would certainly change our conclusions since they might yield additional matter dominated period between the end of inflation and  BBN. Similarly, if there are other gravitationally coupled massive fields (moduli), they might also lead to an intermediate matter dominated period \cite{Kane:2015jia,Giblin:2017wlo}.

Finally, we assumed that the coupling does not introduce complex, coupled dynamics of the inflaton and the daughter fields. This assumption was made because if the dynamics of the daughter field-inflaton system is complex, one can get non-trivial equations of state which is neither matter dominated nor radiation dominated \cite{Dufaux:2006ee, Frolov:2008hy}.

\section{Conclusions}

\label{sec:Concl}

We investigated the post-inflationary dynamics of the inflaton field governed by observationally consistent potentials that are sufficiently flat away from the origin ($\phi\gg M$) and going as simple power-laws, $\propto |\phi|^{2n}$, near the origin ($|\phi|\ll M$). A summary of our results for the dynamics, and its implications are shown in Fig. \ref{fig:ST}.
\subsubsection{Self-resonance: Linear Analysis}
We analyzed in detail how self-interactions of the inflation drive the growth of spatial perturbations in the inflaton (self-resonance). We showed that for different parts of parameter space, the inflaton perturbations can grow either through  broad self-resonance ($M\ll\mpl$) or interestingly, a narrow self-resonance when $M\sim \mpl$ and $n>1$.\footnote{We note that for ease of exposition, we use $M\sim \mpl$ and $M\ll \mpl$ as binary cases. In reality, the boundary between the different regimes is weakly model dependent, and typically lies around $M\sim 0.1\mpl$.} We pointed out and explained the surprising generality and importance of particle production from the first narrow resonance band for all $n>1$. 

Based on our linear analysis of the instabilities in an expanding universe, we provided analytic estimates of the time taken for backreaction as a function of parameters of the model, and confirmed these estimates using numerical simulations. For $M\ll \mpl$ and $n\ge 1$, backreaction takes place with $\mathcal{O}[1]$ $e$-folds after the end of inflation. Whereas for $M\sim \mpl$, this backreaction can take several $e$-folds. For $n>1$, this backreaction time is given by eqs. \eqref{eq:DeltaNbr} and \eqref{eq:DeltaNbrQuart}. 
\subsubsection{Oscillons and Transients}
In the case with $M\ll \mpl$, backreaction is following by complete fragmentation of the inflaton into dense, localized objects, that are long-lived for $n=1$ and short-lived for $n>1$. The long-lived objects are oscillons and dominate the energy budget of the universe for many $e$-folds. Since they collectively behave as dust, they yield an equation of state $w\approx0$. The short-lived objects (transients) also behave as dust and can survive for up to $\mathcal{O}[1]$ {\it e}-folds of expansion. As the transients decay away, the universe becomes radiation dominated. 
\subsubsection{Slow Fragmentation}
For $M\sim \mpl$, slow and steady particle production from narrow resonance (only in the $n>1$ case) eventually leads to backreaction and fragmentation. In the power spectra of the field perturbations, we typically see a peak from the first narrow resonance band, which then multiplies to multiple peaks via re-scattering. 

We do not observe the formation of any transient objects in this case. For $n=1$, in absence of gravitational interactions, no growth of perturbations is seen and the condensate remains intact.
\subsubsection{Eq. of State and Duration to Radiation Domination}
When $n>1$, and for any $M$ we find that at late times the field evolves in a turbulent manner \cite{Khlebnikov:1996mc,Micha:2004bv,Micha:2002ey} and that the relativistic modes dominate the energy budget of the universe. This leads to a radiation-dominated period of expansion, $w\rightarrow w_{\rm{rad}}=1/3$. Note that this is a general result using the fully nonlinear dynamics of the fragmented field and differs from the expectation of the homogeneous field \cite{Turner:1983he}. 

We estimated the duration to radiation domination after inflation, $\Delta N_{\rm rad}$. For $M\ll\mpl$  (and $n>1$), the duration $\Delta N_{\rm rad}\lesssim \mathcal{O}[1]$ $e$-folds.
For $M\sim \mpl$, the duration to radiation domination and backreaction are similar: $\Delta N_{\rm br}\sim \Delta N_{\rm rad}$ and can be many $e$-folds. The estimate was verified by numerical simulations. Note that before reaching radiation domination, the condensate maintains the homogeneous equation of state $w_{\rm hom}=(n-1)/(n+1)$ in this case. 

For $n=1$, the equation of state remains close to $w\approx0$ for $M\ll\mpl$ (when oscillons form) and $M\sim \mpl$ (when the condensate does not fragment).
\subsubsection{Implications for Inflationary Observables}
We showed that our results for $\Delta N_{\rm{rad}}$, under the stated assumptions, lead to a substantial reduction in the uncertainties in the predictions for inflationary observables such as $n_{\rm{s}}$ and $r$ (see Fig. \ref{fig:Planck}).  The fact that a completely decoupled inflaton  attains a radiation-like equation of state (for $n>1$ case) allows us to put an upper bound on $\Delta N_{\rm rad}$, which is the key result for reducing the uncertainty in $r$ and $n_{\rm s}$. If we introduce interactions with other light fields (we assume perturbative decay to relativistic fields), $\Delta N_{\rm rad}$ can only decrease.

For the $n=1$ case (quadratic minimum), no matter what the value of $M$, the inflaton always ends up with a matter-like equation of state, $w=0$. To ensure the transition to a radiation-dominated state of expansion, required as an initial condition for primordial nucleosynthesis, we need to introduce couplings to other light fields \cite{Podolsky:2005bw}. We plan to return to this issue in a future work. 

\section{Acknowledgements}
We acknowledge and thank J. Braden, J. Chluba, E. Copeland,  A. Frolov,  M. Garcia, J. Kang, E. Komatsu, E. Lim and F. Takahashi for fruitful discussions.
The simulations were performed on the COSMOS Shared Memory system at DAMTP, operated by U. of Cambridge on behalf of the STFC DiRAC HPC Facility. We thank D. Sijacki for her generosity regarding the use of her computational resources under the Cambridge COSMOS Consortium. 
MA acknowledges support from the US Dept. of Energy Grant DE-SC0018216. This work was done in part at the Aspen Center for Physics, which is supported by National Science Foundation grant PHY-106629.

\bibliographystyle{apsrev}
\bibliography{bibSelfResonanceAfterInflation}{}

\appendix
\newpage
\section{Reheating Temperature}
\label{sec:RehT}
\begin{figure*}[t!] 
   \centering
   \hspace{-0.15in}   
   \raisebox{-0.5\height}{\includegraphics[height=1.73in]{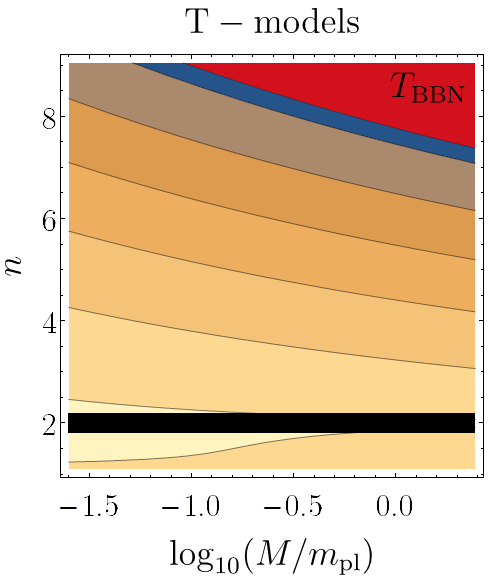}} 
   \hspace{-0.03in}
   \raisebox{-0.5\height}{\includegraphics[clip, trim=1.5cm 0cm 0cm 0cm, height=1.73in]{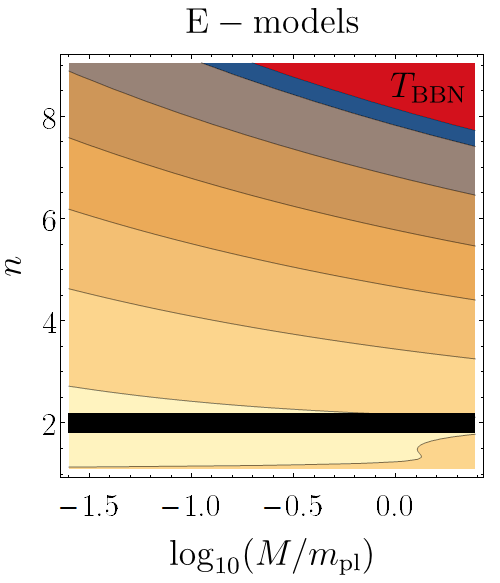}}
   \hspace{-0.03in}
   \raisebox{-0.5\height}{\includegraphics[clip, trim=1.5cm 0cm 0cm 0cm,height=1.73in]{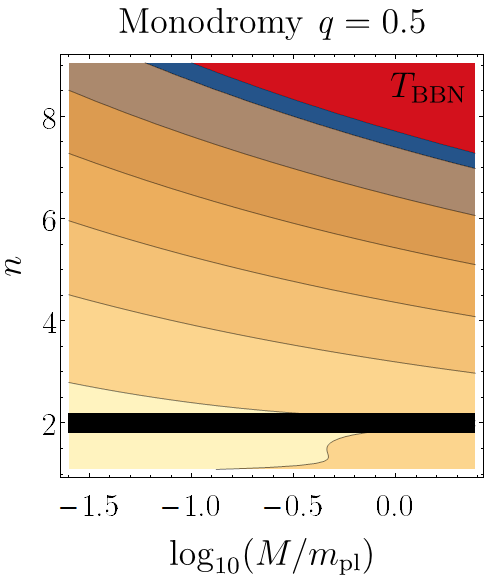}}
   \hspace{-0.03in}
   \raisebox{-0.5\height}{\includegraphics[clip, trim=1.5cm 0cm 0cm 0cm,height=1.73in]{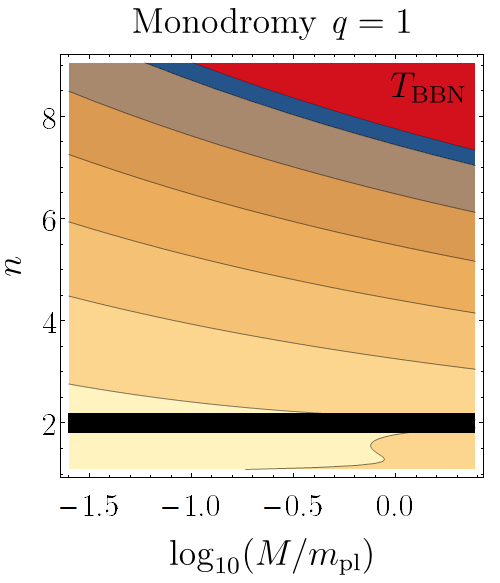}} 
   \hspace{-0.1in}
   \raisebox{-0.51\height}{\includegraphics[width=0.8in]{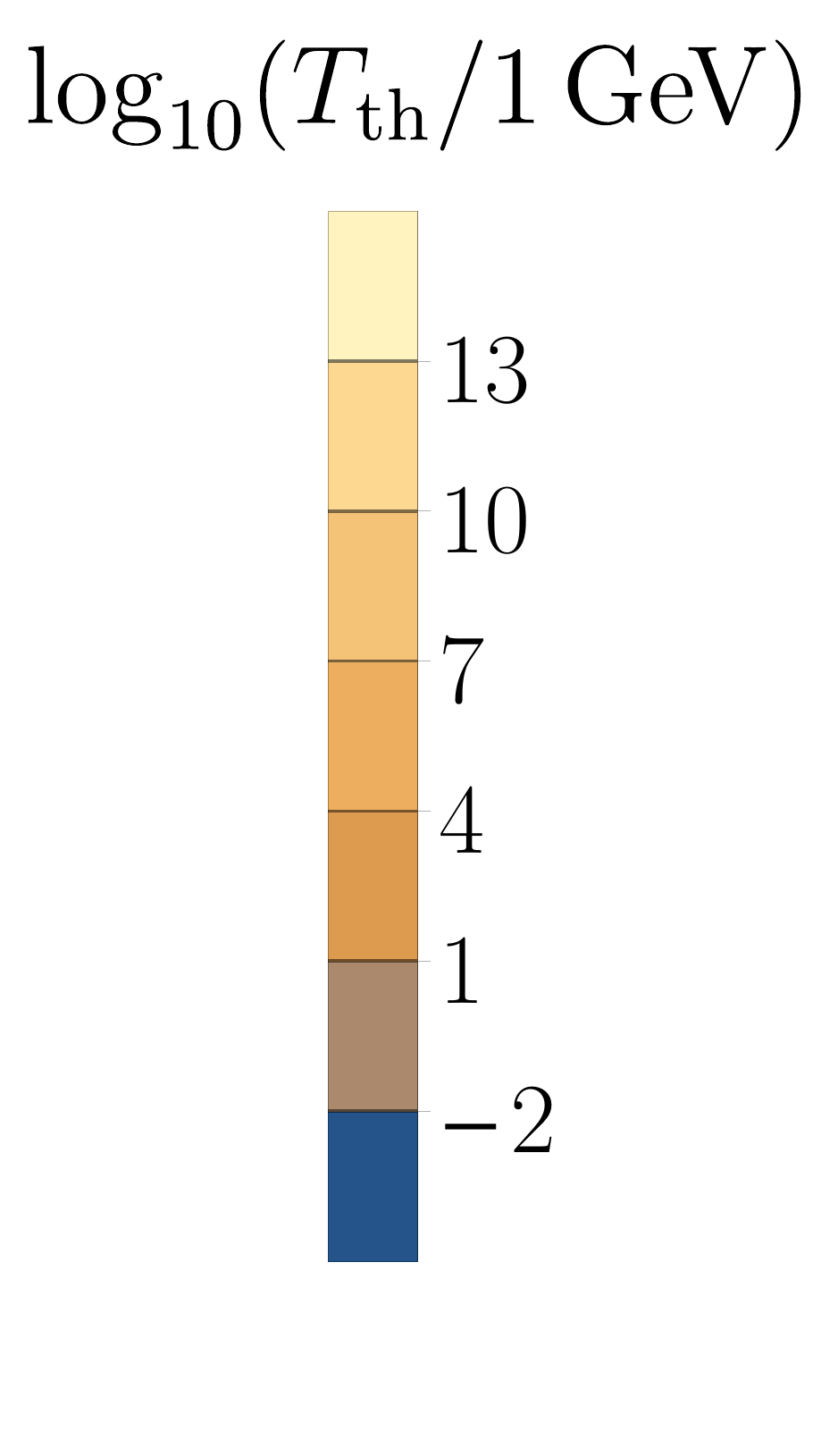}}\\
   \caption{The upper bound on the reheating temperature, $T_{\rm{th}}$, as a function of $M$ and $n$ in the limit when the self-couplings of the inflaton dominate over its couplings to other species of matter. The red areas in the upper right corners in each panel represent regions in parameter space for which the upper bound on $T_{\rm{th}}$ for an isolated inflaton is less than the lower bound ($1\,\rm{MeV}$) imposed by the big bang nucleosynthesis scenario. The horizontal black bands near $n=2$ remind us that we cannot arrive at an upper bound on the reheating temperature on the basis of the expansion history alone for a quartic potential. The monotonic decrease of $T_{\rm{th}}$ with $n$ and $M/\mpl$ for $n>2$ can be understood from the duration of the self-resonance due to the narrow instability band. As upon increasing any of the two parameters it takes longer for fragmentation to take place, the energy density would be redshifted to lower values at the time of backreaction, too, hence the observed dependence. }
   \label{fig:Treh}
\end{figure*}

The analysis in  Section \ref{sec:ObsImpl} emphasizes the importance of the expansion history in the determination of the inflationary observables like $n_{\rm{s}}$ and $r$. While it does not tell us much about the thermal history of the universe, it still allows us to calculate an upper bound on $T_{\rm{th}}$ for an `isolated' inflaton (i.e., an inflaton whose couplings to additional light fields can be neglected during $a_{\rm{end}}<a<a_{\rm{rad}}$) when $n>1$ and $n\neq2$. If we assume that soon after the approach to a radiation-dominated state of expansion the universe reaches thermal equilibrium, i.e., $a_{\rm{br}}\lesssim a_{\rm{rad}}\lesssim a_{\rm{th}}$, then
\Beq
\rho_{\rm{th}}=\frac{\pi^2}{30}g_{\rm{th}}T_{\rm{th}}^4\lesssim3\mpl^2H_{\rm{br}}^2\approx\frac{m_{\rm{br}}^2\phi_{\rm{br}}^2}{2n}\,,
\Eeq
whence, for the case when the first narrow instability band plays a major role,
\Beq
\label{eq:TrehUpperBound}
T_{\rm{th}}\lesssim \frac{\sqrt{b\mpl M}}{\left(\pi^2g_{\rm{th}}/30\right)^{1/4}}\left[\frac{\mpl}{M}\frac{\Delta\kappa}{\kappa}d\delta\frac{n+1}{|4-2n|}\right]^{n/2}.
\Eeq
The new parameter appearing under the square root is defined as
\Beq
\label{eq:Definitionb}
b\equiv\frac{m/\sqrt{2n}}{\mpl(\bar{\phi}/M)^{n-1}}\,.
\Eeq
It follows that
\Beq
b&\equiv\begin{cases}
                    \sqrt{3\pi^2A_{\rm{s}}/N_{\star}^2}\,&{\text{T}}\,,\\
                    2\sqrt{3\pi^2A_{\rm{s}}/N_{\star}^2}\,  & \text{E} \,, \\
                    \sqrt{\frac{6q\pi^2A_{\rm{s}}/n}{\left(2qN_{\star}\mpl^2/M^2\right)^{q/4}}}\left(q\mpl/M\right)^2\,&{\text{Mon}}\,.\\
                   \end{cases}
\Eeq
This upper bound on the reheating temperature holds when the thermal equilibrium is reached after the end of self-resonance when $w=1/3$. If the inflaton is coupled strongly enough to additional light fields and $w=1/3$ is reached earlier, due to the inflaton decays into other relativistic degrees of freedom, then the reheating temperature can be even higher, with the upper bound set by the energy scale of inflation. 

On the other hand, BBN provides a lower bound: $T_{\rm{th}}>T_{\rm{BBN}}\sim 1\,\rm{MeV}$. If it saturates the upper one in eq. \eqref{eq:TrehUpperBound}, then we know that significant couplings to additional light fields have to be introduced explicitly, to make sure that energy is transfered early enough from the condensate to relativistic species of matter and the reheating temperature can be raised to the observationally allowed region. In Fig. \ref{fig:Treh} we give $T_{\rm{th}}$ from eq. \eqref{eq:TrehUpperBound} along with the constraints from BBN. They become important only for large $n$ and $M/\mpl$. Qualitatively, this can be understood from the nature of the self-resonance. Since $\Delta N_{\rm{br}}$ increases monotonically with $n$ and/or $M/\mpl$ for $n>2$, see eq. \eqref{eq:DeltaNbr}, then the energy scale at backreaction (and the upper bound on $T_{\rm{th}}$) will decrease because of the prolonged oscillatory period during which the condensate's energy is redshifted. These dependences can be seen directly in the square brackets in eq. \eqref{eq:TrehUpperBound}. Note that the additional $M$ term appearing under the square root in front of the square brackets comes from the energy scale of inflation and has the opposite effect, i.e., the higher $M$ is the greater the energy scale at the end of inflation is and hence the greater $T_{\rm{th}}$ is. However, this effect is not sufficiently strong and the overall dependence on $M$ is determined by the duration of the self-resonance.

\end{document}